\documentclass[12pt]{article}
\pdfoutput=1
\usepackage{amssymb,amsmath}
\usepackage{hyperref}
\usepackage{comment}
\usepackage[dvipdfmx]{graphicx}
\usepackage{bm}
\usepackage{color}
\usepackage{here}
\usepackage{enumerate}
\usepackage{here}
\usepackage{subfigure} 
\newcommand{\Zb}{\mathbb{Z}}

\newcommand{\Acal}{\mathcal{A}}
\newcommand{\Bcal}{\mathcal{B}}
\newcommand{\Ccal}{\mathcal{C}}
\newcommand{\Dcal}{\mathcal{D}}
\newcommand{\Mcal}{\mathcal{M}}
\newcommand{\Ncal}{\mathcal{N}}

\DeclareMathOperator*{\Tr}{{\rm Tr}}
\DeclareMathOperator*{\tr}{{\rm tr}}

\newcommand{\II}{\mathbb{II}}
\newcommand{\ba}{\mathbf{a}}
\newcommand{\bbf}{\mathbf{f}}
\newcommand{\br}{\mathbf{r}}
\newcommand{\bR}{\mathbf{R}}
\newcommand{\bs}{\mathbf{s}}
\newcommand{\bx}{\mathbf{x}}

\definecolor{mygreen}{rgb}{0,0.714,0.286}

\numberwithin{equation}{section}

\begin{document}

%%%%%%%%%%%%%%%%%%%%%%%%%%%%%%%%%%%%%%%%%%%%
\thispagestyle{empty}
\begin{flushright}
DCPT-21/07
\\
%%%%%%%%%%%%%%%%%%%%%%%%%%%%%%%%%%%%%%%%%%%%%%%%%%%%%%%%%%%%%%%%%
%input \\
%%%%%%%%%%%%%%%%%%%%%%%%%%%%%%%%%%%%%%%%%%%%%%%%%%%%%%%%%%%%%%%%%

\end{flushright}
\vskip2cm
\begin{center}
{\Large \bf Seiberg-like dualities for orthogonal and symplectic \
%\vskip0.5cm
3d $\Ncal = 2$ gauge theories with boundaries\\\
%{\Large \bf Seiberg-like dualities with boundaries   \
%\vskip0.5cm
%for orthogonal and symplectic gauge theories\\\
%\vskip0.3cm 
%and 
%\vskip0.5cm 
%Supersymmetric Indices
%\vskip0.25cm
}

\vskip1.5cm
 Tadashi Okazaki\footnote{tadashi.okazaki@durham.ac.uk}
 \\
 \bigskip
 and 
 \\
\bigskip
 Douglas J Smith\footnote{douglas.smith@durham.ac.uk}

\bigskip
{\it Department of Mathematical Sciences, Durham University,\\
Upper Mountjoy Campus, Stockton Road, Durham DH1 3LE, UK}

\end{center}

%%%%%%%%%%%%%%%%%%%%%%%%%%%%%%%%%%%%%%%%%%%%
\vskip1.5cm
\begin{abstract}
We propose dualities of $\mathcal{N}=(0,2)$ supersymmetric boundary conditions 
for 3d $\mathcal{N}=2$ gauge theories with orthogonal and symplectic gauge groups. 
We show that the boundary 't Hooft anomalies and half-indices perfectly match 
for each pair of the proposed dual boundary conditions. 
\end{abstract}
%%%%%%%%%%%%%%%%%%%%%%%%%%%%%%%%%%%%%%%%%%

%%%%%%%%%%%%%%%%%%

\newpage
\tableofcontents
%%%%%%%%%%%%%%%%%%%%%%%%%%

%%%%%%%%%%%%%%%%%%%%%%%%%%%%%%%%%%%%%%%%%%%%%
%%%%%%%%%%%%%%%%%%%%%%%%%%%%%%%%%%%%%%%%%%%%%
\section{Introduction and conclusions}
\label{sec_intro_con}
%%%%%%%%%%%%%%%%%%%%%%%%%%%%%%%%%%%%%%%%%%%%%
%%%%%%%%%%%%%%%%%%%%%%%%%%%%%%%%%%%%%%%%%%%%%
%backgrounds

%Seiberg dualities of 4d N=1
Seiberg duality \cite{Seiberg:1994pq} of 4d $\mathcal{N}=1$ gauge theories 
is the duality in the IR which relates an `electric' $SU(N_c)$ gauge theories with $N_f$ flavours of quarks and antiquarks 
to a `magnetic' $SU(N_f-N_c)$ gauge theories with $N_f$ flavours of quarks and antiquarks together with 
a gauge singlet field coupled through a superpotential 
(See e.g. \cite{Kutasov:1995ve,Kutasov:1995np,Intriligator:1995id,Intriligator:1995au,Intriligator:1995ne,Intriligator:1995ax,Intriligator:1995er,Csaki:1996eu} for various generalisations.). 

%Seiberg-like dualities of 3d N=2
There exists a three-dimensional analogue, \textit{aka} Seiberg-like duality in 3d $\mathcal{N}=2$ gauge theories. 
For the unitary gauge group, the IR duality \cite{Aharony:1997gp} relates $\mathcal{N}=2$ gauge theory with gauge group $U(N_c)$, 
$N_f$ fundamental chiral multiplets and $N_f$ anti-fundamental chiral multiplets 
to $\mathcal{N}=2$ gauge theory with gauge group $U(N_f-N_c)$, $N_f$ fundamental chiral multiplets and $N_f$ anti-fundamental chiral multiplets 
as well as additional gauge singlet chiral multiplets and a superpotential. 
For the symplectic gauge group, the IR duality \cite{Aharony:1997gp} relates $USp(2N_c)$ gauge theory with $2N_f$ fundamental chiral multiplets 
to $USp(2N_f-2N_c-2)$ gauge theory with $2N_f$ fundamental chiral multiplets along with gauge singlets and a superpotential. 
The dualities are generalised to the theories with a Chern-Simons term \cite{Giveon:2008zn,Nii:2020ikd}, 
which can be achieved by adding real masses to flavours, leading to an effective Chern-Simons term at low energy. 
The 3d Seiberg-like dualities are extended to 
the $SU(N_c)$ gauge theories \cite{Aharony:2013dha, Park:2013wta, Aharony:2014uya, Hwang:2015wna,Nii:2019qdx, Nii:2020xgd,Amariti:2020xqm}, 
the orthogonal gauge theories \cite{Kapustin:2011gh, Benini:2011mf, Hwang:2011ht, Aharony:2011ci, Aharony:2013kma,Nii:2019wjz, Nii:2020eui}, 
the $G_2$ gauge theory \cite{Nii:2019wjz}, 
the quiver gauge theories \cite{Amariti:2017gsm,Benvenuti:2020wpc}, 
the gauge theories with arbitrary numbers of fundamental and anti-fundamental matter fields obeying the $\mathbb{Z}_2$ anomaly constraint $k+\frac{N_f-N_a}{2}\in \mathbb{Z}$ \cite{Benini:2011mf}, 
with adjoint matter fields \cite{Niarchos:2008jb,Kim:2013cma,Park:2013wta,Nii:2014jsa,Hwang:2015wna,Hwang:2018uyj,Nii:2019qdx,Amariti:2020xqm}, 
with other tensor matter fields \cite{Kapustin:2011vz, Amariti:2014lla,Csaki:2014cwa,Amariti:2018wht} 
and with a monopole superpotential \cite{Benini:2017dud, Amariti:2018gdc}. 

%dualities of b.c. for 3d N=2
In the presence of a boundary, the dualities become more elaborate
as the bulk fields are subject to certain boundary conditions and they can further couple to additional boundary degrees of freedom. 
The half-BPS boundary conditions preserving $\mathcal{N}=(0,2)$ supersymmetry in 3d $\mathcal{N}=2$ gauge theories have been studied in 
\cite{Gadde:2013wq, Okazaki:2013kaa, Gadde:2013sca, Yoshida:2014ssa, Dimofte:2017tpi, Brunner:2019qyf, Costello:2020ndc, Sugiyama:2020uqh}. 
Simple dualities of $\mathcal{N}=(0,2)$ half-BPS boundary conditions of 3d $\mathcal{N}=2$ gauge theories 
for Abelian gauge theories were proposed in \cite{Okazaki:2013kaa} 
and various dualities of $\mathcal{N}=(0,2)$ half-BPS boundary conditions for 3d $\mathcal{N}=2$ gauge theories with unitary gauge groups were proposed in \cite{Dimofte:2017tpi}. 
\footnote{
See \cite{Costello:2018fnz, Okazaki:2019bok} for the study of dualities of $\mathcal{N}=(0,4)$ boundary conditions 
and \cite{Bullimore:2016nji, Okazaki:2020lfy} for $\mathcal{N}=(2,2)$ boundary conditions in 3d $\mathcal{N}=4$ supersymmetric gauge theories. 
}

Other approaches to describing Chern-Simons theories with boundary have focussed on the relation to CFTs\footnote{In fact, Chern-Simons theories on compact manifolds without boundary are also related to CFTs \cite{Witten:1988hf, Moore:1989yh}.} and specifically WZW models \cite{Witten:1988hf, Moore:1989yh, Elitzur:1989nr}. This has been used to infer possible boundary matter in many examples including CS with various amounts of supersymmetry, including ABJM models \cite{Berman:2009kj, Chu:2009ms, Faizal:2011cd, Niarchos:2015lla, Okazaki:2015fiq, Faizal:2016skd}, while boundary conditions have been analysed for multiple M2-branes in \cite{Berman:2009xd, Chu:2009iv, Chu:2010rb}. In the context of Seiberg-like dualities, several level-rank dualities of gauged WZW models have been proposed \cite{Armoni:2015jsa, Ireson:2015lda} using the relation to Chern-Simons theories and their dualities.

In this paper, we study $\mathcal{N}=(0,2)$ half-BPS boundary conditions for 3d $\mathcal{N}=2$ gauge theories with 
symplectic and orthogonal gauge groups and propose dualities of these boundary conditions. 
We support our claims by computing boundary 't Hooft anomalies and supersymmetric half-indices 
which perfectly match for the proposed dual pairs of boundary conditions. 
For orthogonal gauge groups the global structure of the group is important
and has been discussed for 3d Chern-Simons theories in the context of
Seiberg-like duality \cite{Aharony:2013kma} and level-rank duality
\cite{Cordova:2017vab}. We show that with boundary conditions there is a similar
set of dualities relating the following groups under Seiberg-like duality:
$SO \leftrightarrow SO$,
$O_{+} \leftrightarrow O_{+}$, $O_{-} \leftrightarrow Spin$ and
$Pin_{\pm} \leftrightarrow Pin_{\pm}$.

The organisation of this article is as follows. 
In section {\ref{Sp_sec}} we discuss the duality we suggest for 
$\mathcal{N}=(0,2)$ half-BPS boundary conditions in 3d $\mathcal{N}=2$ $USp(2N_c)$ gauge theories and test it by computing supersymmetric half-indices. 
In section {\ref{SO_sec}} we propose the dualities of 
$\mathcal{N}=(0,2)$ half-BPS boundary conditions for 3d $\mathcal{N}=2$ $SO(N_c)$ gauge theories. 
We find more extensive identities of half-indices parameterised by two parameters $\zeta$ and $\chi$ corresponding to the global $\mathbb{Z}_2$ symmetries \cite{Aharony:2013kma}. 
In section {\ref{O_sec}} we discuss the dualities of 
$\mathcal{N}=(0,2)$ half-BPS boundary conditions for other orthogonal gauge groups. 
In Appendix \ref{app_GenAnomInd} we present the notation and convention of our tools, including boundary anomalies and supersymmetric indices. 
In Appendix \ref{app_expansion} we show numerical results obtained from Mathematica. 

%%%%%%%%%%%%%%%%%%%%%%%%%%%%%%%%%%%%%%%%%%%
%%%%%%%%%%%%%%%%%%%%%%%%%%%%%%%%%%%%%%%%%%%
\section{$USp(2N_c)$ gauge theories}
\label{Sp_sec}
%%%%%%%%%%%%%%%%%%%%%%%%%%%%%%%%%%%%%%%%%%%
%%%%%%%%%%%%%%%%%%%%%%%%%%%%%%%%%%%%%%%%%%%

%Quantum dynamics of symplectic gauge theory
The quantum dynamics of 3d $\mathcal{N}=2$ supersymmetric gauge theories 
with gauge groups $G=USp(2N_c)$ has been studied in \cite{Aharony:1997bx, Karch:1997ux, Aharony:1997gp}. 
$USp(2N_c)$ is the subgroup of $SU(2N_c)$ that 
keeps an antisymmetric tensor $J^{ab}=(\mathbb{I}_{N_c}\otimes i\sigma_2)^{ab}$ invariant. 
Since the antisymmetric tensor $\epsilon^{i_1\cdots i_{2N_c}}$ breaks up into 
sums of products of the $J^{ab}$, the $USp(2N_c)$ gauge theory has no baryons. 
Also there is no topological current as the gauge group is simple.

%duality
For the symplectic gauge groups, the IR dualities are proposed in \cite{Aharony:1997gp, Karch:1997ux}. 
We will refer to these dual theories as theory A and theory B, but they are also
commonly referred to as the electric and the magnetic theories: 

\begin{itemize}

\item Theory A: $USp(2N_c)$ gauge theory with $2N_f$ chiral multiplets $Q$ in the fundamental representation. 
It contains gauge invariant operators as the meson $M=QQ$ and the monopole operator $V$. 

\item Theory B: $USp(2(N_f-N_c-1))$ gauge theory with $2N_f$ chiral multiplets $q$ in the fundamental representation, 
$N_f (2N_f-1)$ neutral chiral multiplets $M$ in the rank-2 antisymmetric 
representation of $SU(2N_f)$ and a chiral multiplet $V$ 
which has the superpotential 
\begin{align}
\mathcal{W}&=Mqq+V\tilde{V}
\end{align}
where $\tilde{V}$ is the monopole operator.

\end{itemize}

The charges of the chiral multiplets are given by
\begin{align}
\label{sp2N_charge}
\begin{array}{c|c|c|c|c|c|c|c}
&G=USp(2N_c)&\tilde{G}=USp(2(N_f-N_c-1))&SU(2N_f)&U(1)_a&U(1)_R \\ \hline 
Q&{\bf N_c}&{\bf 1}&{\bf 2N_f}&+&r \\ \hline
q&{\bf 1}&{\bf 2(N_f -N_c-1)}&{\bf \overline{2N_f}}&-&1-r \\
M&{\bf 1}&{\bf 1}&{\bf N_f(2N_f-1)}&2&2r \\
V&{\bf 1}&{\bf 1}&{\bf 1}&-2N_f&2N_f -2N_c-2r N_f \\
\end{array}
\end{align}
The quantum numbers crucially depend on the rank of gauge group and the number of flavours.

%%%%%%%%%%%%%%%%%%%%%%%%%%%%%%%%%%%%%%%%%%%
\subsection{$\mathcal{N}=(0,2)$ half-BPS boundary conditions}
%%%%%%%%%%%%%%%%%%%%%%%%%%%%%%%%%%%%%%%%%%%
We introduce a boundary to the 3d $\Ncal = 2$ theories in such a way as to
preserve $\Ncal = (0,2)$ supersymmetry in 2d. For the 3d bulk fields we must
impose boundary conditions, and basic $\mathcal{N}=(0,2)$ boundary conditions \cite{Okazaki:2013kaa} impose either Neumann or Dirichlet boundary conditions on chiral multiplet 
(which we denote by $\mathrm{N}$ or by $\mathrm{D}$)
and either Neumann or Dirichlet boundary conditions on vector multiplet (VM)
(which we denote by $\mathcal{N}$ or by $\mathcal{D}$) 
\footnote{One can also consider singular boundary conditions. 
Although it is intriguing to reduce them to the B-type boundary conditions found in \cite{Okazaki:2020pbb}, we leave it to future work.}
which is compatible with the decomposition of the 3d $\Ncal = 2$ supermultiplets into 2d
$\Ncal = (0,2)$ supermultiplets on the boundary. 
The choice of Neumann or
Dirichlet boundary conditions projects out specific $\Ncal = (0,2)$ supermultiplets.

On the 2d boundary we can have anomalies and these give two important
constraints. If we have a gauge symmetry, which will be the case for Neumann
boundary conditions for bulk vector multiplets, we require gauge anomaly
cancellation. On the other hand, for global symmetries, including gauge symmetry
broken by Dirichlet boundary conditions for the vector multiplet, we do not
require cancellation of the anomalies. Instead, we get  constraint on any
proposed duality that the 't Hooft anomalies must match, i.e.\ the anomaly
polynomials for the two theories must be equal.

The method to calculate the 2d boundary anomaly polynomial was given in
\cite{Dimofte:2017tpi}. There are three types of contribution to the anomalies:
3d bulk supermultiplets projected onto the boundary with Neumann of Dirichlet
boundary conditions; 2d supermultiplets introduced on the boundary;
background Chern-Simons or FI terms. We summarise the general results presented
in \cite{Dimofte:2017tpi} in appendix~\ref{Gen2dAnom}. The specific results
for the fields in the $USp(2N_c)-[2N_f]$ theory A and
$USp(2(N_c - N_f - 1))-[2N_f]$ theory B we consider are
\begin{align}
\Acal^{\textrm{VM}} = & -(N_c + 1) \Tr (\bs^2) - \frac{N_c}{2}(2N_c + 1) \br^2, \\
\Acal^{Q} = & N_f \Tr (\bs^2) + N_c \Tr (\bx^2) + 2N_c N_f(\ba + (r-1) \br)^2, \\
\Acal^{q} = & N_f \Tr (\tilde{\bs}^2) + (N_f - N_c - 1) \Tr (\bx^2) + 2(N_f - N_c - 1) N_f(-\ba - r \br)^2, \\
\Acal^{M} = & (N_f - 1) \Tr (\bx^2) + \frac{1}{2} N_f (2N_f - 1)(2 \ba + (2r-1) \br)^2, \\
\Acal^{V} = & \frac{1}{2} (-2N_f \ba + (2(1-r)N_f - 2N_c - 1) \br)^2
\end{align}
for Dirichlet boundary conditions. For Neumann boundary conditions the
contributions are the same but with opposite sign. The notation $\bs$
($\tilde{\bs}$) refers to the gauge field strength in theory A (B), $\bx$ to
the field strength for the global $SU(2N_f)$ flavour symmetry, and $\ba$ and
$\br$ to the field strengths for the global $U(1)_a$ and $U(1)_R$ symmetries.

We also need 2d boundary matter and the only multiplet required for the
examples we consider is a $USp(2N_c) \times USp(2(N_f - N_c - 1))$
bifundamental Fermi multiplet, $\Psi$. Here $USp(2N_c)$ is the 2d gauge group
inherited from the bulk vector multiplet with Neumann boundary conditions
while $USp(2(N_f - N_c - 1))$ is a global flavour symmetry which is identified
with the dual gauge group broken to a global symmetry by the Dirichlet boundary
conditions for the dual vector multiplet. This gives anomaly contribution
\begin{align}
\Acal^{\Psi} = & (N_f - N_c - 1)  \Tr (\bs^2) + N_c \Tr (\tilde{\bs}^2) \; .
\end{align}
Note that this contribution is actually only half of what might be expected for
such a Fermi. However, this is precisely the contribution required for anomaly
cancellation and we interpret this as due to a reality condition on the Fermi.
We will comment again on this when discussing the Fermi contribution to the
half-index.

For now we consider the 3d dualities with vanishing Chern-Simons level, in which
case we have no background Chern-Simons terms.

The boundary conditions we consider are $(\Ncal, \mathrm{N})$ in theory A, referring to
the choice of Neumann boundary conditions for $(\textrm{VM}, Q)$ together with
$(\Dcal, \mathrm{D, N, D})$ in theory B for $(\textrm{VM}, q, M, V)$. With this choice we need to
cancel the gauge anomaly in theory A which we can do by including the
bifundamental Fermi in theory A. This also leads to anomaly matching with
theory B without any further 2d matter. In particular we have
\begin{align}
\Acal^{A, \textrm{bulk}}_{\Ncal, \mathrm{N}} = & -(N_f - N_c - 1)\Tr(\bs^2) - N_c \Tr(\bx^2) - \frac{N_c}{2}(4N_f(1-r)^2 - 2N_c - 1)\br^2 \nonumber \\
 & + 4 (1-r) N_c N_f \ba\br - 2 N_c N_f \ba^2, \\
\Acal^{A, \textrm{bdry}}_{\Ncal, \mathrm{N}} = & \Acal^{\Psi} = (N_f - N_c - 1)\Tr(\bs^2) + N_c \Tr(\tilde{\bs}^2), \\
\Acal^{B, \textrm{bulk}}_{\Dcal, \mathrm{D, N, D}} = & N_c \Tr(\tilde{\bs}^2) - N_c \Tr(\bx^2) - \frac{N_c}{2}
(4N_f(1-r)^2 - 2N_c - 1)\br^2 \nonumber \\
 & + 4 (1-r) N_c N_f \ba\br - 2 N_c N_f \ba^2 \label{AnomBUSp}
\end{align}
and it is easy to see that all dependence on $\bs$ is cancelled and
\begin{equation}
\Acal^{A}_{\Ncal, \mathrm{N}} = \Acal^{A, \textrm{bulk}}_{\Ncal, \mathrm{N}} + \Acal^{A, \textrm{bdry}}_{\Ncal, \mathrm{N}} = \Acal^{B, \textrm{bulk}}_{\Dcal, \mathrm{D, N, D}} = \Acal^{B}_{\Dcal, \mathrm{D, N, D}} \; .
\end{equation}

Taking the opposite choice of boundary conditions for all fields we also get
anomaly matching if we add the bifundamental Fermi to theory B instead of
theory A. This also cancels the gauge anomaly in theory B.

We therefore have the following proposed dualities
\begin{itemize}
\item Theory A with $(\Ncal, \mathrm{N})$ boundary conditions plus Fermi $\Psi$ $\leftrightarrow$
 Theory B with $(\Dcal, \mathrm{D, N, D})$ boundary conditions.
\item Theory A with $(\Dcal, \mathrm{D})$ boundary conditions $\leftrightarrow$
Theory B with $(\Ncal, \mathrm{N, D, N})$ boundary conditions plus Fermi $\Psi$.
\end{itemize}

For these proposed dualities we have only considered two-dimensional charged
Fermi multiplets which cancel the gauge anomaly as minimal choices of boundary
degrees of freedom. However, it would be interesting to allow
%investigate by taking
two-dimensional vector and charged chiral multiplets to investigate other dual boundary
conditions. We leave this to future work.

%%%%%%%%%%%%%%%%%%%%%%%%%%%%%%%%%%%%%%%%%%%
\subsection{Supersymmetric indices}
%%%%%%%%%%%%%%%%%%%%%%%%%%%%%%%%%%%%%%%%%%%
The supersymmetric ``full-index'' of 3d $\mathcal{N}=2$ supersymmetric theories 
can be defined as a trace over the states on $S^2\times \mathbb{R}$ \cite{Bhattacharya:2008zy, Kim:2009wb, Imamura:2011su, Kapustin:2011jm, Dimofte:2011py}. 
It can be evaluated as a partition function on $S^2\times S^1$ from the UV description  
and the UV formula was obtained in \cite{Kim:2009wb} for the theory with canonical conformal dimension  
and in \cite{Imamura:2011su} for the theory with any conformal dimension via
the localisation technique. 

The supersymmetric ``half-index'' of 3d $\mathcal{N}=2$ supersymmetric theory $\mathcal{T}$ obeying the $\mathcal{N}=(0,2)$ half-BPS boundary condition $\mathcal{B}$
can similarly be defined as a trace over the states that correspond to half-BPS local operators on the boundary
\cite{Gadde:2013wq, Gadde:2013sca, Yoshida:2014ssa, Dimofte:2017tpi}
\begin{align}
\mathbb{II}_{\mathcal{B}}^{\mathcal{T}}&={\Tr}_{\mathrm{Op}} (-1)^F q^{J+\frac{R}{2}} x^{f}
\end{align}
where $F$ is the Fermion number operator,
$J$ is the generator of the $Spin(2)\cong U(1)_J$ rotational symmetry of the two-dimensional plane where the boundary local operators are supported,
$R$ is the R-charge
and $f$ is the Cartan generators of the other global symmetry group.

One can compute the half-index as a partition function on $HS^2\times S^1$ 
which encodes the $\mathcal{N}=(0,2)$ half-BPS boundary conditions on $\partial(HS^2\times S^1)=S^1\times S^1$ 
where $HS^2$ is a hemisphere. 
The UV formula for the half-index of the Neumann boundary condition for the gauge multiplet was derived in \cite{Gadde:2013wq, Gadde:2013sca, Yoshida:2014ssa} 
and the formula for the Dirichlet boundary conditions for the gauge multiplet was proposed in \cite{Dimofte:2017tpi}. 
The half-index has applications in the context of quantum K-theory \cite{Jockers:2021omw} and can also realise the holomorphic block \cite{Pasquetti:2011fj, Beem:2012mb}, 
the $q$-series 3-manifold invariant $\hat{Z}$ \cite{Gukov:2016gkn} which has a number of applications to topology and number theory 
(see e.g. \cite{Gukov:2017kmk, Cheng:2018vpl, Gukov:2019mnk, Chun:2019mal, Chung:2019khu, Ekholm:2020lqy}) and
the 4-manifold invariant \cite{Vafa:1994tf} (see e.g. \cite{Feigin:2018bkf}).

We summarise the general results here for the full-indices and half-indices,
using notation $USp(2N_c)-[2N_f]^A$ for theory A
with gauge group $USp(2N_c)$ and $2N_f$ fundamental chiral multiplets, and
$USp(2\tilde{N}_c)-[2N_f]^B$ for the dual theory B.

%%%%%%%%%%%%%%%%%%%%%%%%%%%%%%%%%%%%%%%%%%%
\subsubsection{Full-index}
%%%%%%%%%%%%%%%%%%%%%%%%%%%%%%%%%%%%%%%%%%%
The test of the IR duality for the symplectic gauge group 
were performed by computing supersymmetric indices in \cite{Bashkirov:2011vy,Hwang:2011ht}. 

The full-index of theory A takes the form
\begin{align}
\label{sp2N_i}
&
I^{USp(2N_c)-[2N_f]^A}
\nonumber\\
&=\frac{1}{2^{N_c} N_c !}
\sum_{m_1, \cdots, m_{N_c}\in \mathbb{Z}}
\oint \prod_{i=1}^{N_c}
\frac{ds_i}{2\pi is_i}
(1-q^{|m_i|} s_{i}^{\pm 2})
\prod_{1\le i<j\le N_c}
(1-q^{\frac{|m_i \pm m_j|}{2}} s_i s_j^{\pm})
(1-q^{\frac{|-m_i \pm m_j|}{2}} s_i^{-1} s_j^{\pm})
\nonumber\\
&\times 
\prod_{i=1}^{N_c}
\prod_{\alpha=1}^{2N_f}
\frac{
(q^{1-\frac{r}{2}+\frac{|m_i|}{2}} a^{-1}s_{i}^{\pm}x_{\alpha}^{-1};q)_{\infty}
}
{
(q^{\frac{r}{2}+\frac{|m_i|}{2}} a s_i^{\pm}x_{\alpha};q)_{\infty}
}
q^{N_f (1-r)\sum_{i=1}^{N_c}|m_i|-\sum_{i=1}^{N_c}|m_i|-\frac12\sum_{i<j}|m_i\pm m_j| }
a^{-2N_f \sum_{i=1}^{N_c}|m_i|}.
\end{align}
where we note that $a$ is the fugacity for the $U(1)_a$ symmetry which can be
combined with the $SU(2N_f)$ flavour symmetry to form a $U(2N_f)$ flavour symmetry. Here we have set the $SU(2N_f)$ flavour fugacities $x_{\alpha} \rightarrow 1$ for simplicity.

The full-index of theory B, which is the dual of theory A with gauge group
$USp(2N_c)$ so has gauge group $USp(2\tilde{N}_c = 2N_f - 2N_c - 2)$, is 
\begin{align}
\label{sp2Nmag_i}
&
I^{USp(2\tilde{N}_c)-[2N_f]^{B}}
\nonumber\\
&=
\frac{1}{2^{\tilde{N}_c} (\tilde{N}_c)!}
\sum_{m_1, \cdots, m_{\tilde{N}_c}\in \mathbb{Z}}
\oint \prod_{i=1}^{\tilde{N}_c}
\frac{ds_i}{2\pi is_i}
\nonumber\\
&\times 
(1-q^{|m_i|} s_{i}^{\pm 2})
\prod_{1\le i<j\le \tilde{N}_c}
(1-q^{\frac{|m_i \pm m_j|}{2}} s_i s_j^{\pm})
(1-q^{\frac{|-m_i \pm m_j|}{2}} s_i^{-1} s_j^{\pm})
\nonumber\\
&\times 
\prod_{i=1}^{\tilde{N}_c}
\prod_{\alpha=1}^{2N_f}
\frac{
(q^{\frac{1+r+|m_i|}{2}} as_{i}^{\pm}x_{\alpha};q)_{\infty}
}
{
(q^{\frac{1-r+|m_i|}{2}}a^{-1}s_{i}^{\pm}x_{\alpha}^{-1};q)_{\infty}
}
q^{N_f r \sum_{i=1}^{\tilde{N}_c}|m_i|-\sum_{i=1}^{\tilde{N}_c}|m_i| -\frac12 \sum_{i<j}|m_i\pm m_j|}a^{2N_f \sum_{i=1}^{\tilde{N}_c}|m_i|}
\nonumber\\
&\times 
\prod_{\alpha<\beta}
\frac{
(q^{1-r}a^{-2}x_{\alpha}^{-1}x_{\beta}^{-1};q)_{\infty}
}
{
(q^{r}a^2 x_{\alpha}x_{\beta};q)_{\infty}
}
\frac{
(q^{-\tilde{N}_c +r N_f}a^{2N_f};q)_{\infty}
}
{
(q^{\tilde{N}_c + 1 -rN_f}a^{-2N_f};q)_{\infty}
}. 
\end{align}

%%%%%%%%%%%%%%%%%%%%%%%%%%%%%%%%%%%%%%%%%%%
\subsubsection{Half-index}
%%%%%%%%%%%%%%%%%%%%%%%%%%%%%%%%%%%%%%%%%%%
We can construct the half-indices for the 3d $\mathcal{N}=2$ gauge theories with Dirichlet boundary condition $\mathcal{D}$ for the vector multiplet using
the expressions in \cite{Dimofte:2017tpi}. We briefly review the construction
for both Neumann and Dirichlet boundary conditions,
and describe some details for symplectic gauge groups in appendix~\ref{GenIndices}. We find for Dirichlet gauge field boundary
conditions in theory A
\begin{align}
&
\II_{\Dcal}^{USp(2N_c)^A} \nonumber\\
& = \frac{1}{(q)_{\infty}^{N_c}} \sum_{m_i \in \Zb}
 \frac{(-q^{1/2})^{k_{eff}[m, m]} u^{k_{eff}[m, -]}}{\prod_{i \ne j}^{N_c}
 (q^{1 + m_i - m_j}u_i u_j^{-1}; q)_{\infty}
 \prod_{i \le j} (q^{1 \pm (m_i + m_j)}u_i^{\pm} u_j^{\pm}; q)_{\infty} }
  \II^{\mathrm{matter}}(s_i \rightarrow q^{m_i}u_i)
\end{align}
where the effective CS coupling $k_{eff}$ is determined by the 't Hooft anomaly.
As we are not turning on background fluxes, the only term relevant is the gauge
field contribution which is $\tilde{N}_c \tr(\bs^2)$ for Dirichlet boundary
conditions in theory A. This gives
\begin{align}
k_{eff}[m, m] = & 2 \tilde{N}_c \sum_{i=1}^{N_c} m_i^2, \\
(-q^{1/2})^{k_{eff}[m, m]} = & q^{\tilde{N}_c \sum_{i=1}^{N_c} m_i^2} , \\
u^{k_{eff}[m, -]} = & \prod_{i=1}^{N_c} u_i^{2 \tilde{N}_c m_i}
\end{align}
where we note that relative to the unitary cases \cite{Dimofte:2017tpi} and
the orthogonal cases we discuss later, there is a factor of 2 when mapping the
anomaly polynomial terms to the effective Chen-Simons coupling contribution.
This factor arises from the conversion of the magnetic charge $m \in cochar(G)$
to the electric charge $km \in weight(G)$, noting that Cartan-Killing form for
$G = USp(2N_c)$ differs by a factor of 2 from that for the unitary or orthogonal
groups -- in particular, the weights of $USp(2N_c)$ have length squared
$\frac{1}{2}$, not $1$.

For Neumann gauge field boundary conditions in theory A we have
\begin{align}
\II_{\Ncal}^{USp(2N_c)^A} & = \frac{(q)_{\infty}^{N_c}}{2^{N_c} N_c!}
 \left( \prod_{i = 1}^{N_c} \oint \frac{ds_i}{2\pi i s_i} \right)
 \left( \prod_{i \ne j} (s_i s_j^{-1}; q)_{\infty} \right)
 \left( \prod_{i \le j} (s_i^{\pm} s_j^{\pm}; q)_{\infty} \right)
        \II^{\textrm{matter}} I^{\Psi}. 
\end{align}

In both cases the matter contributions are given by the product of
contributions for each chiral multiplet in the theory, depending on the chosen
boundary conditions. For the bulk chiral multiplets in theories A and B we find the following contributions to $\mathbb{II}^{\textrm{matter}}$: 
\begin{align}
\II_{\mathrm{N}}^{Q} & = \prod_{\alpha=1}^{2N_f} \frac{1}{\prod_{i=1}^{N_c} (q^{\frac{r}{2}} s_i a x_{\alpha}; q)_{\infty} ( q^{\frac{r}{2}} s_i^{-1} a x_{\alpha}; q)_{\infty}}, \\
\II_{\mathrm{D}}^{Q} & = \prod_{\alpha=1}^{2N_f} \prod_{i=1}^{N_c} ( q^{1-\frac{r}{2}} s_i a^{-1} x_{\alpha}^{-1}; q)_{\infty} (q^{1-\frac{r}{2}} s_i^{-1} a^{-1} x_{\alpha}^{-1}; q)_{\infty}, \\
\II_{\mathrm{N}}^{q} & = \prod_{\alpha=1}^{2N_f} \frac{1}{\prod_{i=1}^{\tilde{N}_c} (q^{\frac{1-r}{2}}s_i a^{-1} x_{\alpha}^{-1}; q)_{\infty} (s_i^{-1} a^{-1} x_{\alpha}^{-1} q^{(1-r)/2}; q)_{\infty}}, \\
\II_{\mathrm{D}}^{q} & = \prod_{\alpha=1}^{2N_f} \prod_{i=1}^{\tilde{N}_c} (s_i a x_{\alpha} q^{(1+r)/2}; q)_{\infty} (q^{\frac{1-r}{2}} s_i^{-1} a x_{\alpha}; q)_{\infty}, \\
\II_{\mathrm{N}}^{M} & = \prod_{\alpha < \beta} \frac{1}{(q^r x_{\alpha} x_{\beta} a^2 ; q)_{\infty}}, \\
\II_{\mathrm{D}}^{M} & = \prod_{\alpha < \beta} (x_{\alpha}^{-1} x_{\beta}^{-1} a^{-2} q^{1-r}; q)_{\infty}, \\
\II_{\mathrm{N}}^{V} & = \frac{1}{(q^{(1-r) N_f - N_c}a^{-2N_f} ; q)_{\infty}}, \\
\II_{\mathrm{D}}^{V} & = (q^{N_c - (1-r) N_f + 1} a^{2N_f} ; q)_{\infty}. 
\end{align}

The bifundamental Fermi would be expected to give contribution
\begin{align}
I^{2d} & = \prod_{i=1}^{N_c} \prod_{j=1}^{\tilde{N}_c} F(s_i^{\pm} u_j^{\pm} q^{1/2}; q) F(s_i^{\pm} u_j^{\mp} q^{1/2}; q), \\
F(s_i^{\pm} u_j^{\pm} q^{1/2}; q) F(s_i^{\pm} u_j^{\mp} q^{1/2}; q) & = \left( (s_i u_j q^{1/2}; q)_{\infty} (s_i u_j^{-1} q^{1/2}; q)_{\infty} (s_i^{-1} u_j q^{1/2}; q)_{\infty} (s_i^{-1} u_j^{-1} q^{1/2}; q)_{\infty} \right)^2, 
\end{align}
but as noted for the anomaly contribution we should impose a reality condition.
This will half the number of degrees of freedom so the contribution to the
2d index will be the square root of this quantity\footnote{There is no sign ambiguity as the contribution must be a $q$-series starting with $+1$ rather than $-1$.}.
As it is a perfect square this gives a consistent contribution to the index
\begin{align}
I^{\Psi} & = \prod_{i=1}^{N_c} \prod_{j=1}^{\tilde{N}_c} (s_i u_j^{\pm} q^{1/2}; q)_{\infty} (s_i^{-1} u_j^{\pm} q^{1/2}; q)_{\infty}. 
\end{align}

The half-indices for theory B are constructed in exactly the same way but
exchanging $N_c \leftrightarrow \tilde{N}_c$ except for the matter
contributions which are already written specifically for each theory. Note that
since the dualities
we consider are between theories with opposite boundary conditions for the
vector multiplets it is consistent to use fugacities $s_i$ for the preserved
gauge symmetry in whichever theory
has Neumann boundary conditions, and fugacities
$u_i$ in the other theory with Dirichlet boundary conditions breaking the gauge symmetry to a global symmetry.

We now present some examples which we have checked numerically to find matching indices. For calculational purposes these are all checked with flavour fugacities $x_{\alpha} = 1$. After presenting these examples we will construct the half-indices with non-zero Chern-Simons level.

%%%%%%%%%%%%%%%%%%%%%%%%%%%%%%%%%%%%%%%%%%%
\subsection{$N_c=1, N_f=3$ $(USp(2)-[6])$}
%%%%%%%%%%%%%%%%%%%%%%%%%%%%%%%%%%%%%%%%%%%
We start with the simplest example with $N_c=1, N_f=3$, 
where Theory A is $USp(2)=SU(2)$ gauge theory with six fundamental chiral multiplets $Q$. 

The full-index is
\begin{align}
\label{sp2nf3_i}
I^{USp(2)-[6]^A}
&=
\frac12 \sum_{m\in \mathbb{Z}}
\oint \frac{ds}{2\pi is}
(1-q^{|m|}s^{\pm 2})
\prod_{\alpha=1}^{6}
\frac{
(q^{1-\frac{r}{2}+\frac{|m|}{2}}a^{-1}s^{\pm}x_{\alpha}^{-1};q)_{\infty}
}
{
(q^{\frac{r}{2}+\frac{|m|}{2}} a s^{\pm} x_{\alpha};q)_{\infty}
}
q^{3(1-r)|m|-|m|}
a^{-6|m|}. 
\end{align}

Theory B is $USp(2)=SU(2)$ gauge theory with six fundamental chiral multiplets $q$ 
as well as gauge singlets $M$ and $V$. 
The index (\ref{sp2nf3_i}) agrees with the index 
\begin{align}
\label{sp2nf3mag_i}
I^{USp(2)-[6]^B}
&=
\frac12 \sum_{m\in \mathbb{Z}}
\oint \frac{ds}{2\pi is}(1-q^{|m|}s^{\pm 2})
\prod_{\alpha=1}^6 
\frac{
(q^{\frac{1+r+|m|}{2}} as^{\pm}x_{\alpha};q)_{\infty}
}
{
(q^{\frac{1-r+|m|}{2}} a^{-1}s^{\pm}x_{\alpha}^{-1};q)_{\infty}
}
q^{3r|m|-|m|}a^{6|m|}
\nonumber\\
&\times 
\prod_{\alpha<\beta}
\frac{
(q^{1-r}a^{-2}x_{\alpha}^{-1}x_{\beta}^{-1};q)_{\infty}
}
{
(q^{r}a^{2}x_{\alpha}x_{\beta};q)_{\infty}
}
\frac{
(q^{3r-1}a^6;q)_{\infty}
}
{
(q^{2-3r}a^{-6};q)_{\infty}
}
\end{align}
of Theory B. 
See \cite{Bashkirov:2011vy} 
\footnote{
There is a typo in eq.(6) of \cite{Bashkirov:2011vy}. 
}
and Appendix~\ref{Sp_check} for the expansions of indices.

%%%%%%%%%%%%%%%%%%%%%%%%%%%%%%%%%%%%%%%%%%%
\subsubsection{$USp(2)-[6]$ with $\mathcal{N}, (\mathrm{N,N,N;N,N,N})+\mathrm{Fermis}$}
%%%%%%%%%%%%%%%%%%%%%%%%%%%%%%%%%%%%%%%%%%%

The half-index of Neumann b.c. $(\mathcal{N},\mathrm{N})$ together with charged Fermi multiplets for Theory A reads
\begin{align}
\label{sp2nf3N_h}
\mathbb{II}_{\mathcal{N}, \mathrm{N}+\Psi}^{USp(2)-[6]^A}
&=
\frac12 (q)_{\infty}\oint \frac{ds}{2\pi is}
(s^2;q)_{\infty}
(s^{-2};q)_{\infty}
\frac{
(q^{\frac12}s^{\pm}u^{\pm};q)_{\infty}
(q^{\frac12}s^{\pm}u^{\mp};q)_{\infty}
}
{
(q^{\frac{r}{2}}as^{\pm};q)_{\infty}^6
}. 
\end{align}
We have checked that 
the half-index (\ref{sp2nf3N_h}) agrees with the half-index 
\begin{align}
\label{sp2nf3magD_h}
\mathbb{II}_{\mathcal{D}, \mathrm{D,N,D}}^{USp(2)-[6]^B}
&=
\frac{1}{(q)_{\infty}}
\sum_{m\in \mathbb{Z}}
\frac{1}{(q^{1+2m} u^2;q)_{\infty}
(q^{1-2m} u^{-2};q)_{\infty}
}
(q^{\frac{1+r\pm m}{2}} au^{\pm};q)_{\infty}^6
q^{m^2} u^{2m}
\nonumber\\
&\times 
\frac{1}{(q^r a^2;q)_{\infty}^{15}}(q^{3r-1}a^6;q)_{\infty}
\end{align}
of Dirichlet b.c. $(\mathcal{D}, \mathrm{D,N,D})$ for Theory B. 
See Appendix \ref{Sp_N_check} for the expansion of indices.

%%%%%%%%%%%%%%%%%%%%%%%%%%%%%%%%%%%%%%%%%%%
\subsubsection{$USp(2)-[6]$ with $\mathcal{D}, (\mathrm{D,D,D;D,D,D})$}
%%%%%%%%%%%%%%%%%%%%%%%%%%%%%%%%%%%%%%%%%%%

The half-index of Dirichlet b.c. $(\mathcal{D},\mathrm{D})$ for Theory A takes the form
\begin{align}
\label{sp2nf3D_h}
\mathbb{II}_{\mathcal{D},\mathrm{D}}^{USp(2)-[6]^A}
&=
\frac{1}{(q)_{\infty}}
\sum_{m\in \mathbb{Z}}
\frac{1}
{
(q^{1+2m}u^2;q)_{\infty}
(q^{1-2m}u^{-2};q)_{\infty}
}
(q^{1-\frac{r}{2}\pm m} a^{-1} u^{\pm};q)_{\infty}^6
q^{m^2} u^{2m}. 
\end{align}

On the other hand, the half-index 
of Neumann b.c. $(\mathcal{N}, \mathrm{N,D,N})$ plus charged Fermi multiplets for Theory B is 
\begin{align}
\label{sp2nf3magN_h}
\mathbb{II}_{\mathcal{N},\mathrm{N,D,N}+\Psi }^{USp(2)-[6]^B}
&=
\frac12 (q)_{\infty}
\oint \frac{ds}{2\pi is}
(s^2;q)_{\infty} (s^{-2};q)_{\infty}
\nonumber\\
&\times 
\frac{1}{(q^{\frac{1-r}{2}} a^{-1}s^{\pm};q)_{\infty}^6}
(q^{\frac12} s^{\pm} u^{\pm};q)_{\infty}
(q^{\frac12} s^{\pm} u^{\mp};q)_{\infty}
(q^{1-r}a^{-2};q)_{\infty}^{15}
\frac{1}
{(q^{2-3r}a^{-6};q)_{\infty}}. 
\end{align}
We have confirmed that 
the half-indices (\ref{sp2nf3D_h}) and (\ref{sp2nf3magN_h}) perfectly match. 
See Appendix \ref{Sp_D_check} for the expansion of indices.

%%%%%%%%%%%%%%%%%%%%%%%%%%%%%%%%%%%%%%%%%%%
\subsection{$N_c=2, N_f=5$ $(USp(4)-[10])$}
%%%%%%%%%%%%%%%%%%%%%%%%%%%%%%%%%%%%%%%%%%%
Next example is the case with $N_c=2, N_f=5$, 
where Theory A has gauge group $USp(4)$ and ten fundamental flavours.  

The full-index of Theory A is 
\begin{align}
\label{sp4nf5_i}
I^{USp(4)-[10]^A}
&=
\frac{1}{2^2\cdot 2} \sum_{m_1, m_2\in \mathbb{Z}}
\oint \prod_{i=1}^2 \frac{ds_i}{2\pi is_i}
(1-q^{|m_i|}s_i^{\pm 2})
(1-q^{\frac{|m_1 \pm m_2|}{2}} s_1 s_2^{\pm})
(1-q^{\frac{|-m_1 \pm m_2|}{2}} s_1^{-1} s_2^{\pm})
\nonumber\\
&\times 
\prod_{i=1}^2
\prod_{\alpha=1}^{10}
\frac{
(q^{1-\frac{r}{2}+\frac{|m_i|}{2}}a^{-1}s_i^{\pm}x_{\alpha}^{-1};q)_{\infty}
}
{
(q^{\frac{r}{2}+\frac{|m_i|}{2}} a s_i^{\pm} x_{\alpha};q)_{\infty}
}
q^{5(1-r)\sum_{i=1}^2|m_i|-\sum_{i=1}^2|m_i|-\frac12|m_1\pm m_2|}
a^{-10\sum_{i=1}^{2}|m_i|}. 
\end{align}

Unlike the previous example, 
Theory B has a different gauge group, i.e. $USp(4)$ as well as ten fundamental chiral multiplets and neutral chiral multiplets. 

We have the full-index
\begin{align}
\label{sp4nf5mag_i}
I^{USp(4)-[10]^B}
&=
\frac{1}{2^2\cdot 2} \sum_{m_1, m_2\in \mathbb{Z}}
\oint \prod_{i=1}^2 \frac{ds_i}{2\pi is_i}
(1-q^{|m_i|}s_i^{\pm 2})
(1-q^{\frac{|m_1 \pm m_2|}{2}} s_1 s_2^{\pm})
(1-q^{\frac{|-m_1 \pm m_2|}{2}} s_1^{-1} s_2^{\pm})
\nonumber\\
&\times 
\prod_{i=1}^2
\prod_{\alpha=1}^{10}
\frac{
(q^{\frac{1+r+|m|}{2}} as_i^{\pm}x_{\alpha};q)_{\infty}
}
{
(q^{\frac{1-r+|m|}{2}} a^{-1}s_i^{\pm}x_{\alpha}^{-1};q)_{\infty}
}
q^{5r\sum_{i=1}^{2}|m_i|-\sum_{i=1}^2|m_i| -\frac12|m_1\pm m_2|}
a^{10\sum_{i=1}^2|m_i|}
\nonumber\\
&\times 
\prod_{\alpha<\beta}
\frac{
(q^{1-r}a^{-2}x_{\alpha}^{-1}x_{\beta}^{-1};q)_{\infty}
}
{
(q^{r}a^{2}x_{\alpha}x_{\beta};q)_{\infty}
}
\frac{
(q^{5r-2}a^{10};q)_{\infty}
}
{
(q^{3-5r}a^{-10};q)_{\infty}
}
\end{align}
of Theory B. 
One finds the agreement of 
indices (\ref{sp4nf5_i}) and (\ref{sp4nf5mag_i}). 
See \cite{Bashkirov:2011vy} and Appendix~\ref{Sp_check} 
for the expansion of indices.

%%%%%%%%%%%%%%%%%%%%%%%%%%%%%%%%%%%%%%%%%%%
\subsubsection{$USp(4)-[10]$ with $\mathcal{N}, (\mathrm{N,N,N,N,N;N,N,N,N,N})+\mathrm{Fermis}$}
%%%%%%%%%%%%%%%%%%%%%%%%%%%%%%%%%%%%%%%%%%%
The half-index of Neumann b.c. $(\mathcal{N},\mathrm{N})$ together with the charged Fermi multiplets for Theory A is calculated as
\begin{align}
\label{sp4nf5N_h}
\mathbb{II}_{\mathcal{N},\mathrm{N}+\Psi}^{USp(4)-[10]^A}
&=
\frac{1}{2^2\cdot 2}
(q)_{\infty}^2 
\oint \frac{ds_1}{2\pi is_1}
\frac{ds_2}{2\pi is_2}
(s_{1}^{\pm 2};q)_{\infty}
(s_2^{\pm 2};q)_{\infty}
(s_1 s_{2}^{\pm};q)_{\infty}
(s_1^{-1} s_{2}^{\pm};q)_{\infty}
\nonumber\\
&\times 
\prod_{i=1}^2
\frac{1}{(q^{\frac{r}{2}} a s_i^{\pm};q)_{\infty}^{10}}
\prod_{i,j=1}^{2}
(q^{\frac12}s_{i}^{\pm} u_{j}^{\pm};q)_{\infty}
(q^{\frac12}s_{i}^{\pm} u_{j}^{\mp};q)_{\infty}. 
\end{align}

The half-index of Dirichlet b.c. $(\mathcal{D}, \mathrm{D,N,D})$ for Theory B is 
\begin{align}
\label{sp4nf5magD_h}
&
\mathbb{II}_{\mathcal{D},\mathrm{D,N,D}}^{USp(4)-[10]^B}
\nonumber\\
&=
\frac{1}{(q)_{\infty}^2}
\sum_{m_1, m_2\in \mathbb{Z}}
\frac{1}
{
\prod_{i=1}^2
(q^{1\pm 2m_i} u_i^{\pm 2};q)_{\infty}
(q^{1+m_1\pm m_2}u_1 u_2^{\pm};q)_{\infty}
(q^{1-m_1\pm m_2}u_1^{-1} u_2^{\pm};q)_{\infty}
}
\nonumber\\
&\times 
\prod_{i=1}^2 
(q^{\frac{1+r\pm m_i}{2}} au_i^{\pm};q)_{\infty}^{10} 
q^{2\sum_{i=1}^2 m_i^2} u^{4\sum_{i=1}^2 m_i}
\frac{1}{(q^r a^2;q)_{\infty}^{45}}
(q^{5r-2}a^{10};q)_{\infty}. 
\end{align}
As shown in Appendix \ref{Sp_N_check}, 
we find that 
the half-indices (\ref{sp4nf5N_h}) and (\ref{sp4nf5magD_h}) precisely match. 

%%%%%%%%%%%%%%%%%%%%%%%%%%%%%%%%%%%%%%%%%%%
\subsubsection{$USp(4)-[10]$ with $\mathcal{D}, (\mathrm{D,D,D,D,D;D,D,D,D,D})$}
%%%%%%%%%%%%%%%%%%%%%%%%%%%%%%%%%%%%%%%%%%%
For Dirichlet b.c. $(\mathcal{D},\mathrm{D})$, 
we have the half-index of Theory A: 
\begin{align}
\label{sp4nf5D_h}
&
\mathbb{II}_{\mathcal{D},\mathrm{D}}^{USp(4)-[10]^A}
\nonumber\\
&=
\frac{1}{(q)_{\infty}^2}\sum_{m_1, m_2\in \mathbb{Z}}
\frac{1}
{
\prod_{i=1}^2 (q^{1\pm 2m_{i}} u_i^{\pm 2} ;q)_{\infty}
(q^{1+m_1\pm m_2}u_1 u_2^{\pm};q)_{\infty}
(q^{1-m_1\pm m_2}u_1^{-1} u_2^{\pm};q)_{\infty}
}
\nonumber\\
&\times 
\prod_{i=1}^2 (q^{1-\frac{r}{2}\pm m_{i}} a^{-1} u_{i}^{\pm};q)_{\infty}^{10}
q^{2\sum_{i=1}^2 m_{i}^2}
u^{4\sum_{i=1}^2 m_i}. 
\end{align}

We have confirmed that the half-index (\ref{sp4nf5N_h}) agrees with 
the half-index 
\begin{align}
\label{sp4nf5magN_h}
\mathbb{II}_{\mathcal{N},\mathrm{N,D,N}+\Psi}^{USp(4)-[10]^B}
&=
\frac{1}{2^2\cdot 2}
(q)_{\infty}^2 
\oint \frac{ds_1}{2\pi is_1}\frac{ds_2}{2\pi is_2}
(s_1^{\pm 2};q)_{\infty}
(s_2^{\pm 2};q)_{\infty}
(s_1 s_2^{\pm};q)_{\infty}
(s_1^{-1} s_2^{\pm};q)_{\infty}
\nonumber\\
&\times 
\prod_{i=1}^2 
\frac{1}
{
(q^{\frac{1-r}{2}} a^{-1}s^{\pm};q)_{\infty}^{10}
}
\prod_{i,j=1}^{2}
(q^{\frac12} s_i^{\pm} u_j^{\pm};q)_{\infty}
(q^{\frac12} s_i^{\pm} u_j^{\mp};q)_{\infty}
\nonumber\\
&\times 
(q^{1-r}a^{-2};q)_{\infty}^{45}
\frac{1}
{
(q^{3-5r}a^{-10};q)_{\infty}
}
\end{align}
of Neumann b.c. $(\mathcal{N}, \mathrm{N,D,N})$ for Theory B which involves the charged Fermi multiplets. 
See \ref{Sp_D_check} for the expansion of indices. 

%%%%%%%%%%%%%%%%%%%%%%%%%%%%%%%%%%%%%%%%%%%
\subsection{Chern-Simons level $k \ne 0$}
\label{USpCSlevel}
%%%%%%%%%%%%%%%%%%%%%%%%%%%%%%%%%%%%%%%%%%%
As shown in \cite{Benini:2011mf} we can induce a non-zero Chern-Simons coupling starting
from the case of vanishing Chern-Simons level with $N_F + 2|k|$ fundamental
chirals and giving masses to $2|k|$ of them. Taking the masses to $\pm \infty$
we integrate out these $2|k|$ chirals and are left with $N_F$
flavours and Chern-Simons coupling $k = \pm |k|$. Here the $\pm$ signs are the
same for the masses and the sign of $k$. Note that we only require
$2|k| \in \Zb$ so although $N_F + 2|k|$ is an even integer (which we can label
$2N_f$ as in the case of $k=0$), $N_F$ is only
constrained to be integer. In the dual theory B the corresponding chirals get
masses of the opposite sign and consequently theory B gets a Chern-Simons level $-k$. This produces a duality $USp(2N_c)_{k} \leftrightarrow USp(2|k| + N_F - 2N_c - 2)_{-k}$ with $N_F$ fundamental chirals in both theories.

As shown in \cite{Dimofte:2017tpi}, chiral edge modes are present for bulk
chirals with Neumann boundary conditions and positive masses, and with
Dirichlet boundary conditions and negative masses. These edge modes introduce
additional 2d chiral or Fermi multiplets. On the other hand, with negative masses for Neumann and positive masses for Dirichlet boundary conditions there are
no edge modes and the dualities follow through without additional 2d multiplets.
We focus on the latter cases for simplicity.

To derive the half-indices with $k \ne 0$ we can take limits of the fugacities
following the procedure described in \cite{Benini:2011mf} for the full indices, and somewhat similar
to the process outlined in \cite{Dimofte:2017tpi} for unitary half-indices. This will
result in the matching of half-indices
\begin{align}
\II_{\Dcal, \mathrm{D}}^{USp(2N_c)_{k}-[N_F]^A} = & \II_{\Ncal, \mathrm{N, D}+\Psi}^{USp(2\tilde{N}_c)_{-k}-[N_F]^B} & k > 0 \\
\II_{\Ncal, \mathrm{N}+\Psi}^{USp(2N_c)_{k}-[N_F]^A} = & \II_{\Dcal, \mathrm{D, N}}^{USp(2\tilde{N}_c)_{-k}-[N_F]^B} & k < 0
\end{align}
where we note the singlet $V$ is also integrated out for $k \ne 0$.

These half-indices are given by a simple prescription of removing the
$\II_V$ contribution which is present in the half-indices for the $k=0$ cases and noting that
$\tilde{N}_c = |k| + N_F/2 - N_c -1$.

A derivation follows from the limit taking
masses to infinity as explained in \cite{Benini:2011mf}. First write $SU(2N_f) \times U(1)_a$ as $U(2N_f)$ with fugacities $\hat{x}_{\alpha} = a x_{\alpha}$ and in
$\II^V$ replace $a^{2N_f} \rightarrow \prod_{\alpha=1}^{2N_f} \hat{x}_{\alpha}$.
Note that while $\prod_{\alpha=1}^{2N_f} x_{\alpha} = 1$ for the
$SU(2N_f)$ fugacities $x_{\alpha}$, there is no such constraint on the
$U(2N_f)$ fugacities $\hat{x}_{\alpha}$. Now, for the $2|k|$ values of $\alpha$
corresponding to massive chirals we take the limit $\hat{x}_{\alpha} \rightarrow 0$
for negative mass in theory A and $\hat{x}_{\alpha} \rightarrow \infty$
for positive mass in theory A. For the simple choice of boundary
conditions we are considering it is easy to see that this simply removes
(set to $1$) the contributions to the half-indices from those $2|k|$
fundamental chirals in theories A and B and at the same time
removes the contribution from the corresponding parts of $M$ and the
contribution of $V$. Hence, in the cases we consider, this is a well-defined
limit and the matching of half-indices follows for the matching for $k=0$,
assuming that holds for arbitrary flavour fugacities. We can then write the
remaining $U(N_F)$ flavour symmetry as $SU(N_F) \times U(1)_a$ and replace the
remaining $N_F$ fugacities $\hat{x}_{\alpha}$ with $a x_{\alpha}$, now with the
constraint on the $SU(N_F)$ fugacities $\prod_{\alpha = 1}^{N_F} x_{\alpha} = 1$. This gives precisely the half-indices following the simple prescription above of simply omitting the contributions from the massive fields. Note that if we
took the limits with the `wrong' boundary conditions, the individual
contributions would not have well-defined limits, and indeed we would need to
introduce appropriate additional 2d chirals or Fermis to get well-defined matching half-indices.

One additional point to note is that if we calculate the anomaly polynomials
for the duals with $k \ne 0$ simply by including the CS contribution
$\pm k \Tr(\bs^2)$ for theory A and $\mp k \Tr(\tilde{\bs}^2)$ for theory B, we would naively not get matching results. This is because there are background Chern-Simons terms generated by integrating out the massive fields and these contributions only taken into account those associated to the gauge groups. Including these background Chern-Simons levels we do have matching
anomaly polynomials and indeed we can see these Chern-Simons levels by simply
taking the difference in the anomaly polynomial before and after integrating out the massive fields in each theory. As we have not turned on background fluxes,
our half-indices are not sensitive to these other background Chern-Simons levels.
In particular we have the following contributions from background Chern-Simons levels, in addition to the $\pm k \Tr(\bs^2)$ or $\mp k \Tr(\tilde{\bs}^2)$ contributions
\begin{align}
\Acal_{CS \; \Ncal, \mathrm{N}}^{USp(2N_c)_{k<0}-[N_F]^A} 
& = \Acal_{\Ncal, \mathrm{N}}^{USp(2N_c)-[N_F + 2|k|]^A} - \Acal_{\Ncal, \mathrm{N}}^{USp(2N_c)_{k<0}-[N_F]^A} \nonumber \\
 & = 2kN_c\ba^2 + 4kN_c(r-1)\ba \br + 2kN(r-1)^2 \br^2, \\
\Acal_{CS \; \Dcal, \mathrm{D, N}}^{USp(2\tilde{N}_c)_{k<0}-[N_F]^B} & = \Acal_{\Dcal, \mathrm{D, N, D}}^{USp(2\tilde{N}_c)-[N_F + 2|k|]^B} - \Acal_{\Dcal, \mathrm{D, N}}^{USp(2\tilde{N}_c)_{k<0}-[N_F]^B} \nonumber \\
 & = k \Tr(\bx^2) + 2 \left(k (N_c + N_f) + N_f^2\right) \ba^2 \nonumber \\ 
 & + (4 k ((r-1) N_c + r N_f) + 2 N_f (2 N_c + 2 N_f (r-1)+1)) \ba \br \nonumber \\
 & + \frac{1}{2} \left(2 k^2+k \left(4 N \left(r^2-2 r+2\right)+4 N_f \left(r^2-1\right)+3\right)+(2 N+2 N_f (r-1)+1)^2\right) \br^2, \\
\Acal_{CS \; \Dcal, \mathrm{D}}^{USp(2N_c)_{k>0}-[N_F]^A} & = \Acal_{\Dcal, \mathrm{D}}^{USp(2N_c)-[N_F + 2|k|]^A} - \Acal_{\Dcal, \mathrm{D}}^{USp(2N_c)_{k>0}-[N_F]^A} \nonumber \\
 & = 2kN_c\ba^2 + 4kN_c(r-1)\ba \br + 2kN(r-1)^2 \br^2, \\
\Acal_{CS \; \Ncal, \mathrm{N, D}}^{USp(2\tilde{N}_c)_{k>0}-[N_F]^B} & = \Acal_{\Ncal, \mathrm{N, D, N}}^{USp(2\tilde{N}_c)-[N_F + 2|k|]^B} - \Acal_{\Ncal, \mathrm{N, D}}^{USp(2\tilde{N}_c)_{k>0}-[N_F]^B} \nonumber \\
 & = k \Tr(\bx^2) + 2 \left(k (N_c + N_f) - N_f^2\right) \ba^2 \nonumber \\
 & + (4 k ((r-1) N_c + r N_f) - 2 N_f (2 N_c + 2 N_f (r-1)+1)) \ba \br \nonumber \\
 & - \frac{1}{2} \left(2 k^2 - k \left(4 N \left(r^2-2 r+2\right)+4 N_f \left(r^2-1\right)+3\right)+(2 N+2 N_f (r-1)+1)^2\right) \br^2. 
\end{align}

In the following, we explicitly show examples of the half-indices with CS level $k\neq 0$ 
for which we have checked the precise agreement. 
%%%%%%%%%%%%%%%%%%%%%%%%%%%%%%%%%%%%%%%%%%%
\subsubsection{$USp(2)_{\pm k}-[6-2k]$}
%%%%%%%%%%%%%%%%%%%%%%%%%%%%%%%%%%%%%%%%%%%
For $k \in \{ \frac12, 1, \frac32, \ldots 3 \}$ we have the matching pairs of half-indices
\begin{align}
\II_{\Ncal, \mathrm{N}+\Psi}^{USp(2)_{-k}-[6-2k]^A} = & \frac{1}{2} (q)_{\infty} \oint \frac{ds}{2\pi i s} (s^{\pm 2}; q)_{\infty} \frac{(q^{1/2} s^{\pm} u^{\pm}; q)_{\infty} (q^{1/2} s^{\pm} u^{\mp}; q)_{\infty}}{(q^{r/2} a s^{\pm}; q)_{\infty}^{6-2k}}, \\
\II_{\Dcal, \mathrm{D, N, D}}^{USp(2)_k-[6-2k]^B} = & \frac{1}{(q)_{\infty}} \sum_{m \in \Zb} \frac{q^{m^2} u^{2}}{(q^{1 \pm 2m}u^{\pm 2}; q)_{\infty}} (q^{(1 + r)/2 \pm m} a u^{\pm}; q)_{\infty}^{6-2k} \frac{1}{(q^r a^2; q)_{\infty}^{(6-2k)(5-2k)/2}},
\end{align}
and
\begin{align}
\II_{\Dcal, \mathrm{D}}^{USp(2)_k-[6-2k]^A} = & \frac{1}{(q)_{\infty}} \sum_{m \in \Zb
} \frac{q^{m^2} u^{2}}{(q^{1 \pm 2m}u^{\pm 2}; q)_{\infty}} (q^{1 - r/2 \pm m} a^{-1} u^{\pm}; q)_{\infty}^{6-2k}, \\
\II_{\Ncal, \mathrm{N, D, N}+\Psi}^{USp(2)_{-k}-[6-2k]^B} = & \frac{1}{2} (q)_{\infty} \oint \frac{ds}{2\pi i s} (s^{\pm 2}; q)_{\infty} \frac{(q^{1/2} s^{\pm} u^{\pm}; q)_{\infty} (q^{1/2} s^{\pm} u^{\mp}; q)_{\infty}}{(q^{(1-r)/2} a^{-1} s^{\pm}; q)_{\infty}^{6-2k}} (q^{1-r} a^{-2}; q)_{\infty}^{(6-2k)(5-2k)/2} \; .
\end{align}

%%%%%%%%%%%%%%%%%%%%%%%%%%%%%%%%%%%%%%%%%%%
%%%%%%%%%%%%%%%%%%%%%%%%%%%%%%%%%%%%%%%%%%%
\section{$SO(N_c)$ gauge theories}
\label{SO_sec}
%%%%%%%%%%%%%%%%%%%%%%%%%%%%%%%%%%%%%%%%%%%
%%%%%%%%%%%%%%%%%%%%%%%%%%%%%%%%%%%%%%%%%%%
%Quantum dynamics of orthogonal gauge theory
The quantum dynamics of 3d $\mathcal{N}=2$ supersymmetric gauge theories 
with orthogonal gauge groups which is more subtle than with the unitary or symplectic gauge groups 
has been studied in \cite{Aharony:2011ci, Aharony:2013kma}.

%duality 
The IR dualities associated with the Lie algebra $\mathfrak{g}=\mathfrak{so}(N_c)$ of gauge symmetry are derived in \cite{Aharony:2013kma} 
by taking an appropriate limit of the 4d $\mathcal{N}=1$ gauge theories on a circle. 
There are three distinct 4d gauge theories for $\mathfrak{g}=\mathfrak{so}(N_c)$ 
with gauge group $SO(N_c)_+$, $SO(N_c)_-$ or $Spin(N_c)$ 
where $SO(N_c)_{\pm}$ are distinguished by $\mathbb{Z}_2^{\mathcal{M}}$, 
a $\mathbb{Z}_2$ global symmetry that changes the sign of the non-trivial line operator \cite{Aharony:2013hda}.

The 4d duality between $SO(N_c)_+$ gauge theories leads to the 3d IR duality \cite{Aharony:2013kma}: 
\begin{itemize}

\item Theory A: $SO(N_c)$ gauge theory with $N_f$ chiral multiplets $Q$ in the vector representation.

\item Theory B: $SO(\tilde{N}_c=N_f-N_c+2)$ gauge theory with $N_f$ chiral multiplets $q$ in the vector representation, 
$N_f (N_f+1)/2$ neutral chiral multiplets $M$ in the rank-2 symmetric
representation of $SU(N_f)$ and a chiral multiplet $V$
which has the superpotential
\begin{align}
\label{suppot_SO}
\mathcal{W}&=\frac12 Mqq+\frac{i^{N_f -N_c}}{4}V\tilde{V}.
\end{align}

\end{itemize}

The charges of the chiral multiplets are 
\begin{align}
\label{soN_charge}
\begin{array}{c|c|c|c|c|c|c|c|c|c|c|c}
&G=SO(N_c)&\tilde{G}=SO(\tilde{N}_c)&SU(N_f)&U(1)_a&U(1)_R
&\mathbb{Z}_2^{\mathcal{C}}&\mathbb{Z}_2^{\mathcal{M}}
&\mathbb{Z}_2^{\tilde{\mathcal{C}}}&\mathbb{Z}_2^{\tilde{\mathcal{M}}} \\ \hline 
Q&{\bf N_c}&{\bf 1}&{\bf N_f}&+&r&0&0&0&0 \\ \hline
q&{\bf 1}&{\bf N_f -N_c +2}&{\bf N_f}&-&1-r&0&0&0&0 \\
M&{\bf 1}&{\bf 1}&{\bf \frac{N_f (N_f+1)}{2}}&2&2r&0&0&+&+ \\
V&{\bf 1}&{\bf 1}&{\bf 1}&-N_f&\tilde{N}_c - r N_f&0&0&+&- \\
\end{array}
\end{align}
Here $\mathbb{Z}_2^{\mathcal{C}}$ (resp.\ $\mathbb{Z}_2^{\tilde{\mathcal{C}}}$) is the charge conjugation symmetry in theory A (resp.\ theory B). 
$\mathbb{Z}_2^{\mathcal{M}}$ (resp.\ $\mathbb{Z}_2^{\tilde{\mathcal{M}}}$) is the magnetic symmetry in theory A (resp.\ theory B) \cite{Cordova:2017vab}.

%%%%%%%%%%%%%%%%%%%%%%%%%%%%%%%%%%%%%%%%%%%
\subsection{$\mathcal{N}=(0,2)$ half-BPS boundary conditions}
%%%%%%%%%%%%%%%%%%%%%%%%%%%%%%%%%%%%%%%%%%%

The requirement of gauge anomaly cancellation and 't Hooft anomaly matching are the same as for symplectic gauge theories. Again, using
the general results presented
in \cite{Dimofte:2017tpi} and summarised in appendix~\ref{Gen2dAnom} we find
the following specific results
for the fields in the $SO(N_c)-[N_f]$ theory A and
$SO(N_f - N_c + 2)-[N_f]$ theory B
\begin{align}
\Acal^{\textrm{VM}} = & -(N_c - 2) \Tr (\bs^2) - \frac{N_c}{4}(N_c - 1) \br^2, \\
\Acal^{Q} = & N_f \Tr (\bs^2) + \frac{N_c}{2} \Tr (\bx^2) + \frac{1}{2}N_c N_f(\ba + (r-1) \br)^2, \\
\Acal^{q} = & N_f \Tr (\tilde{\bs}^2) + \frac{1}{2}(N_f - N_c + 2) \Tr (\bx^2) + \frac{1}{2}(N_f - N_c + 2) N_f(-\ba - r \br)^2, \\
\Acal^{M} = & \frac{1}{2}(N_f + 2) \Tr (\bx^2) + \frac{1}{4} N_f (N_f + 1)(2 \ba + (2r-1) \br)^2, \\
\Acal^{V} = & \frac{1}{2} (-N_f \ba + ((1-r)N_f - N_c + 2) \br)^2
\end{align}
for Dirichlet boundary conditions. For Neumann boundary conditions the
contributions are the same but with opposite sign. As for the symplectic case,
the notation $\bs$
($\tilde{\bs}$) refers to the gauge field strength in theory A (B), $\bx$ to
the field strength for the global $SU(N_f)$ flavour symmetry, and $\ba$ and
$\br$ to the field strengths for the global $U(1)_a$ and $U(1)_R$ symmetries.

We also need 2d boundary matter and the only multiplet required for the
examples we consider is an $SO(N_c) \times SO(N_f - N_c + 2)$
bifundamental Fermi. Analogously to the symplectic case, here
$SO(N_c)$ is the 2d gauge group
inherited from the bulk vector multiplet with Neumann boundary conditions 
while $SO(N_f - N_c + 2)$ is a global flavour symmetry which is identified 
with the dual gauge group broken to a global symmetry by the Dirichlet boundary
conditions for the dual vector multiplet. This gives anomaly contribution
\begin{align}
\Acal^{\Psi} = & (N_f - N_c + 2)  \Tr (\bs^2) + N_c \Tr (\tilde{\bs}^2) \; .
\end{align}
Note that as for the symplectic case this contribution is actually only half of what might be expected for
such a Fermi. However, this is precisely the contribution required for anomaly
cancellation and we interpret this as due to a reality condition on the Fermi.
We will comment again on this when discussing the Fermi contribution to the
half-index.

For now we consider the 3d dualities with vanishing Chern-Simons level, in which
case we have no background Chern-Simons terms.

The boundary conditions we consider are $(\Ncal, \mathrm{N})$ in theory A, referring to
the choice of Neumann boundary conditions for $(\textrm{VM}, Q)$ together with
$(\Dcal, \mathrm{D, N, D})$ in theory B for $(\textrm{VM}, q, M, V)$. With this choice we need to
cancel the gauge anomaly in theory A which we can do by including the
bifundamental Fermi in theory A. This also leads to anomaly matching with
theory B without any further 2d matter. In particular we have
\begin{align}
\Acal^{A, \textrm{bulk}}_{\Ncal, \mathrm{N}} = & -(N_f - N_c + 2)\Tr(\bs^2) - \frac{N_c}{2} \Tr(\bx^2) - \frac{N_c}{4}(2N_f(1-r)^2 - N_c + 1)\br^2 \nonumber \\
 & +  (1-r) N_c N_f \ba\br - \frac{1}{2} N_c N_f \ba^2, \\
\Acal^{A, \textrm{bdry}}_{\Ncal, \mathrm{N}} = & \Acal^{\Psi} = (N_f - N_c + 2)\Tr(\bs^2) + N_c \Tr(\tilde{\bs}^2), \\
\Acal^{B, \textrm{bulk}}_{\Dcal, \mathrm{D, N, D}} = & N_c \Tr(\tilde{\bs}^2) - \frac{N_c}{2} \Tr(\bx^2) - \frac{N_c}{4}
(2N_f(1-r)^2 - N_c + 1)\br^2 \nonumber \\
 & + (1-r) N_c N_f \ba\br - \frac{1}{2} N_c N_f \ba^2 \label{AnomBSO}, 
\end{align}
and it is easy to see that all dependence on $\bs$ is cancelled and
\begin{equation}
\Acal^{A}_{\Ncal, \mathrm{N}} = \Acal^{A, \textrm{bulk}}_{\Ncal, \mathrm{N}} + \Acal^{A, \textrm{bdry}}_{\Ncal, \mathrm{N}} = \Acal^{B, \textrm{bulk}}_{\Dcal, \mathrm{D, N, D}} = \Acal^{B}_{\Dcal, \mathrm{D, N, D}} \; .
\end{equation}

Taking the opposite choice of boundary conditions for all fields we also get
anomaly matching if we add the bifundamental Fermi to theory B instead of
theory A. This also cancels the gauge anomaly in theory B.

We therefore have the following proposed dualities
\begin{itemize}
\item Theory A with $(\Ncal, \mathrm{N})$ boundary conditions plus Fermi $\Psi$ $\leftrightarrow$
 Theory B with $(\Dcal, \mathrm{D, N, D})$ boundary conditions.
\item Theory A with $(\Dcal, \mathrm{D})$ boundary conditions $\leftrightarrow$
Theory B with $(\Ncal, \mathrm{N, D, N})$ boundary conditions plus Fermi $\Psi$.
\end{itemize}
and in all cases we have an identification of discrete fugacities (defined in the following section) $\tilde{\zeta}$ and $\tilde{\chi}$ in theory B in terms of those in theory A, $\tilde{\zeta} = \zeta$ and
$\tilde{\chi} = \zeta \chi$.

%%%%%%%%%%%%%%%%%%%%%%%%%%%%%%%%%%%%%%%%%%%
\subsection{Supersymmetric indices}
%%%%%%%%%%%%%%%%%%%%%%%%%%%%%%%%%%%%%%%%%%%

%%%%%%%%%%%%%%%%%%%%%%%%%%%%%%%%%%%%%%%%%%%
\subsubsection{Full-index}
%%%%%%%%%%%%%%%%%%%%%%%%%%%%%%%%%%%%%%%%%%%
For the orthogonal gauge groups the Seiberg-like dualities have been tested 
from the evaluation of supersymmetric indices in \cite{Hwang:2011ht,Hwang:2011qt,Aharony:2013kma}. 
The index depends on the global structure of the gauge group. As shown in
\cite{Aharony:2013kma} all the indices can be constructed from the $SO(N_c)$
indices provided we include discrete fugacities $\zeta$ and $\chi$ for the $\Zb_2^{\Mcal}$ and
$\Zb_2^{\Ccal}$ groups.

For $\chi=+1$ or $N_c$ being odd the full-index of theory A with gauge group $G=SO(N_c=2N+\epsilon)$ for $\epsilon \in \{ 0, 1 \}$ reads
\begin{align}
\label{soN_gi1}
&
I^{SO(N_c = 2N+\epsilon)-[N_f]_{\zeta,\chi}^A}
\nonumber\\
&=
\frac{1}{2^{N+\epsilon-1}N!}
\sum_{m_1, \cdots, m_N}
\oint 
\prod_{i=1}^{N}
\frac{ds_i}{2\pi is_i}
\left[
\prod_{i=1}^N 
(1-\chi q^{\frac{|m_i|}{2}} s_i)
(1-\chi q^{\frac{|m_i|}{2}} s_i^{-1})
\right]^{\epsilon}
\nonumber\\
&\times 
\prod_{i<j}
(1-q^{\frac{|m_i -m_j|}{2}} s_i s_j^{-1})
(1-q^{\frac{|m_i -m_j|}{2}} s_i^{-1} s_j)
(1-q^{\frac{|m_i +m_j|}{2}} s_i s_j)
(1-q^{\frac{|m_i +m_j|}{2}} s_i^{-1} s_j^{-1})
\nonumber\\
&\times 
\left[
\prod_{\alpha=1}^{N_f}
\frac{
(\chi^{-1} q^{1-\frac{r}{2}} a^{-1}x_{\alpha}^{-1};q)_{\infty}
}
{
(\chi q^{\frac{r}{2}}ax_{\alpha};q)_{\infty}
}
\right]^{\epsilon}
\prod_{i=1}^{N}
\prod_{\alpha=1}^{N_f}
\frac{
(q^{1- \frac{r+|m|}{2}} a^{-1}s_{i}^{\mp}x_{\alpha}^{-1};q)_{\infty}
}
{
(q^{\frac{r+|m|}{2}} as_{i}^{\pm}x_{\alpha};q)_{\infty}
}
\nonumber\\
&\times 
q^{N_f \sum_{i=1}^{N}\frac{1-r}{2} |m_i|
-\sum_{i=1}^{N}\frac{|m_i|}{2}\epsilon
-\sum_{i<j}\frac{|m_i -m_j|}{2}
-\sum_{i<j}\frac{|m_i+m_j|}{2}
}
a^{-\sum_{i=1}^{N}|m_i|}
\zeta^{\sum_{i=1}^{N}m_i}. 
\end{align}
For $\chi=-1$ and $N_c$ even the full-index of theory A is 
\begin{align}
\label{soN_gi2}
&
I^{SO(N_c = 2N)-[N_f]_{\zeta,\chi=-}^A}
\nonumber\\
&=
\frac{1}{2^{N-1}(N-1)!}
\sum_{m_1, \cdots, m_{N-1}}
\oint 
\prod_{i=1}^{N-1}
\frac{ds_i}{2\pi is_i}
\prod_{i=1}^{N-1} 
(1-q^{|m_i|} s_i^2)
(1-q^{|m_i|} s_i^{-2})
\nonumber\\
&\times 
\prod_{i<j}
(1-q^{\frac{|m_i -m_j|}{2}} s_i s_j^{-1})
(1-q^{\frac{|m_i -m_j|}{2}} s_i^{-1} s_j)
(1-q^{\frac{|m_i +m_j|}{2}} s_i s_j)
(1-q^{\frac{|m_i +m_j|}{2}} s_i^{-1} s_j^{-1})
\nonumber\\
&\times 
\prod_{\alpha=1}^{N_f}
\frac{
(\pm q^{1-\frac{r}{2}} a^{-1}x_{\alpha}^{-1};q)_{\infty}
}
{
(\pm q^{\frac{r}{2}}ax_{\alpha};q)_{\infty}
}
\prod_{i=1}^{N-1}
\prod_{\alpha=1}^{N_f}
\frac{
(q^{1- \frac{r+|m|}{2}} a^{-1}s_{i}^{\mp}x_{\alpha}^{-1};q)_{\infty}
}
{
(q^{\frac{r+|m|}{2}} as_{i}^{\pm}x_{\alpha};q)_{\infty}
}
\nonumber\\
&\times 
q^{N_f \sum_{i=1}^{N-1}\frac{1-r}{2} |m_i|
-\sum_{i=1}^{N}|m_i|
-\sum_{i<j}\frac{|m_i -m_j|}{2}
-\sum_{i<j}\frac{|m_i+m_j|}{2}
}
a^{-\sum_{i=1}^{N-1}|m_i|}
\zeta^{\sum_{i=1}^{N-1}m_i}.
\end{align}

In theory B there are additional gauge singlets $M$ and $V$. 
As shown in (\ref{soN_charge}), the chiral superfield $V$ is odd under $\mathbb{Z}_2^{\mathcal{M}}$ 
so that the index of $V$ depends on the fugacity $\zeta$. 
For $\chi=+1$ or $N_c$ being odd the full-index of theory B is
\begin{align}
\label{soN_gi3}
&
I^{SO(\tilde{N}_c = 2\tilde{N}+\tilde{\epsilon})-[N_f]_{\zeta,\chi}^B}
\nonumber\\
&=
\frac{1}{2^{\tilde{N}+\tilde{\epsilon}-1}\tilde{N}!}
\sum_{m_1, \cdots, m_{\tilde{N}}}
\oint 
\prod_{i=1}^{\tilde{N}}
\frac{ds_i}{2\pi is_i}
\left[
\prod_{i=1}^{\tilde{N}} 
(1-\chi q^{\frac{|m_i|}{2}} s_i)
(1-\chi q^{\frac{|m_i|}{2}} s_i^{-1})
\right]^{\tilde{\epsilon}}
\nonumber\\
&\times 
\prod_{i<j}
(1-q^{\frac{|m_i -m_j|}{2}} s_i s_j^{-1})
(1-q^{\frac{|m_i -m_j|}{2}} s_i^{-1} s_j)
(1-q^{\frac{|m_i +m_j|}{2}} s_i s_j)
(1-q^{\frac{|m_i +m_j|}{2}} s_i^{-1} s_j^{-1})
\nonumber\\
&\times 
\left[
\prod_{\alpha=1}^{\tilde{N}_f}
\frac{
(\chi^{-1} q^{\frac{1+r}{2}} ax_{\alpha}^{-1};q)_{\infty}
}
{
(\chi q^{\frac{1-r}{2}}a^{-1}x_{\alpha};q)_{\infty}
}
\right]^{\tilde{\epsilon}}
\prod_{i=1}^{\tilde{N}}
\prod_{\alpha=1}^{N_f}
\frac{
(q^{\frac{1+r+|m|}{2}} as_{i}^{\mp}x_{\alpha}^{-1};q)_{\infty}
}
{
(q^{\frac{1-r+|m|}{2}} a^{-1}s_{i}^{\pm}x_{\alpha};q)_{\infty}
}
\nonumber\\
&\times 
q^{N_f \sum_{i=1}^{\tilde{N}}\frac{1-r}{2} |m_i|
-\sum_{i=1}^{\tilde{N}}\frac{|m_i|}{2}\tilde{\epsilon}
-\sum_{i<j}\frac{|m_i -m_j|}{2}
-\sum_{i<j}\frac{|m_i+m_j|}{2}
}
a^{\sum_{i=1}^{N}|m_i|}
\zeta^{\sum_{i=1}^{N}m_i}
\nonumber\\
&\times 
\prod_{\alpha\le \beta}
\frac{
(q^{1-r}a^{-2}x_{\alpha}^{-1}x_{\beta}^{-1};q)_{\infty}
}
{
(q^{r}a^2 x_{\alpha}x_{\beta};q)_{\infty}
}
\frac{
(\zeta q^{\frac{N_c+rN_f-N_f}{2}} a^{N_f};q)_{\infty}
}
{
(\zeta^{-1}q^{1+\frac{N_f-rN_f-N_c}{2}} a^{-N_f};q)_{\infty}
}. 
\end{align}

For $\chi=-1$ and $N_c$ even the full-index of theory B is
\footnote{
We have an extra overall factor 2 compared to the formula in \cite{Aharony:2013kma}. 
}
\begin{align}
\label{soN_gi4}
&
I^{SO(\tilde{N}_c = 2\tilde{N})-[N_f]_{\zeta,\chi=-}^B}
\nonumber\\
&=
\frac{1}{2^{\tilde{N}-1}(\tilde{N}-1)!}
\sum_{m_1, \cdots, m_{\tilde{N}-1}}
\oint 
\prod_{i=1}^{\tilde{N}-1}
\frac{ds_i}{2\pi is_i}
\prod_{i=1}^{\tilde{N}-1} 
(1-q^{|m_i|} s_i^2)
(1-q^{|m_i|} s_i^{-2})
\nonumber\\
&\times 
\prod_{i<j}
(1-q^{\frac{|m_i -m_j|}{2}} s_i s_j^{-1})
(1-q^{\frac{|m_i -m_j|}{2}} s_i^{-1} s_j)
(1-q^{\frac{|m_i +m_j|}{2}} s_i s_j)
(1-q^{\frac{|m_i +m_j|}{2}} s_i^{-1} s_j^{-1})
\nonumber\\
&\times 
\prod_{\alpha=1}^{N_f}
\frac{
(\pm q^{\frac{1+r}{2}} ax_{\alpha}^{-1};q)_{\infty}
}
{
(\pm q^{\frac{1-r}{2}}a^{-1}x_{\alpha};q)_{\infty}
}
\prod_{i=1}^{\tilde{N}-1}
\prod_{\alpha=1}^{N_f}
\frac{
(q^{\frac{1+r+|m|}{2}} as_{i}^{\mp}x_{\alpha}^{-1};q)_{\infty}
}
{
(q^{\frac{1-r+|m|}{2}} a^{-1}s_{i}^{\pm}x_{\alpha};q)_{\infty}
}
\nonumber\\
&\times 
q^{N_f \sum_{i=1}^{\tilde{N}-1}\frac{1-r}{2} |m_i|
-\sum_{i=1}^{\tilde{N}}|m_i|
-\sum_{i<j}\frac{|m_i -m_j|}{2}
-\sum_{i<j}\frac{|m_i+m_j|}{2}
}
a^{\sum_{i=1}^{\tilde{N}-1}|m_i|}
\zeta^{\sum_{i=1}^{\tilde{N}-1}m_i}
\nonumber\\
&\times 
\prod_{\alpha\le \beta}
\frac{
(q^{1-r}a^{-2}x_{\alpha}^{-1}x_{\beta}^{-1};q)_{\infty}
}
{
(q^{r}a^2 x_{\alpha}x_{\beta};q)_{\infty}
}
\frac{
(\zeta q^{\frac{N_c+rN_f-N_f}{2}} a^{N_f};q)_{\infty}
}
{
(\zeta^{-1} q^{1+\frac{N_f-rN_f-N_c}{2}} a^{-N_f};q)_{\infty}
}. 
\end{align}

The dualities lead to the identities of full-indices (\ref{soN_gi1})-(\ref{soN_gi4}) \cite{Aharony:2013kma}: 
\begin{align}
\label{so_equal}
I^{SO(N_c)-[N_f]^A_{\zeta,\chi}}&=I^{SO(\tilde{N}_c)-[N_f]^B_{\zeta,\zeta \chi}}. 
\end{align}

%%%%%%%%%%%%%%%%%%%%%%%%%%%%%%%%%%%%%%%%%%%
\subsubsection{Half-index}
%%%%%%%%%%%%%%%%%%%%%%%%%%%%%%%%%%%%%%%%%%%
There are four half-indices for each choice of boundary conditions, labelled by the values of
$\zeta$ and $\chi$. The standard $SO(N_c)$ index is given by $\zeta = \chi = +1$.
The half-indices for the theory A $\II^{SO(N_c)-[N_f]^A_{\zeta,\chi}}$ match the
half-indices for the dual theory B $\II^{SO(\tilde{N}_c)-[N_f]_{\tilde{\zeta} \tilde{\chi}}^{B}}$
with $\tilde{\zeta} = \zeta$ and $\tilde{\chi} = \zeta \chi$. 

Using the expressions in \cite{Dimofte:2017tpi} for general gauge group and
choosing group $SO(N_c)$ (see appendix~\ref{GenIndices} for further details)
where $N_c = 2N + \epsilon$ for $\epsilon \in \{ 0, 1 \}$
since we often have to distinguish between $N_c$ even or odd, we
have for Dirichlet gauge field
boundary conditions in the cases with $N_c$ being even with $\chi = +1$
or $N_c$ being odd with $\chi = \pm 1$
\begin{align}
&
\II_{\Dcal}^{SO(N_c)_{\zeta \chi}} 
\nonumber\\
& = \frac{1}{(q)_{\infty}^{N}} \sum_{m_i \in \Zb}
 \frac{(\zeta\chi)^{\sum_i m_i} (-q^{1/2})^{k_{eff}[m, m]} u^{k_{eff}[m, -]}}{
  \prod_{i=1}^{N} (\chi q^{1 \pm m_i} u_i^{\pm}; q)_{\infty}^{\epsilon}
 \prod_{i \ne j} (q^{1 + m_i - m_j}u_i u_j^{-1}; q)_{\infty}
 \prod_{i<j} (q^{1 \pm m_i \pm m_j}u_i^{\pm} u_j^{\pm}; q)_{\infty} }
% (q^{1 - m_i - m_j}u_i^{-1} u_j^{-1}; q)_{\infty}}
\nonumber\\
&\times 
        \II^{\textrm{matter}}(s_i \rightarrow q^{m_i}u_i)
\end{align}
where the effective CS coupling $k_{eff}$ is determined by the 't Hooft anomaly
and the indices $i$ and $j$ take values from $1$ to $N$. The relevant anomaly
contribution is $\tilde{N}_c \Tr(\bs^2)$ giving
\begin{align}
k_{eff}[m, m] = & \tilde{N}_c \sum_{i=1}^{N} m_i^2, \\
(-q^{1/2})^{k_{eff}[m, m]} = & (-1)^{\tilde{N}_c \sum_{i=1}^{N} m_i} q^{\tilde{N}_c \sum_{i=1}^{N} m_i^2} , \\
u^{k_{eff}[m, -]} = & \prod_{i=1}^{N} u_i^{\tilde{N}_c m_i}.
\end{align}

In the case
$\chi = -1$ with $N_c$ even we have instead
\begin{align}
\hspace*{-1.0cm}
&
\II_{\Dcal}^{SO(N_c)_{\zeta -}} 
\nonumber\\
& = \frac{1}{(q)_{\infty}^{N - 1} (-q; q)_{\infty}} \sum_{m_i \in \Zb}
 \frac{(-\zeta)^{\sum_i m_i} (-q^{1/2})^{k_{eff}[m, m]} u^{k_{eff}[m, -]}}{
 \prod_{i} (u_i^{\pm} q^{1 \pm m_i}; q)_{\infty}
 \prod_{i} (-u_i^{\pm} q^{1 \pm m_i}; q)_{\infty} } \nonumber \\
  & \times \frac{1}{ \prod_{i \ne j} (q^{1 + m_i - m_j}u_i u_j^{-1}; q)_{\infty}
 \prod_{i<j} (q^{1 \pm m_i \pm m_j}u_i^{\pm} u_j^{\pm}; q)_{\infty} }
% (q^{1 - m_i - m_j}u_i^{-1} u_j^{-1}; q)_{\infty}}
        \II^{\textrm{matter}}(s_i \rightarrow q^{m_i}u_i)
\end{align}
where now $i$ and $j$ take values from $1$ to $N - 1$.

For Neumann gauge field boundary conditions we have, excluding the case of
$N_c$ even with $\chi = -1$
\begin{align}
&
\II_{\Ncal}^{SO(N_c)_{\zeta \chi}} \nonumber\\
& = \frac{(q)_{\infty}^{N}}{2^{N - 1 + \epsilon} N!}
 \left( \prod_{i = 1}^{N} \oint \frac{ds_i}{2\pi i s_i} \right)
 \left( \prod_{i \ne j} (s_i s_j^{-1}; q)_{\infty} \right)
 \left( \prod_{i<j} (s_i^{\pm} s_j^{\pm}; q)_{\infty} \right)
 \left( \prod_{i} (\chi s_i^{\pm}; q)_{\infty} \right)^{\epsilon}
 \nonumber\\
 &\times \II^{\textrm{matter}} I^{\Psi}
\end{align}
while for $N_c$ even with $\chi = -1$ we have
\begin{align}
\hspace*{-1.0cm}
\II_{\Ncal}^{SO(N_c)_{\zeta -}} & = \frac{(q)_{\infty}^{N-1}(-q; q)_{\infty}}{2^{N - 1} (N-1)!}
 \left( \prod_{i = 1}^{N-1} \oint \frac{ds_i}{2\pi i s_i} \right)
 \left( \prod_{i = 1} (s_i^{\pm}; q)_{\infty} (-s_i^{\pm}; q)_{\infty} \right)
 \nonumber \\
 & \times \left( \prod_{i \ne j} (s_i s_j^{-1}; q)_{\infty} \right)
 \left( \prod_{i<j} (s_i^{\pm} s_j^{\pm}; q)_{\infty} \right)
        \II^{\textrm{matter}} I^{\Psi}. 
\end{align}
Note that the symmetry factor is $2^{N-1+\epsilon}N!$, the order of the Weyl
group for $SO(N_c)$, except for the case of $N_c$ even and $\chi = -1$. In the
latter case it is $2^{N-1}(N-1)!$ which is twice the order of the Weyl group
of $SO(N_c-2)$. This result can be found in \cite{Kim:2014dza}, although note
that they refer to the `Weyl group' of $O(N_c)$ which have an order twice that
of the Weyl groups for $SO(N_c)$.

The half-indices for theory B are given by the same expressions after replacing
$N_c = 2N + \epsilon$ with $\tilde{N}_c = 2\tilde{N} + \tilde{\epsilon}$.

For the 3d fields in theory A with gauge group $SO(N_c = 2N + \epsilon)$ and
theory B with gauge group
$SO(\tilde{N}_c = 2\tilde{N} + \tilde{\epsilon} = N_f - N_c + 2)$ with Neumann
or Dirichlet boundary conditions we have the following contributions to the
half-indices
\begin{align}
\II_{\mathrm{N}}^{Q} & = \prod_{\alpha=1}^{N_f} \frac{1}{ (\chi q^{\frac{r}{2}}a x_{\alpha} ; q)_{\infty}^{\epsilon} \prod_{i=1}^{N} (q^{\frac{r}{2}} s_i^{\pm} a x_{\alpha} ; q)_{\infty}}, \\
\II_{\mathrm{D}}^{Q} & = \prod_{\alpha=1}^{N_f} (\chi q^{1-\frac{r}{2}} a^{-1} x_{\alpha}^{-1} ; q)_{\infty}^{\epsilon} \prod_{i=1}^{N} (q^{1-\frac{r}{2}} s_i^{\pm} a^{-1} x_{\alpha}^{-1} ; q)_{\infty}, \\
\II_{\mathrm{N}}^{q} & = \prod_{\alpha=1}^{N_f} \frac{1}{ (\tilde{\chi} q^{\frac{1-r}{2}} a^{-1} x_{\alpha}^{-1} ; q)_{\infty}^{\tilde{\epsilon}} 
\prod_{i=1}^{\tilde{N}} (q^{\frac{1-r}{2}} s_i^{\pm} a^{-1} x_{\alpha}^{-1} ; q)_{\infty}}, \\
\II_{\mathrm{D}}^{q} & = \prod_{\alpha=1}^{N_f} (\tilde{\chi} q^{\frac{1+r}{2}} a x_{\alpha} ; q)_{\infty}^{\tilde{\epsilon}} 
\prod_{i=1}^{\tilde{N}} (q^{\frac{1+r}{2}} s_i^{\pm} a x_{\alpha} ; q)_{\infty}, \\
\II_{\mathrm{N}}^{M} & = \prod_{\alpha \le \beta} \frac{1}{(q^r x_{\alpha} x_{\beta} a^2; q)_{\infty}}, \\
\II_{\mathrm{D}}^{M} & = \prod_{\alpha \le \beta} (q^{1-r} x_{\alpha}^{-1} x_{\beta}^{-1} a^{-2} ; q)_{\infty}, \\
\II_{\mathrm{N}}^{V} & = \frac{1}{\left( \zeta q^{\frac{(1-r) N_f - N_c + 2}{2}} a^{-N_f} ; q\right)_{\infty}}, \\
\II_{\mathrm{D}}^{V} & = \left(\zeta q^{\frac{N_c - (1-r) N_f}{2}} a^{N_f} ; q\right)_{\infty},
\end{align}
where for $\chi = -1$ and $N_c$ even we send $s_{N} \rightarrow 1$ and
$s^{-1}_{N} \rightarrow -1$ in the contributions from $Q$, while
for $\tilde{\chi} = -1$ and $\tilde{N}_c$ even we send $s_{\tilde{N}} \rightarrow 1$
and $s^{-1}_{\tilde{N}} \rightarrow -1$ in the contributions from $q$.

The bifundamental Fermi, again with a reality condition as in the symplectic
case, gives contribution to the 2d index
\begin{align}
I^{\Psi} & = \left( \prod_{i=1}^{N} \prod_{j=1}^{\tilde{N}} F_{00} \right)
 \left( \prod_{j=1}^{\tilde{N}} F_{10} \right)^{\epsilon}
 \left( \prod_{i=1}^{N} F_{01} \right)^{\tilde{\epsilon}}
 F_{11}^{\epsilon\tilde{\epsilon}}, \\
F_{00} & = (z_i \tilde{z}_j q^{1/2}; q)_{\infty} (z_i \tilde{z}_j^{-1} q^{1/2}; q)_{\infty} (z_i^{-1} \tilde{z}_j q^{1/2}; q)_{\infty} (z_i^{-1} \tilde{z}_j^{-1} q^{1/2}; q)_{\infty}, \\
F_{10} & = (\chi \tilde{z}_j q^{1/2}; q)_{\infty} (\chi \tilde{z}_j^{-1} q^{1/2}; q)_{\infty}, \\
F_{01} & = (\tilde{\chi} z_i q^{1/2}; q)_{\infty} (\tilde{\chi} z_i^{-1} q^{1/2}; q)_{\infty}, \\
F_{11} & = (\chi \tilde{\chi} q^{1/2}; q)_{\infty},
\end{align}
where $F_{10}$ is included only if $N_c$ is odd, $F_{01}$ is included only if
$\tilde{N_c}$ is odd and $F_{11}$ is included only if both $N_c$ and $\tilde{N}_c$
are odd. 
In the case of $N_c$ even and $\chi = -1$ we replace $z_{N} \rightarrow 1$ and
$z^{-1}_{N} \rightarrow -1$. Similarly in the case of even $\tilde{N}_c$
and $\tilde{\chi} = -1$
we replace $\tilde{z}_{N} \rightarrow 1$ and $\tilde{z}^{-1}_{N} \rightarrow -1$.

Then in the half-index for theory A with $\Ncal$ boundary conditions we include
$I^{\Psi}$ with the replacements $z_i \rightarrow s_i$ and
$\tilde{z}_i \rightarrow u_i$. If instead we have $\Ncal$ boundary conditions
in theory B we replace $z_i \rightarrow u_i$ and $\tilde{z}_i \rightarrow s_i$.

We now list some explicit examples using notation $SO(N_c)-[N_f]_{\zeta\chi}^A$
to refer to theory A with gauge group $SO(N_c)$ with $N_f$ fundamental chirals
and discrete fugacities $\zeta$ and $\chi$. This is dual to theory B with
gauge group $SO(\tilde{N}_c)$ with $N_f$ fundamental chirals and discrete
fugacities $\tilde{\zeta} = \zeta$ and $\tilde{\chi} = \zeta \chi$ which we
label $SO(\tilde{N}_c)-[N_f]_{\tilde{\zeta}\tilde{\chi}}^B$.

We first consider some examples with gauge group $SO(1)$ in theory A. These are free theories
with $N_f$ chirals but they are dual to interacting $SO(N_f+1)-[N_f]$ theories.
The indices and half-indices for $N_f$ free chirals are very simple so this
leads to an interesting set of $q$-series identities (assuming the duality
holds). In some examples the (half-)index of the dual theory also takes a
simple form and the matching of (half-)indices can easily be checked
analytically. Following these $SO(1)$ examples we consider some cases with
gauge group $SO(2)$, equivalent to $U(1)$ with $N_f$ chirals of charge $+1$
and $N_f$ of charge $-1$, or gauge group $SO(3)$. The latter differs from $SU(2)$ examples in that the matter is in the triplet representation corresponding to adjoint, not fundamental, of $SU(2)$. As there are significant differences between $SO(N_c)$ with $N_c$ odd or even, our examples cover all combinations with
$(N_c, \tilde{N}_c)$ both even, both odd or one even and one odd.

%%%%%%%%%%%%%%%%%%%%%%%%%%%%%%%%%%%%%%%%%%%
\subsection{$N_c=1, N_f=1$ $(SO(1)-[1])$}
%%%%%%%%%%%%%%%%%%%%%%%%%%%%%%%%%%%%%%%%%%%

For theory A we have the full-indices
\begin{align}
\label{so1nf1_i++}
I^{SO(1)-[1]_{\pm +}^A}
&=
\frac{
(q^{1-\frac{r}{2}} a^{-1};q)_{\infty}
}
{
(q^{\frac{r}{2}} a;q)_{\infty}
}, \\
\label{so1nf1_i+-}
I^{SO(1)-[1]_{\pm -}^A}
&=
\frac{
(-q^{1-\frac{r}{2}} a^{-1};q)_{\infty}
}
{
(-q^{\frac{r}{2}} a;q)_{\infty}
}.
\end{align}
These indices agree with the indices of theory B
\begin{align}
\label{so1nf1mag_i++}
I^{SO(2)-[1]_{++}^B}
&=
\sum_{m\in \mathbb{Z}}
\oint \frac{ds}{2\pi is}
\frac{
(q^{\frac{1+r+|m|}{2}} as^{\pm};q)_{\infty}
}
{
(q^{\frac{1-r+|m|}{2}} a^{-1}s^{\pm};q)_{\infty}
}
q^{\frac{r|m|}{2}} a^{|m|}
\frac{
(q^{1-r}a^{-2};q)_{\infty}
}
{
(q^{r}a^{2};q)_{\infty}
}
\frac{
(q^{\frac{r}{2}}a;q)_{\infty}
}
{
(q^{1-\frac{r}{2}}a^{-1};q)_{\infty}
}, \\
\label{so1nf1mag_i--}
I^{SO(2)-[1]_{--}^B}
&=
\frac{
(\pm q^{\frac{1+r}{2}}a;q)_{\infty}
}
{
(\pm q^{\frac{1-r}{2}}a^{-1};q)_{\infty}
}
\frac{
(q^{1-r}a^{-2};q)_{\infty}
}
{
(q^{r}a^{2};q)_{\infty}
}
\frac{
(-q^{\frac{r}{2}}a;q)_{\infty}
}
{
(-q^{1-\frac{r}{2}}a^{-1};q)_{\infty}
}, \\
\label{so1nf1mag_i+-}
I^{SO(2)-[1]_{+-}^B}
&=
\frac{
(\pm q^{\frac{1+r}{2}}a;q)_{\infty}
}
{
(\pm q^{\frac{1-r}{2}}a^{-1};q)_{\infty}
}
\frac{
(q^{1-r}a^{-2};q)_{\infty}
}
{
(q^{r}a^{2};q)_{\infty}
}
\frac{
(q^{\frac{r}{2}}a;q)_{\infty}
}
{
(q^{1-\frac{r}{2}}a^{-1};q)_{\infty}
}, \\
\label{so1nf1mag_i-+}
I^{SO(2)-[1]_{-+}^B}
&=
\sum_{m\in \mathbb{Z}}
\oint \frac{ds}{2\pi is}
\frac{
(q^{\frac{1+r+|m|}{2}} as^{\pm};q)_{\infty}
}
{
(q^{\frac{1-r+|m|}{2}} a^{-1}s^{\pm};q)_{\infty}
}
q^{\frac{r|m|}{2}} a^{|m|} (-1)^m 
\frac{
(q^{1-r}a^{-2};q)_{\infty}
}
{
(q^{r}a^{2};q)_{\infty}
}
\frac{
(-q^{\frac{r}{2}}a;q)_{\infty}
}
{
(-q^{1-\frac{r}{2}}a^{-1};q)_{\infty}
} \; .
\end{align}

Note that in these examples, since the half-index of theory A is not sensitive
to the value of $\zeta$ we have
\begin{align}
I^{SO(2)-[1]_{++}^B} = & I^{SO(1)-[1]_{\pm+}^A} = I^{SO(2)-[1]_{--}^B}, \\
I^{SO(2)-[1]_{+-}^B} = & I^{SO(1)-[1]_{\pm-}^A} = I^{SO(2)-[1]_{-+}^B}.
\end{align}
This is a particular example of the more general case
\begin{align}
I^{SO(N_f - 1)-[N_f]_{++}^B} = & I^{SO(1)-[N_f]_{\pm+}^A} = I^{SO(N_f - 1)-[N_f]_{--}^B}, \\
I^{SO(N_f - 1)-[N_f]_{+-}^B} = & I^{SO(1)-[N_f]_{\pm-}^A} = I^{SO(N_f - 1)-[N_f]_{-+}^B}.
\end{align}
We will see the cases of $N_f = 2, 3$ below but not specifically comment again
on this point.

%%%%%%%%%%%%%%%%%%%%%%%%%%%%%%%%%%%%%%%%%%%
\subsubsection{$SO(1)-[1]$ with $\mathcal{N}, \mathrm{N}+\mathrm{Fermi}$}
%%%%%%%%%%%%%%%%%%%%%%%%%%%%%%%%%%%%%%%%%%%

We now present the half-indices with Neumann boundary conditions in theory A
\begin{align}
\label{so1nf1_Nn++}
\mathbb{II}_{\mathcal{N},\mathrm{N}+\Psi}^{SO(1)-[1]_{++}^A}
&=
\frac{1}
{(q^{\frac{r}{2}} a;q)_{\infty}}
(q^{\frac12}u^{\pm};q)_{\infty}, \\
\label{so1nf1_Nn-+}
\mathbb{II}_{\mathcal{N},\mathrm{N}+\Psi}^{SO(1)-[1]_{-+}^A}
&=
\frac{1}
{(q^{\frac{r}{2}} a;q)_{\infty}}
(\pm q^{\frac12};q)_{\infty}, \\
\label{so1nf1_Nn+-}
\mathbb{II}_{\mathcal{N},\mathrm{N}+\Psi}^{SO(1)-[1]_{+-}^A}
&=
\frac{1}
{(-q^{\frac{r}{2}} a;q)_{\infty}}
(\pm q^{\frac12};q)_{\infty}, \\
\label{so1nf1_Nn--}
\mathbb{II}_{\mathcal{N},\mathrm{N}+\Psi}^{SO(1)-[1]_{--}^A}
&=
\frac{1}
{(-q^{\frac{r}{2}} a;q)_{\infty}}
(-q^{\frac12}u^{\pm};q)_{\infty} \; .
\end{align}
These half-indices agree with the theory B half-indices
\begin{align}
\label{so1nf1mag_Ddnd++}
\mathbb{II}_{\mathcal{D},\mathrm{D,N,D}}^{SO(2)-[1]_{++}^B}
&=
\frac{1}{(q)_{\infty}}
\sum_{m\in \mathbb{Z}}
(q^{\frac{1+r}{2}\pm m}au^{\pm};q)_{\infty}
(-1)^m q^{\frac{m^2}{2}}u^{m}
\frac{1}
{(q^r a^2;q)_{\infty}}
(q^{\frac{r}{2}} a;q)_{\infty}, \\
\label{so1nf1mag_Ddnd--}
\mathbb{II}_{\mathcal{D},\mathrm{D,N,D}}^{SO(2)-[1]_{--}^B}
&=
\frac{1}{(-q;q)_{\infty}}
(\pm q^{\frac{1+r}{2}} a;q)_{\infty}
\frac{1}
{(q^r a^2;q)_{\infty}}
(-q^{\frac{r}{2}} a;q)_{\infty}, \\
\label{so1nf1mag_Ddnd+-}
\mathbb{II}_{\mathcal{D},\mathrm{D,N,D}}^{SO(2)-[1]_{+-}^B}
&=
\frac{1}{(-q;q)_{\infty}}
(\pm q^{\frac{1+r}{2}} a;q)_{\infty}
\frac{1}
{(q^r a^2;q)_{\infty}}
(q^{\frac{r}{2}} a;q)_{\infty}, \\
\label{so1nf1mag_Ddnd-+}
\mathbb{II}_{\mathcal{D},\mathrm{D,N,D}}^{SO(2)-[1]_{-+}^B}
&=
\frac{1}{(q)_{\infty}}
\sum_{m\in \mathbb{Z}}
(q^{\frac{1+r}{2}\pm m}au^{\pm};q)_{\infty}
q^{\frac{m^2}{2}}u^{m}
\frac{1}
{(q^r a^2;q)_{\infty}}
(-q^{\frac{r}{2}} a;q)_{\infty}.
\end{align}

Note that,unlike for the full-indices, these theory A half-indices are
sensitive to the value of $\zeta$ due to the bifundamental Fermi multiplet.

%%%%%%%%%%%%%%%%%%%%%%%%%%%%%%%%%%%%%%%%%%%
\subsubsection{$SO(1)-[1]$ with $\mathcal{D}, \mathrm{D} $}
%%%%%%%%%%%%%%%%%%%%%%%%%%%%%%%%%%%%%%%%%%%

Instead, taking Dirichlet boundary conditions in theory A we have
\begin{align}
\label{so1nf1_Dd++}
\mathbb{II}_{\mathcal{D},\mathrm{D}}^{SO(1)-[1]_{\pm +}^A}
&=
(q^{1-\frac{r}{2}}a^{-1};q)_{\infty}, \\
\label{so1nf1_Dd+-}
\mathbb{II}_{\mathcal{D},\mathrm{D}}^{SO(1)-[1]_{\pm -}^A}
&=
(-q^{1-\frac{r}{2}}a^{-1};q)_{\infty}. 
\end{align}
These agree with the theory B half-indices
\begin{align}
\label{so1nf1mag_Nndn++}
\mathbb{II}_{\mathcal{N},\mathrm{N,D,N}+\Psi}^{SO(2)-[1]_{++}^B}
&=
(q)_{\infty}\oint \frac{ds}{2\pi is}
\frac{1}
{(q^{\frac{1-r}{2}} a^{-1}s^{\pm};q)_{\infty}}
(q^{\frac12}s^{\pm};q)_{\infty}
(q^{1-r}a^{-2};q)_{\infty}
\frac{1}
{(q^{1-\frac{r}{2}}a^{-1};q)_{\infty}}, \\
\label{so1nf1mag_Nndn--}
\mathbb{II}_{\mathcal{N},\mathrm{N,D,N}+\Psi}^{SO(2)-[1]_{--}^B}
&=
(-q;q)_{\infty}
\frac{1}
{(\pm q^{\frac{1-r}{2}} a^{-1};q)_{\infty}}
(\pm q^{\frac12};q)_{\infty}
(q^{1-r}a^{-2};q)_{\infty}
\frac{1}
{(-q^{1-\frac{r}{2}}a^{-1};q)_{\infty}}, \\
\label{so1nf1mag_Nndn+-}
\mathbb{II}_{\mathcal{N},\mathrm{N,D,N}+\Psi}^{SO(2)-[1]_{+-}^B}
&=
(-q;q)_{\infty}
\frac{1}
{(\pm q^{\frac{1-r}{2}} a^{-1};q)_{\infty}}
(\pm q^{\frac12};q)_{\infty}
(q^{1-r}a^{-2};q)_{\infty}
\frac{1}
{(q^{1-\frac{r}{2}}a^{-1};q)_{\infty}}, \\
\label{so1nf1mag_Nndn-+}
\mathbb{II}_{\mathcal{N},\mathrm{N,D,N}+\Psi}^{SO(2)-[1]_{-+}^B}
&=
(q)_{\infty}\oint \frac{ds}{2\pi is}
\frac{1}
{(q^{\frac{1-r}{2}} a^{-1}s^{\pm};q)_{\infty}}
(-q^{\frac12}s^{\pm};q)_{\infty}
(q^{1-r}a^{-2};q)_{\infty}
\frac{1}
{(-q^{1-\frac{r}{2}}a^{-1};q)_{\infty}}. 
\end{align}

Note that, as for the full-indices, the half-indices of theory A are not
sensitive
to the value of $\zeta$ so we have
\begin{align}
\II_{\mathcal{N},\mathrm{N,D,N}+\Psi}^{SO(2)-[1]_{++}^B} = & \II_{\Dcal, D}^{SO(1)-[1]_{\pm+}^A} = \II_{\mathcal{N},\mathrm{N,D,N}+\Psi}^{SO(2)-[1]_{--}^B}, \\
\II_{\mathcal{N},\mathrm{N,D,N}+\Psi}^{SO(2)-[1]_{+-}^B} = & \II_{\Dcal, D}^{SO(1)-[1]_{\pm-}^A} = \II_{\mathcal{N},\mathrm{N,D,N}+\Psi}^{SO(2)-[1]_{-+}^B} \; .
\end{align}
This is a particular example of the more general case
\begin{align}
\II_{\mathcal{N},\mathrm{N,D,N}+\Psi}^{SO(N_f + 1)-[N_f]_{++}^B} = & \II_{\Dcal, D}^{SO(1)-[N_f]_{\pm+}^A} = \II_{\mathcal{N},\mathrm{N,D,N}+\Psi}^{SO(N_f + 1)-[N_f]_{--}^B}, \\
\II_{\mathcal{N},\mathrm{N,D,N}+\Psi}^{SO(N_f + 1)-[N_f]_{+-}^B} = & \II_{\Dcal, D}^{SO(1)-[N_f]_{\pm-}^A} = \II_{\mathcal{N},\mathrm{N,D,N}+\Psi}^{SO(N_f + 1)-[N_f]_{-+}^B} \; .
\end{align}
We will see the cases for $N_f = 2, 3$ below.

%%%%%%%%%%%%%%%%%%%%%%%%%%%%%%%%%%%%%%%%%%%
\subsection{$N_c=1, N_f=2$ $(SO(1)-[2])$}
%%%%%%%%%%%%%%%%%%%%%%%%%%%%%%%%%%%%%%%%%%%

In theory A we have the full-indices
\begin{align}
\label{so1nf2_i++}
I^{SO(1)-[2]_{\pm +}^A}
&=
\frac{
(q^{1-\frac{r}{2}} a^{-1};q)_{\infty}^2
}
{
(q^{\frac{r}{2}} a;q)_{\infty}^2
}, \\
\label{so1nf2_i+-}
I^{SO(1)-[2]_{\pm -}^A}
&=
\frac{
(-q^{1-\frac{r}{2}} a^{-1};q)_{\infty}^2
}
{
(-q^{\frac{r}{2}} a;q)_{\infty}^2
},
\end{align}
which agree with the indices of theory B
\begin{align}
\label{so1nf2mag_i++}
I^{SO(3)-[2]_{++}^B}
&=
\frac12 
\sum_{m\in \mathbb{Z}}
\oint \frac{ds}{2\pi is}
(1-q^{\frac{|m|}{2}}s)
(1-q^{\frac{|m|}{2}}s^{-1})
\nonumber\\
&\times 
\frac{
(q^{\frac{1+r+|m|}{2}} as^{\pm};q)_{\infty}^2
}
{
(q^{\frac{1-r+|m|}{2}} a^{-1}s^{\pm};q)_{\infty}^2
}
\frac{
(q^{\frac{1+r}{2}} a;q)_{\infty}^2
}
{
(q^{\frac{1-r}{2}} a^{-1};q)_{\infty}^2
}
q^{r|m|-\frac{|m|}{2}} a^{2|m|}
\frac{
(q^{1-r}a^{-2};q)_{\infty}^3
}
{
(q^{r}a^{2};q)_{\infty}^3
}
\frac{
(q^{-\frac12+r}a^2;q)_{\infty}
}
{
(q^{\frac32 -r}a^{-2};q)_{\infty}
}, \\
\label{so1nf2mag_i--}
I^{SO(3)-[2]_{--}^B}
&=
\frac12 
\sum_{m\in \mathbb{Z}}
\oint \frac{ds}{2\pi is}
(1+q^{\frac{|m|}{2}}s)
(1+q^{\frac{|m|}{2}}s^{-1})
\nonumber\\
&\times 
\frac{
(q^{\frac{1+r+|m|}{2}} as^{\pm};q)_{\infty}^2
}
{
(q^{\frac{1-r+|m|}{2}} a^{-1}s^{\pm};q)_{\infty}^2
}
\frac{
(-q^{\frac{1+r}{2}} a;q)_{\infty}^2
}
{
(-q^{\frac{1-r}{2}} a^{-1};q)_{\infty}^2
}
q^{r|m|-\frac{|m|}{2}} a^{2|m|} (-1)^m 
\frac{
(q^{1-r}a^{-2};q)_{\infty}^3
}
{
(q^{r}a^{2};q)_{\infty}^3
}
\frac{
(-q^{-\frac12+r}a^2;q)_{\infty}
}
{
(-q^{\frac32 -r}a^{-2};q)_{\infty}
}, \\
\label{so1nf2mag_i+-}
I^{SO(3)-[2]_{+-}^B}
&=
\frac12 
\sum_{m\in \mathbb{Z}}
\oint \frac{ds}{2\pi is}
(1+q^{\frac{|m|}{2}}s)
(1+q^{\frac{|m|}{2}}s^{-1})
\nonumber\\
&\times 
\frac{
(q^{\frac{1+r+|m|}{2}} as^{\pm};q)_{\infty}^2
}
{
(q^{\frac{1-r+|m|}{2}} a^{-1}s^{\pm};q)_{\infty}^2
}
\frac{
(-q^{\frac{1+r}{2}} a;q)_{\infty}^2
}
{
(-q^{\frac{1-r}{2}} a^{-1};q)_{\infty}^2
}
q^{r|m|-\frac{|m|}{2}} a^{2|m|}
\frac{
(q^{1-r}a^{-2};q)_{\infty}^3
}
{
(q^{r}a^{2};q)_{\infty}^3
}
\frac{
(q^{-\frac12+r}a^2;q)_{\infty}
}
{
(q^{\frac32 -r}a^{-2};q)_{\infty}
}, \\
\label{so1nf2mag_i-+}
I^{SO(3)-[2]_{-+}^B}
&=
\frac12 
\sum_{m\in \mathbb{Z}}
\oint \frac{ds}{2\pi is}
(1-q^{\frac{|m|}{2}}s)
(1-q^{\frac{|m|}{2}}s^{-1})
\nonumber\\
&\times 
\frac{
(q^{\frac{1+r+|m|}{2}} as^{\pm};q)_{\infty}^2
}
{
(q^{\frac{1-r+|m|}{2}} a^{-1}s^{\pm};q)_{\infty}^2
}
\frac{
(q^{\frac{1+r}{2}} a;q)_{\infty}^2
}
{
(q^{\frac{1-r}{2}} a^{-1};q)_{\infty}^2
}
q^{r|m|-\frac{|m|}{2}} a^{2|m|}(-1)^m
\frac{
(q^{1-r}a^{-2};q)_{\infty}^3
}
{
(q^{r}a^{2};q)_{\infty}^3
}
\frac{
(-q^{-\frac12+r}a^2;q)_{\infty}
}
{
(-q^{\frac32 -r}a^{-2};q)_{\infty}
} \; .
\end{align}

%%%%%%%%%%%%%%%%%%%%%%%%%%%%%%%%%%%%%%%%%%%
\subsubsection{$SO(1)-[2]$ with $\mathcal{N}, (\mathrm{N,N})+\mathrm{Fermis}$}
%%%%%%%%%%%%%%%%%%%%%%%%%%%%%%%%%%%%%%%%%%%

For the Neumann half-indices in theory A we have
\begin{align}
\label{so1nf2_Nn++}
\mathbb{II}_{\mathcal{N},\mathrm{N}+\Psi}^{SO(1)-[2]_{++}^A}
&=
\frac{1}
{(q^{\frac{r}{2}} a;q)_{\infty}^2}
(q^{\frac12}u^{\pm};q)_{\infty}
(q^{\frac12};q)_{\infty}, \\
\label{so1nf2_Nn-+}
\mathbb{II}_{\mathcal{N},\mathrm{N}+\Psi}^{SO(1)-[2]_{-+}^A}
&=
\frac{1}
{(q^{\frac{r}{2}} a;q)_{\infty}^2}
(q^{\frac12}u^{\pm};q)_{\infty}
(-q^{\frac12};q)_{\infty}, \\
\label{so1nf2_Nn+-}
\mathbb{II}_{\mathcal{N},\mathrm{N}+\Psi}^{SO(1)-[2]_{+-}^A}
&=
\frac{1}
{(-q^{\frac{r}{2}} a;q)_{\infty}^2}
(-q^{\frac12}u^{\pm};q)_{\infty}
(q^{\frac12};q)_{\infty}, \\
\label{so1nf2_Nn--}
\mathbb{II}_{\mathcal{N},\mathrm{N}+\Psi}^{SO(1)-[2]_{--}^A}
&=
\frac{1}
{(-q^{\frac{r}{2}} a;q)_{\infty}^2}
(-q^{\frac12}u^{\pm};q)_{\infty}
(-q^{\frac12};q)_{\infty}.
\end{align}

These agrees with the theory B half-indices
\begin{align}
\label{so1nf2mag_Ddnd++}
\mathbb{II}_{\mathcal{D},\mathrm{D,N,D}}^{SO(3)-[2]_{++}^B}
&=
\frac{1}{(q)_{\infty}}
\sum_{m\in \mathbb{Z}}
\frac{1}
{(q^{1\pm m}u^{\pm};q)_{\infty}}
(q^{\frac{1+r}{2}\pm m}au^{\pm};q)_{\infty}^2
(q^{\frac{1+r}{2}}a;q)_{\infty}^2
(-1)^m q^{\frac{m^2}{2}}u^{m}
\nonumber\\
&\times 
\frac{1}
{(q^r a^2;q)_{\infty}^3}
(q^{-\frac12+r} a^2;q)_{\infty}, \\
\label{so1nf2mag_Ddnd--}
\mathbb{II}_{\mathcal{D},\mathrm{D,N,D}}^{SO(3)-[2]_{--}^B}
&=
\frac{1}{(q)_{\infty}}
\sum_{m\in \mathbb{Z}}
\frac{1}
{(-q^{1\pm m}u^{\pm};q)_{\infty}}
(q^{\frac{1+r}{2}\pm m}au^{\pm};q)_{\infty}^2
(-q^{\frac{1+r}{2}}a;q)_{\infty}^2
(-1)^m q^{\frac{m^2}{2}}u^{m}
\nonumber\\
&\times 
\frac{1}
{(q^r a^2;q)_{\infty}^3}
(-q^{-\frac12+r} a^2;q)_{\infty}, \\
\label{so1nf2mag_Ddnd+-}
\mathbb{II}_{\mathcal{D},\mathrm{D,N,D}}^{SO(3)-[2]_{+-}^B}
&=
\frac{1}{(q)_{\infty}}
\sum_{m\in \mathbb{Z}}
\frac{1}
{(-q^{1\pm m}u^{\pm};q)_{\infty}}
(q^{\frac{1+r}{2}\pm m}au^{\pm};q)_{\infty}^2
(-q^{\frac{1+r}{2}}a;q)_{\infty}^2
q^{\frac{m^2}{2}}u^{m}
\nonumber\\
&\times 
\frac{1}
{(q^r a^2;q)_{\infty}^3}
(q^{-\frac12+r} a^2;q)_{\infty}, \\
\label{so1nf2mag_Ddnd-+}
\mathbb{II}_{\mathcal{D},\mathrm{D,N,D}}^{SO(3)-[2]_{-+}^B}
&=
\frac{1}{(q)_{\infty}}
\sum_{m\in \mathbb{Z}}
\frac{1}
{(q^{1\pm m}u^{\pm};q)_{\infty}}
(q^{\frac{1+r}{2}\pm m}au^{\pm};q)_{\infty}^2
(q^{\frac{1+r}{2}}a;q)_{\infty}^2
q^{\frac{m^2}{2}}u^{m}
\nonumber\\
&\times 
\frac{1}
{(q^r a^2;q)_{\infty}^3}
(-q^{-\frac12+r} a^2;q)_{\infty}. 
\end{align}

%%%%%%%%%%%%%%%%%%%%%%%%%%%%%%%%%%%%%%%%%%%
\subsubsection{$SO(1)-[2]$ with $\mathcal{D}, (\mathrm{D,D})$}
%%%%%%%%%%%%%%%%%%%%%%%%%%%%%%%%%%%%%%%%%%%

The theory A Dirichlet half-indices are
\begin{align}
\label{so1nf2_Dd++}
\mathbb{II}_{\mathcal{D},\mathrm{D}}^{SO(1)-[2]_{\pm +}^A}
&=(q^{1-\frac{r}{2}}a^{-1};q)_{\infty}^2, \\
\label{so1nf2_Dd+-}
\mathbb{II}_{\mathcal{D},\mathrm{D}}^{SO(1)-[2]_{\pm -}^A}
&=(-q^{1-\frac{r}{2}}a^{-1};q)_{\infty}^2,
\end{align}
and the matching theory B half-indices are
\begin{align}
\label{so1nf2mag_Nndn++}
\mathbb{II}_{\mathcal{N},\mathrm{N,D,N}+\Psi}^{SO(3)-[2]_{++}^B}
&=
\frac12 (q)_{\infty}\oint \frac{ds}{2\pi is}
(s^{\pm};q)_{\infty}
\frac{1}
{
(q^{\frac{1-r}{2}}a^{-1}s^{\pm};q)_{\infty}^2
(q^{\frac{1-r}{2}}a^{-1};q)_{\infty}^2
}
(q^{\frac12}s^{\pm};q)_{\infty}
(q^{\frac12};q)_{\infty}
\nonumber\\
&\times 
(q^{1-r}a^{-2};q)_{\infty}^3 
\frac{1}
{(q^{\frac{3}{2}-r}a^{-2};q)_{\infty}}, \\
\label{so1nf2mag_Nndn--}
\mathbb{II}_{\mathcal{N},\mathrm{N,D,N}+\Psi}^{SO(3)-[2]_{--}^B}
&=
\frac12 (q)_{\infty}\oint \frac{ds}{2\pi is}
(-s^{\pm};q)_{\infty}
\frac{1}
{
(q^{\frac{1-r}{2}}a^{-1}s^{\pm};q)_{\infty}^2
(-q^{\frac{1-r}{2}}a^{-1};q)_{\infty}^2
}
(q^{\frac12}s^{\pm};q)_{\infty}
(-q^{\frac12};q)_{\infty}
\nonumber\\
&\times
(q^{1-r}a^{-2};q)_{\infty}^3
\frac{1}
{(-q^{\frac{3}{2}-r}a^{-2};q)_{\infty}}, \\
\label{so1nf2mag_Nndn+-}
\mathbb{II}_{\mathcal{N},\mathrm{N,D,N}+\Psi}^{SO(3)-[2]_{+-}^B}
&=
\frac12 (q)_{\infty}\oint \frac{ds}{2\pi is}
(-s^{\pm};q)_{\infty}
\frac{1}
{
(q^{\frac{1-r}{2}}a^{-1}s^{\pm};q)_{\infty}^2
(-q^{\frac{1-r}{2}}a^{-1};q)_{\infty}^2
}
(-q^{\frac12}s^{\pm};q)_{\infty}
(q^{\frac12};q)_{\infty}
\nonumber\\
&\times
(q^{1-r}a^{-2};q)_{\infty}^3
\frac{1}
{(q^{\frac{3}{2}-r}a^{-2};q)_{\infty}}, \\
\label{so1nf2mag_Nndn-+}
\mathbb{II}_{\mathcal{N},\mathrm{N,D,N}+\Psi}^{SO(3)-[2]_{-+}^B}
&=
\frac12 (q)_{\infty}\oint \frac{ds}{2\pi is}
(s^{\pm};q)_{\infty}
\frac{1}
{
(q^{\frac{1-r}{2}}a^{-1}s^{\pm};q)_{\infty}^2
(q^{\frac{1-r}{2}}a^{-1};q)_{\infty}^2
}
(-q^{\frac12}s^{\pm};q)_{\infty}
(-q^{\frac12};q)_{\infty}
\nonumber\\
&\times
(q^{1-r}a^{-2};q)_{\infty}^3
\frac{1}
{(-q^{\frac{3}{2}-r}a^{-2};q)_{\infty}}.
\end{align}

%%%%%%%%%%%%%%%%%%%%%%%%%%%%%%%%%%%%%%%%%%%
\subsection{$N_c=1, N_f=3$ $(SO(1)-[3])$}
%%%%%%%%%%%%%%%%%%%%%%%%%%%%%%%%%%%%%%%%%%

For theory A the full-indices
\begin{align}
\label{so1nf3_i++}
I^{SO(1)-[3]_{\pm +}^A}
&=
\frac{
(q^{1-\frac{r}{2}} a^{-1};q)_{\infty}^3
}
{
(q^{\frac{r}{2}} a;q)_{\infty}^3
}, \\
\label{so1nf3_i+-}
I^{SO(1)-[3]_{\pm -}^A}
&=
\frac{
(-q^{1-\frac{r}{2}} a^{-1};q)_{\infty}^3
}
{
(-q^{\frac{r}{2}} a;q)_{\infty}^3
}
\end{align}
agree with the full-indices in theory B given by
\begin{align}
\label{so1nf3mag_i++}
&I^{SO(4)-[3]_{++}^B}
\nonumber\\
&=
\frac{1}{2\cdot 2}
\sum_{m_1, m_2 \in \mathbb{Z}}
\oint \frac{ds_1}{2\pi is_1}\frac{ds_2}{2\pi is_2}
(1-q^{\frac{|m_1-m_2|}{2}}s_1^{\pm} s_2^{\mp})
(1-q^{\frac{|m_1-m_2|}{2}}s_1^{\pm} s_2^{\pm})
\nonumber\\
&\times 
\prod_{i=1}^{3}
\frac{
(q^{\frac{1+r+|m|}{2}} as_i^{\pm};q)_{\infty}^3
}
{
(q^{\frac{1-r+|m|}{2}} a^{-1}s_i^{\pm};q)_{\infty}^3
}
q^{\sum_{i=1}^3 \frac{3r}{2}|m_i|-\sum_{i<j}\frac{|m_i -m_j|}{2}-\sum_{i<j}\frac{|m_i+m_j|}{2}} 
a^{3\sum_{i=1}^3|m|}
\nonumber\\
&\times 
\frac{
(q^{1-r}a^{-2};q)_{\infty}^6
}
{
(q^{r}a^{2};q)_{\infty}^6
}
\frac{
(q^{-1+\frac{3r}{2}}a^3;q)_{\infty}
}
{
(q^{2-\frac{3r}{2}}a^{-3};q)_{\infty}
}, \\
\label{so1nf3mag_i--}
I^{SO(4)-[3]_{--}^B}
&=
\frac12 
\sum_{m\in \mathbb{Z}}
\oint \frac{ds}{2\pi is}
(1-q^{|m|}s^2)
(1-q^{|m|}s^{-2})
\nonumber\\
&\times 
\frac{
(q^{\frac{1+r+|m|}{2}} as^{\pm};q)_{\infty}^3
}
{
(q^{\frac{1-r+|m|}{2}} a^{-1}s^{\pm};q)_{\infty}^3
}
\frac{
(\pm q^{\frac{1+r}{2}} a;q)_{\infty}^3
}
{
(\pm q^{\frac{1-r}{2}} a^{-1};q)_{\infty}^3
}
q^{\frac{3r}{2}|m|-|m|} a^{3|m|} (-1)^m 
\nonumber\\
&\times 
\frac{
(q^{1-r}a^{-2};q)_{\infty}^6
}
{
(q^{r}a^{2};q)_{\infty}^6
}
\frac{
(-q^{-1+\frac{3r}{2}}a^3;q)_{\infty}
}
{
(-q^{2-\frac{3r}{2}}a^{-3};q)_{\infty}
}, \\
\label{so1nf3mag_i+-}
I^{SO(4)-[3]_{+-}^B}
&=
\frac12 
\sum_{m\in \mathbb{Z}}
\oint \frac{ds}{2\pi is}
(1-q^{|m|}s^2)
(1-q^{|m|}s^{-2})
\nonumber\\
&\times 
\frac{
(q^{\frac{1+r+|m|}{2}} as^{\pm};q)_{\infty}^3
}
{
(q^{\frac{1-r+|m|}{2}} a^{-1}s^{\pm};q)_{\infty}^3
}
\frac{
(\pm q^{\frac{1+r}{2}} a;q)_{\infty}^3
}
{
(\pm q^{\frac{1-r}{2}} a^{-1};q)_{\infty}^3
}
q^{\frac{3r}{2}|m|-|m|} a^{3|m|} 
\nonumber\\
&\times 
\frac{
(q^{1-r}a^{-2};q)_{\infty}^6
}
{
(q^{r}a^{2};q)_{\infty}^6
}
\frac{
(q^{-1+\frac{3r}{2}}a^3;q)_{\infty}
}
{
(q^{2-\frac{3r}{2}}a^{-3};q)_{\infty}
}, \\
\label{so1nf3mag_i-+}
&I^{SO(4)-[3]_{-+}^B}
\nonumber\\
&=
\frac{1}{2\cdot 2}
\sum_{m_1, m_2 \in \mathbb{Z}}
\oint \frac{ds_1}{2\pi is_1}\frac{ds_2}{2\pi is_2}
(1-q^{\frac{|m_1-m_2|}{2}}s_1^{\pm} s_2^{\mp})
(1-q^{\frac{|m_1-m_2|}{2}}s_1^{\pm} s_2^{\pm})
\nonumber\\
&\times 
\prod_{i=1}^{3}
\frac{
(q^{\frac{1+r+|m|}{2}} as_i^{\pm};q)_{\infty}^3
}
{
(q^{\frac{1-r+|m|}{2}} a^{-1}s_i^{\pm};q)_{\infty}^3
}
q^{\sum_{i=1}^3 \frac{3r}{2}|m_i|-\sum_{i<j}\frac{|m_i -m_j|}{2}-\sum_{i<j}\frac{|m_i+m_j|}{2}} 
a^{3\sum_{i=1}^3|m|}
(-1)^{\sum_{i=1}^3 m_i}
\nonumber\\
&\times 
\frac{
(q^{1-r}a^{-2};q)_{\infty}^6
}
{
(q^{r}a^{2};q)_{\infty}^6
}
\frac{
(-q^{-1+\frac{3r}{2}}a^3;q)_{\infty}
}
{
(-q^{2-\frac{3r}{2}}a^{-3};q)_{\infty}
} \; .
\end{align}

%%%%%%%%%%%%%%%%%%%%%%%%%%%%%%%%%%%%%%%%%%%
\subsubsection{$SO(1)-[3]$ with $\mathcal{N}, (\mathrm{N,N,N})+\mathrm{Fermis}$}
%%%%%%%%%%%%%%%%%%%%%%%%%%%%%%%%%%%%%%%%%%%

The half-indices with Neumann boundary conditions in theory A are
\begin{align}
\label{so1nf3_Nn++}
\mathbb{II}_{\mathcal{N},\mathrm{N}+\Psi}^{SO(1)-[3]_{++}^A}
&=
\frac{1}
{(q^{\frac{r}{2}} a;q)_{\infty}^3}
(q^{\frac12}u_1^{\pm};q)_{\infty}
(q^{\frac12}u_2^{\pm};q)_{\infty}, \\
\label{so1nf3_Nn-+}
\mathbb{II}_{\mathcal{N},\mathrm{N}+\Psi}^{SO(1)-[3]_{-+}^A}
&=
\frac{1}
{(q^{\frac{r}{2}} a;q)_{\infty}^3}
(q^{\frac12}u^{\pm};q)_{\infty}
(\pm q^{\frac12};q)_{\infty}, \\
\label{so1nf3_Nn+-}
\mathbb{II}_{\mathcal{N},\mathrm{N}+\Psi}^{SO(1)-[3]_{+-}^A}
&=
\frac{1}
{(-q^{\frac{r}{2}} a;q)_{\infty}^3}
(-q^{\frac12}u^{\pm};q)_{\infty}
(\pm q^{\frac12};q)_{\infty}, \\
\label{so1nf3_Nn--}
\mathbb{II}_{\mathcal{N},\mathrm{N}+\Psi}^{SO(1)-[3]_{--}^A}
&=
\frac{1}
{(-q^{\frac{r}{2}} a;q)_{\infty}^3}
(-q^{\frac12}u_1^{\pm};q)_{\infty}
(-q^{\frac12}u_2^{\pm};q)_{\infty}
\end{align}
and the matching theory B half-indices are
\begin{align}
\label{so1nf3mag_Ddnd++}
\mathbb{II}_{\mathcal{D},\mathrm{D,N,D}}^{SO(4)-[3]_{++}^B}
&=
\frac{1}{(q)_{\infty}^2}
\sum_{m_1, m_2 \in \mathbb{Z}}
\frac{1}
{
(q^{1\pm m_1\mp m_2}u_1^{\pm}u_2^{\mp};q)_{\infty}
(q^{1\pm m_1\pm m_2}u_1^{\pm}u_2^{\pm};q)_{\infty}
}
\nonumber\\
&\times 
\prod_{i=1}^2
(q^{\frac{1+r}{2}\pm m_i}au_i^{\pm};q)_{\infty}^3
(-1)^{m_1+m_2} q^{\frac{m_1^2+m_2^2}{2}}u_1^{m_1}u_2^{m_2}
\nonumber\\
&\times 
\frac{1}
{(q^r a^2;q)_{\infty}^6}
(q^{-1+\frac{3r}{2}} a^3;q)_{\infty}, \\
\label{so1nf3mag_Ddnd--}
\mathbb{II}_{\mathcal{D},\mathrm{D,N,D}}^{SO(4)-[3]_{--}^B}
&=
\frac{1}{(q)_{\infty} (-q;q)_{\infty}}
\sum_{m\in \mathbb{Z}}
\frac{1}
{
(q^{1\pm m}u^{\pm};q)_{\infty}
(-q^{1\pm m}u^{\pm};q)_{\infty}
}
\nonumber\\
&\times 
(q^{\frac{1+r}{2}\pm m}au^{\pm};q)_{\infty}^3
(\pm q^{\frac{1+r}{2}}a;q)_{\infty}^3
(-1)^{m} q^{\frac{m^2}{2}}u^{m}
\nonumber\\
&\times 
\frac{1}
{(q^r a^2;q)_{\infty}^6}
(-q^{-1+\frac{3r}{2}} a^3;q)_{\infty}, \\
\label{so1nf3mag_Ddnd+-}
\mathbb{II}_{\mathcal{D},\mathrm{D,N,D}}^{SO(4)-[3]_{+-}^B}
&=
\frac{1}{(q)_{\infty} (-q;q)_{\infty}}
\sum_{m\in \mathbb{Z}}
\frac{1}
{
(q^{1\pm m}u^{\pm};q)_{\infty}
(-q^{1\pm m}u^{\pm};q)_{\infty}
}
\nonumber\\
&\times 
(q^{\frac{1+r}{2}\pm m}au^{\pm};q)_{\infty}^3
(\pm q^{\frac{1+r}{2}}a;q)_{\infty}^3
q^{\frac{m^2}{2}}u^{m}
\nonumber\\
&\times 
\frac{1}
{(q^r a^2;q)_{\infty}^6}
(q^{-1+\frac{3r}{2}} a^3;q)_{\infty}, \\
\label{so1nf3mag_Ddnd-+}
\mathbb{II}_{\mathcal{D},\mathrm{D,N,D}}^{SO(4)-[3]_{-+}^B}
&=
\frac{1}{(q)_{\infty}^2}
\sum_{m_1, m_2 \in \mathbb{Z}}
\frac{1}
{
(q^{1\pm m_1\mp m_2}u_1^{\pm}u_2^{\mp};q)_{\infty}
(q^{1\pm m_1\pm m_2}u_1^{\pm}u_2^{\pm};q)_{\infty}
}
\nonumber\\
&\times 
\prod_{i=1}^2
(q^{\frac{1+r}{2}\pm m_i}au_i^{\pm};q)_{\infty}^3
q^{\frac{m_1^2+m_2^2}{2}}u_1^{m_1}u_2^{m_2}
\nonumber\\
&\times 
\frac{1}
{(q^r a^2;q)_{\infty}^6}
(-q^{-1+\frac{3r}{2}} a^3;q)_{\infty}.
\end{align}

%%%%%%%%%%%%%%%%%%%%%%%%%%%%%%%%%%%%%%%%%%%
\subsubsection{$SO(1)-[3]$ with $\mathcal{D}, (\mathrm{D,D,D})$}
%%%%%%%%%%%%%%%%%%%%%%%%%%%%%%%%%%%%%%%%%%%

With Dirichlet conditions the theory A half-indices are
\begin{align}
\label{so1nf3_Dd++}
\mathbb{II}_{\mathcal{D},\mathrm{D}}^{SO(1)-[3]_{++}^A}
&=
(q^{1-\frac{r}{2}}a^{-1};q)_{\infty}^3, \\
\label{so1nf3_Dd+-}
\mathbb{II}_{\mathcal{D},\mathrm{D}}^{SO(1)-[3]_{\pm -}^A}
&=
(-q^{1-\frac{r}{2}}a^{-1};q)_{\infty}^3 \; .
\end{align}
The matching theory B half-indices are
\begin{align}
\label{so1nf3mag_Nndn++}
\mathbb{II}_{\mathcal{N},\mathrm{N,D,N}+\Psi}^{SO(4)-[3]_{++}^B}
&=
\frac{1}{2\cdot 2}
\oint \frac{ds_1}{2\pi is_1}\frac{ds_2}{2\pi is_2}
(s_1^{\pm}s_2^{\mp};q)_{\infty}
(s_1^{\pm}s_2^{\pm};q)_{\infty}
\nonumber\\
&\times 
\prod_{i=1}^2 
\frac{1}
{(q^{\frac{1-r}{2}} a^{-1}s_i^{\pm};q)_{\infty}^3}
(q^{\frac12}s_i^{\pm};q)_{\infty}
(q^{1-r}a^{-2};q)_{\infty}^6
\frac{1}{(q^{2-\frac{3r}{2}}a^{-3};q)_{\infty}}, \\
\label{so1nf3mag_Nndn--}
\mathbb{II}_{\mathcal{N},\mathrm{N,D,N}+\Psi}^{SO(4)-[3]_{--}^B}
&=
\frac{1}{2}
(q)_{\infty}
(-q;q)_{\infty}
\oint \frac{ds}{2\pi is}
(s^{\pm};q)_{\infty}
(-s^{\pm};q)_{\infty}
\nonumber\\
&\times 
\frac{1}
{
(q^{\frac{1-r}{2}} a^{-1}s^{\pm};q)_{\infty}^3
(\pm q^{\frac{1-r}{2}} a^{-1};q)_{\infty}^3
}
(q^{\frac12}s^{\pm};q)_{\infty}
(\pm q^{\frac12};q)_{\infty}
\nonumber\\
&\times 
(q^{1-r}a^{-2};q)_{\infty}^6
\frac{1}{(-q^{2-\frac{3r}{2}}a^{-3};q)_{\infty}}, \\
\label{so1nf3mag_Nndn+-}
\mathbb{II}_{\mathcal{N},\mathrm{N,D,N}+\Psi}^{SO(4)-[3]_{+-}^B}
&=
\frac{1}{2}
(q)_{\infty}
(-q;q)_{\infty}
\oint \frac{ds}{2\pi is}
(s^{\pm};q)_{\infty}
(-s^{\pm};q)_{\infty}
\nonumber\\
&\times 
\frac{1}
{
(q^{\frac{1-r}{2}} a^{-1}s^{\pm};q)_{\infty}^3
(\pm q^{\frac{1-r}{2}} a^{-1};q)_{\infty}^3
}
(-q^{\frac12}s^{\pm};q)_{\infty}
(\pm q^{\frac12};q)_{\infty}
\nonumber\\
&\times 
(q^{1-r}a^{-2};q)_{\infty}^6
\frac{1}{(q^{2-\frac{3r}{2}}a^{-3};q)_{\infty}}, \\
\label{so1nf3mag_Nndn-+}
\mathbb{II}_{\mathcal{N},\mathrm{N,D,N}+\Psi}^{SO(4)-[3]_{-+}^B}
&=
\frac{1}{2\cdot 2}
\oint \frac{ds_1}{2\pi is_1}\frac{ds_2}{2\pi is_2}
(s_1^{\pm}s_2^{\mp};q)_{\infty}
(s_1^{\pm}s_2^{\pm};q)_{\infty}
\nonumber\\
&\times 
\prod_{i=1}^2 
\frac{1}
{(q^{\frac{1-r}{2}} a^{-1}s_i^{\pm};q)_{\infty}^3}
(-q^{\frac12}s_i^{\pm};q)_{\infty}
(q^{1-r}a^{-2};q)_{\infty}^6
\frac{1}{(-q^{2-\frac{3r}{2}}a^{-3};q)_{\infty}} \; .
\end{align}

%%%%%%%%%%%%%%%%%%%%%%%%%%%%%%%%%%%%%%%%%%%
\subsection{$N_c=2, N_f=2$ $(SO(2)-[2])$}
%%%%%%%%%%%%%%%%%%%%%%%%%%%%%%%%%%%%%%%%%%%

We now consider an example with $SO(2)$ gauge group in theory A. This is
equivalent to $U(1)$ with each fundamental chiral of $SO(2)$ corresponding
to two chirals of $U(1)$ with charges $\pm 1$. Note that the cases with
$\chi = -1$ in theory A are not sensitive to the value of $\zeta$ so we have
\begin{align}
I^{SO(N_f)-[N_f]_{+-}^B} = & I^{SO(2)-[N_f]_{\pm -}^A} = I^{SO(N_f)-[N_f]_{--}^B}.
\end{align}
Here we present only the cases with $N_f = 2$ which gives gauge group $SO(2)$
also in theory B.

The theory A full-indices are
\begin{align}
\label{so2nf2_i++}
I^{SO(2)-[2]_{++}^A}
&=
\sum_{m\in \mathbb{Z}}
\oint \frac{ds}{2\pi is} 
\frac
{
(q^{1-\frac{r+|m|}{2}} a^{-1}s^{-1};q)_{\infty}^2
(q^{1-\frac{r+|m|}{2}} a^{-1}s;q)_{\infty}^2
}
{
(q^{\frac{r+|m|}{2}} as;q)_{\infty}^2
(q^{\frac{r+|m|}{2}} as^{-1};q)_{\infty}^2
}
q^{ (1-r)|m| }
a^{-2|m|}, \\
\label{so2nf2_i-+}
I^{SO(2)-[2]_{-+}^A}
&=
\sum_{m\in \mathbb{Z}}
\oint \frac{ds}{2\pi is}
\frac
{
(q^{1-\frac{r+|m|}{2}} a^{-1}s^{-1};q)_{\infty}^2
(q^{1-\frac{r+|m|}{2}} a^{-1}s;q)_{\infty}^2
}
{
(q^{\frac{r+|m|}{2}} as;q)_{\infty}^2
(q^{\frac{r+|m|}{2}} as^{-1};q)_{\infty}^2
}
q^{ (1-r)|m| }
a^{-2|m|}
(-1)^m, \\
\label{so2nf2_i+-}
I^{SO(2)-[2]_{\pm-}^A}
&=
\frac
{
(q^{1-\frac{r}{2}} a^{-1};q)_{\infty}^2
(-q^{1-\frac{r}{2}} a^{-1};q)_{\infty}^2
}
{
(q^{\frac{r}{2}} a;q)_{\infty}^2
(-q^{\frac{r}{2}} a;q)_{\infty}^2
}.
\end{align}
The matching theory B indices are
\begin{align}
\label{so2nf2mag_i++}
I^{SO(2)-[2]_{++}^B}
&=
\sum_{m\in \mathbb{Z}}
\oint \frac{ds}{2\pi is}
\frac{
(q^{\frac{1+r+|m|}{2}}as^{-1};q)_{\infty}^2
(q^{\frac{1+r+|m|}{2}}as;q)_{\infty}^2
}
{
(q^{\frac{1-r+|m|}{2}}a^{-1}s;q)_{\infty}^2
(q^{\frac{1-r+|m|}{2}}a^{-1}s^{-1};q)_{\infty}^2
}
q^{r|m|} a^{2|m|}
\nonumber\\ 
\times &
\frac{
(q^{1-r}a^{-2};q)_{\infty}^3
}
{
(q^{r}a^2;q)_{\infty}^3
}
\frac{
(q^{r} a^2;q)_{\infty}
}
{
(q^{1-r} a^{-2};q)_{\infty}
}, \\
\label{so2nf2mag_i--}
I^{SO(2)-[2]_{--}^B}
&=
\frac{
(q^{\frac{1+r}{2}}a;q)_{\infty}^2
(-q^{\frac{1+r}{2}}a;q)_{\infty}^2
}
{
(q^{\frac{1-r}{2}}a^{-1};q)_{\infty}^2
(-q^{\frac{1-r}{2}}a^{-1};q)_{\infty}^2
}
\frac{
(q^{1-r}a^{-2};q)_{\infty}^3
}
{
(q^{r}a^2;q)_{\infty}^3
}
\frac{
(-q^{r} a^2;q)_{\infty}
}
{
(-q^{1-r} a^{-2};q)_{\infty}
}, \\
\label{so2nf2mag_i+-}
I^{SO(2)-[2]_{+-}^B}
&=
\frac{
(q^{\frac{1+r}{2}}a;q)_{\infty}^2
(-q^{\frac{1+r}{2}}a;q)_{\infty}^2
}
{
(q^{\frac{1-r}{2}}a^{-1};q)_{\infty}^2
(-q^{\frac{1-r}{2}}a^{-1};q)_{\infty}^2
}
\frac{
(q^{1-r}a^{-2};q)_{\infty}^3
}
{
(q^{r}a^2;q)_{\infty}^3
}
\frac{
(q^{r} a^2;q)_{\infty}
}
{
(q^{1-r} a^{-2};q)_{\infty}
}, \\
\label{so2nf2mag_i-+}
I^{SO(2)-[2]_{-+}^B}
&=
\sum_{m\in \mathbb{Z}}
\oint \frac{ds}{2\pi is}
\frac{
(q^{\frac{1+r+|m|}{2}}as^{-1};q)_{\infty}^2
(q^{\frac{1+r+|m|}{2}}as;q)_{\infty}^2
}
{
(q^{\frac{1-r+|m|}{2}}a^{-1}s;q)_{\infty}^2
(q^{\frac{1-r+|m|}{2}}a^{-1}s^{-1};q)_{\infty}^2
}
q^{r|m|} a^{2|m|} (-1)^m
\nonumber\\ 
\times &
\frac{
(q^{1-r}a^{-2};q)_{\infty}^3
}
{
(q^{r}a^2;q)_{\infty}^3
}
\frac{
(-q^{r} a^2;q)_{\infty}
}
{
(-q^{1-r} a^{-2};q)_{\infty}
} \; .
\end{align}
In the indices (\ref{so2nf2mag_i++}) and (\ref{so2nf2mag_i+-}) 
the contribution from the gauge singlet $V$ 
is cancelled by that from the meson $M$. 
Accordingly, the index (\ref{so2nf2mag_i++}) can be alternatively interpreted as 
the index of the mirror theory \cite{deBoer:1997ka, Krattenthaler:2011da, Imamura:2011su} 
for the SQED$_2$, i.e. the $U(1)$ gauge theory with two pairs of chiral multiplets of charge $\pm 1$ 
as well as gauge singlets. 
Note that such a cancellation does not occur and the indices 
of theory B and that of the mirror theory are distinguished when one turns on the fugacities for the flavour and topological symmetries. 

%%%%%%%%%%%%%%%%%%%%%%%%%%%%%%%%%%%%%%%%%%%
\subsubsection{$SO(2)-[2]$ with $\mathcal{N}, (\mathrm{N,N})+\mathrm{Fermis}$}
%%%%%%%%%%%%%%%%%%%%%%%%%%%%%%%%%%%%%%%%%%%

The theory A half-indices are
\begin{align}
\label{so2nf2_Nn++}
\mathbb{II}_{\mathcal{N},\mathrm{N}+\Psi}^{SO(2)-[2]_{++}^A}
&=
(q)_{\infty}\oint \frac{ds}{2\pi is}
\frac{1}
{
(q^{\frac{r}{2}}as;q)_{\infty}^2
(q^{\frac{r}{2}}as^{-1};q)_{\infty}^2
}
(q^{\frac12}s^{\pm}u^{\pm};q)_{\infty}
(q^{\frac12}s^{\pm}u^{\mp};q)_{\infty}, \\
\label{so2nf2_Nn-+}
\mathbb{II}_{\mathcal{N},\mathrm{N}+\Psi}^{SO(2)-[2]_{-+}^A}
&=
(q)_{\infty}\oint \frac{ds}{2\pi is}
\frac{1}
{
(q^{\frac{r}{2}}as;q)_{\infty}^2
(q^{\frac{r}{2}}as^{-1};q)_{\infty}^2
}
(\pm q^{\frac12}s;q)_{\infty}
(\pm q^{\frac12}s^{-1};q)_{\infty}, \\
\label{so2nf2_Nn+-}
\mathbb{II}_{\mathcal{N},\mathrm{N}+\Psi}^{SO(2)-[2]_{+-}^A}
&=
(-q;q)_{\infty}
\frac{1}
{
(q^{\frac{r}{2}}a;q)_{\infty}^2
(-q^{\frac{r}{2}}a;q)_{\infty}^2
}
(q^{\frac12};q)_{\infty}^2
(-q^{\frac12};q)_{\infty}^2, \\
\label{so2nf2_Nn--}
\mathbb{II}_{\mathcal{N},\mathrm{N}+\Psi}^{SO(2)-[2]_{--}^A}
&=
(-q;q)_{\infty}
\frac{1}
{
(q^{\frac{r}{2}}a;q)_{\infty}^2
(-q^{\frac{r}{2}}a;q)_{\infty}^2
}
(q^{\frac12}u^{\pm};q)_{\infty}^2
(-q^{\frac12}u^{\pm};q)_{\infty}^2 \;
\end{align}
They agree with theory B indices
\begin{align}
\label{so2nf2mag_Ddnd++}
\mathbb{II}_{\mathcal{D},\mathrm{D,N,D}}^{SO(2)-[2]_{++}^B}
&=
\frac{1}{(q)_{\infty}}
\sum_{m\in \mathbb{Z}}
(q^{\frac{1+r}{2}\pm m}au^{\pm};q)_{\infty}^2
q^{m^2}u^{2m}
\frac{1}
{(q^r a^2;q)_{\infty}^3}
(q^{r} a^2;q)_{\infty}, \\
\label{so2nf2mag_Ddnd--}
\mathbb{II}_{\mathcal{D},\mathrm{D,N,D}}^{SO(2)-[2]_{--}^B}
&=
\frac{1}{(-q;q)_{\infty}}
(q^{\frac{1+r}{2}}a;q)_{\infty}^2
(-q^{\frac{1+r}{2}}a;q)_{\infty}^2
\frac{1}
{(q^r a^2;q)_{\infty}^3}
(-q^{r} a^2;q)_{\infty}, \\
\label{so2nf2mag_Ddnd+-}
\mathbb{II}_{\mathcal{D},\mathrm{D,N,D}}^{SO(2)-[2]_{+-}^B}
&=
\frac{1}{(-q;q)_{\infty}}
(q^{\frac{1+r}{2}}a;q)_{\infty}^2
(-q^{\frac{1+r}{2}}a;q)_{\infty}^2
\frac{1}
{(q^r a^2;q)_{\infty}^3}
(q^{r} a^2;q)_{\infty}, \\
\label{so2nf2mag_Ddnd-+}
\mathbb{II}_{\mathcal{D},\mathrm{D,N,D}}^{SO(2)-[2]_{-+}^B}
&=
\frac{1}{(q)_{\infty}}
\sum_{m\in \mathbb{Z}}
(q^{\frac{1+r}{2}\pm m}a u^{\pm};q)_{\infty}^2
(-1)^m 
q^{m^2}
u^{2m}
\frac{1}
{(q^r a^2;q)_{\infty}^3}
(-q^{r} a^2;q)_{\infty}.
\end{align}

%%%%%%%%%%%%%%%%%%%%%%%%%%%%%%%%%%%%%%%%%%%
\subsubsection{$SO(2)-[2]$ with $\mathcal{D}, (\mathrm{D,D})$}
%%%%%%%%%%%%%%%%%%%%%%%%%%%%%%%%%%%%%%%%%%%

As for the full-indices, for $\chi = -1$ the theory A Dirichlet index is
not sensitive to the value of $\zeta$ so we have identities
\begin{align}
\II_{\mathcal{N},\mathrm{N,D,N}}^{SO(N_f)-[N_f]_{+-}^B} = & \II_{\mathcal{D},\mathrm{D}}^{SO(2)-[N_f]_{\pm -}^A} = \II_{\mathcal{N},\mathrm{N,D,N}}^{SO(N_f)-[N_f]_{--}^B}.
\end{align}

For $N_f = 2$ we have the half-indices in theory A
\begin{align}
\label{so2nf2_Dd++}
\mathbb{II}_{\mathcal{D},\mathrm{D}}^{SO(2)-[2]_{++}^A}
&=
\frac{1}{(q)_{\infty}}
\sum_{m\in \mathbb{Z}}
(q^{1-\frac{r}{2}\pm m}a^{-1}u^{\pm};q)_{\infty}^2 
q^{m^2} u^{2m}, \\
\label{so2nf2_Dd-+}
\mathbb{II}_{\mathcal{D},\mathrm{D}}^{SO(2)-[2]_{-+}^A}
&=
\frac{1}{(q)_{\infty}}
\sum_{m\in \mathbb{Z}}
(q^{1-\frac{r}{2}\pm m}a^{-1}u^{\pm};q)_{\infty}^2
(-1)^m q^{m^2} u^{2m}, \\
\label{so2nf2_Dd+-}
\mathbb{II}_{\mathcal{D},\mathrm{D}}^{SO(2)-[2]_{\pm-}^A}
&=
\frac{1}{(-q;q)_{\infty}}
(q^{1-\frac{r}{2}}a^{-1};q)_{\infty}^2
(-q^{1-\frac{r}{2}}a^{-1};q)_{\infty}^2 \; .
\end{align}
The matching theory B half-indices are
\begin{align}
\label{so2nf2mag_Nndn++}
\mathbb{II}_{\mathcal{N},\mathrm{N,D,N}+\Psi}^{SO(2)-[2]_{++}^B}
&=
(q)_{\infty}
\oint \frac{ds}{2\pi is}
\frac{1}
{(q^{\frac{1-r}{2}} a^{-1}s^{\pm};q)_{\infty}^2}
(q^{\frac12}s^{\pm}u^{\pm};q)_{\infty}
(q^{\frac12}s^{\pm}u^{\mp};q)_{\infty}
\nonumber\\
&\times 
(q^{1-r}a^{-2};q)_{\infty}^3
\frac{1}
{
(q^{1-r}a^{-2};q)_{\infty}
}, \\
\label{so2nf2mag_Nndn--}
\mathbb{II}_{\mathcal{N},\mathrm{N,D,N}+\Psi}^{SO(2)-[2]_{--}^B}
&=
(-q;q)_{\infty}
\frac{1}
{
(q^{\frac{1-r}{2}} a^{-1};q)_{\infty}^2
(-q^{\frac{1-r}{2}} a^{-1};q)_{\infty}^2
}
(q^{\frac12}u^{\pm};q)_{\infty}
(-q^{\frac12}u^{\pm};q)_{\infty}
\nonumber\\
&\times 
(q^{1-r}a^{-2};q)_{\infty}^3
\frac{1}
{
(-q^{1-r}a^{-2};q)_{\infty}
}, \\
\label{so2nf2mag_Nndn+-}
\mathbb{II}_{\mathcal{N},\mathrm{N,D,N}+\Psi}^{SO(2)-[2]_{+-}^B}
&=
(-q;q)_{\infty}
\frac{1}
{
(q^{\frac{1-r}{2}} a^{-1};q)_{\infty}^2
(-q^{\frac{1-r}{2}} a^{-1};q)_{\infty}^2
}
(q^{\frac12};q)_{\infty}^2
(-q^{\frac12};q)_{\infty}^2
\nonumber\\
&\times 
(q^{1-r}a^{-2};q)_{\infty}^3
\frac{1}
{
(q^{1-r}a^{-2};q)_{\infty}
}, \\
\label{so2nf2mag_Nndn-+}
\mathbb{II}_{\mathcal{N},\mathrm{N,D,N}+\Psi}^{SO(2)-[2]_{-+}^B}
&=
(q)_{\infty}
\oint \frac{ds}{2\pi is}
\frac{1}
{
(q^{\frac{1-r}{2}} a^{-1}s^{\pm};q)_{\infty}^2
}
(q^{\frac12}s^{\pm};q)_{\infty}^2
(-q^{\frac12}s^{\pm};q)_{\infty}^2
\nonumber\\
&\times 
(q^{1-r}a^{-2};q)_{\infty}^3
\frac{1}
{
(-q^{1-r}a^{-2};q)_{\infty}
}.
\end{align}

%%%%%%%%%%%%%%%%%%%%%%%%%%%%%%%%%%%%%%%%%%%
\subsection{$N_c=3, N_f=3$ $(SO(3)-[3])$}
%%%%%%%%%%%%%%%%%%%%%%%%%%%%%%%%%%%%%%%%%%%

We now consider examples with non-Abelian gauge groups. In the first case with
$SO(3)-[3]^A$ the dual theory B is still Abelian with gauge group $SO(2)$. In
the following section we will look at the case of $N_f = 4$ where theory B
also has gauge group $SO(3)$.

The theory A indices are
\begin{align}
\label{so3nf3_i++}
I^{SO(3)-[3]_{++}^A}
&=
\frac12 \sum_{m\in \mathbb{Z}}
\oint \frac{ds}{2\pi is}
(1-q^{\frac{|m|}{2}} s)
(1-q^{\frac{|m|}{2}} s^{-1})
\nonumber\\
&\times 
\frac
{
(q^{1-\frac{r+|m|}{2}} a^{-1}s^{-1};q)_{\infty}^3
(q^{1-\frac{r+|m|}{2}} a^{-1}s;q)_{\infty}^3
(q^{1-\frac{r}{2}} a^{-1};q)_{\infty}^3
}
{
(q^{\frac{r+|m|}{2}} as;q)_{\infty}^3
(q^{\frac{r+|m|}{2}} as^{-1};q)_{\infty}^3
(q^{\frac{r}{2}} a;q)_{\infty}^3
}
q^{\frac{3(1-r)|m|}{2}-\frac{|m|}{2}}
a^{-3|m|}, \\
\label{so3nf3_i-+}
I^{SO(3)-[3]_{-+}^A}
&=
\frac12 \sum_{m\in \mathbb{Z}}
\oint \frac{ds}{2\pi is}
(1-q^{\frac{|m|}{2}} s)
(1-q^{\frac{|m|}{2}} s^{-1})
\nonumber\\
&\times
\frac
{
(q^{1-\frac{r+|m|}{2}} a^{-1}s^{-1};q)_{\infty}^3
(q^{1-\frac{r+|m|}{2}} a^{-1}s;q)_{\infty}^3
(q^{1-\frac{r}{2}} a^{-1};q)_{\infty}^3
}
{
(q^{\frac{r+|m|}{2}} as;q)_{\infty}^3
(q^{\frac{r+|m|}{2}} as^{-1};q)_{\infty}^3
(q^{\frac{r}{2}} a;q)_{\infty}^3
}
(-1)^m
q^{\frac{3(1-r)|m|}{2}-\frac{|m|}{2}}
a^{-3|m|}, \\
\label{so3nf3_i+-}
I^{SO(3)-[3]_{+-}^A}
&=
\frac12 \sum_{m\in \mathbb{Z}}
\oint \frac{ds}{2\pi is}
(1+q^{\frac{|m|}{2}} s)
(1+q^{\frac{|m|}{2}} s^{-1})
\nonumber\\
&\times
\frac
{
(q^{1-\frac{r+|m|}{2}} a^{-1}s^{-1};q)_{\infty}^3
(q^{1-\frac{r+|m|}{2}} a^{-1}s;q)_{\infty}^3
(-q^{1-\frac{r}{2}} a^{-1};q)_{\infty}^3
}
{
(q^{\frac{r+|m|}{2}} as;q)_{\infty}^3
(q^{\frac{r+|m|}{2}} as^{-1};q)_{\infty}^3
(-q^{\frac{r}{2}} a;q)_{\infty}^3
}
q^{\frac{3(1-r)|m|}{2}-\frac{|m|}{2}}
a^{-3|m|}, \\
\label{so3nf3_i--}
I^{SO(3)-[3]_{--}^A}
&=
\frac12 \sum_{m\in \mathbb{Z}}
\oint \frac{ds}{2\pi is}
(1+q^{\frac{|m|}{2}} s)
(1+q^{\frac{|m|}{2}} s^{-1})
\nonumber\\
&\times
\frac
{
(q^{1-\frac{r+|m|}{2}} a^{-1}s^{-1};q)_{\infty}^3
(q^{1-\frac{r+|m|}{2}} a^{-1}s;q)_{\infty}^3
(-q^{1-\frac{r}{2}} a^{-1};q)_{\infty}^3
}
{
(q^{\frac{r+|m|}{2}} as;q)_{\infty}^3
(q^{\frac{r+|m|}{2}} as^{-1};q)_{\infty}^3
(-q^{\frac{r}{2}} a;q)_{\infty}^3
}
q^{\frac{3(1-r)|m|}{2}-\frac{|m|}{2}}
a^{-3|m|}
(-1)^m \; .
\end{align}
These indices coincide with theory B indices
\begin{align}
\label{so3nf3mag_i++}
I^{SO(2)-[3]_{++}^B}
&=
\sum_{m\in \mathbb{Z}}
\oint \frac{ds}{2\pi is}
\frac{
(q^{\frac{1+r+|m|}{2}}as^{-1};q)_{\infty}^3
(q^{\frac{1+r+|m|}{2}}as;q)_{\infty}^3
(q^{\frac{1+r}{2}}a;q)_{\infty}^3
}
{
(q^{\frac{1-r+|m|}{2}}a^{-1}s;q)_{\infty}^3
(q^{\frac{1-r+|m|}{2}}a^{-1}s^{-1};q)_{\infty}^3
(q^{\frac{1-r}{2}}a^{-1};q)_{\infty}^3
}
q^{\frac{3r|m|}{2}}a^{3|m|}
\nonumber\\
&\times 
\frac{
(q^{1-r}a^{-2};q)_{\infty}^6
}
{
(q^{r}a^2;q)_{\infty}^6
}
\frac{
(q^{\frac{3r}{2}}a^3;q)_{\infty}
}
{
(q^{1-\frac{3r}{2}} a^{-3};q)_{\infty}
}, \\
\label{so3nf3mag_i--}
I^{SO(2)-[3]_{--}^B}
&=
\frac{
(q^{\frac{1+r}{2}} a;q)_{\infty}^3
(-q^{\frac{1+r}{2}} a;q)_{\infty}^3
}
{
(q^{\frac{1-r}{2}} a^{-1};q)_{\infty}^3
(-q^{\frac{1-r}{2}} a^{-1};q)_{\infty}^3
}
\frac{
(q^{1-r}a^{-2};q)_{\infty}^6
}
{
(q^{r}a^2;q)_{\infty}^6
}
\frac{
(-q^{\frac{3r}{2}}a^3;q)_{\infty}
}
{
(-q^{1-\frac{3r}{2}} a^{-3};q)_{\infty}
}, \\
\label{so3nf3mag_i+-}
I^{SO(2)-[3]_{+-}^B}
&=
\frac{
(q^{\frac{1+r}{2}} a;q)_{\infty}^3
(-q^{\frac{1+r}{2}} a;q)_{\infty}^3
}
{
(q^{\frac{1-r}{2}} a^{-1};q)_{\infty}^3
(-q^{\frac{1-r}{2}} a^{-1};q)_{\infty}^3
}
\frac{
(q^{1-r}a^{-2};q)_{\infty}^6
}
{
(q^{r}a^2;q)_{\infty}^6
}
\frac{
(q^{\frac{3r}{2}}a^3;q)_{\infty}
}
{
(q^{1-\frac{3r}{2}} a^{-3};q)_{\infty}
}, \\
\label{so3nf3mag_i-+}
I^{SO(2)-[3]_{-+}^B}
&=
\sum_{m\in \mathbb{Z}}
\oint \frac{ds}{2\pi is}
\frac{
(q^{\frac{1+r+|m|}{2}}as^{-1};q)_{\infty}^3
(q^{\frac{1+r+|m|}{2}}as;q)_{\infty}^3
}
{
(q^{\frac{1-r+|m|}{2}}a^{-1}s;q)_{\infty}^3
(q^{\frac{1-r+|m|}{2}}a^{-1}s^{-1};q)_{\infty}^3
}
q^{\frac{3r|m|}{2}}a^{3|m|}(-1)^m 
\nonumber\\
&\times 
\frac{
(q^{1-r}a^{-2};q)_{\infty}^6
}
{
(q^{r}a^2;q)_{\infty}^6
}
\frac{
(-q^{\frac{3r}{2}}a^3;q)_{\infty}
}
{
(-q^{1-\frac{3r}{2}} a^{-3};q)_{\infty}
} \; .
\end{align}

%%%%%%%%%%%%%%%%%%%%%%%%%%%%%%%%%%%%%%%%%%%
\subsubsection{$SO(3)-[3]$ with $\mathcal{N}, (\mathrm{N,N,N})+\mathrm{Fermis}$}
%%%%%%%%%%%%%%%%%%%%%%%%%%%%%%%%%%%%%%%%%%%

The Neumann theory A half-indices are
\begin{align}
\label{so3nf3_Nn++}
\mathbb{II}_{\mathcal{N},\mathrm{N}+\Psi}^{SO(3)-[3]_{++}^A}
&=
\frac12 (q)_{\infty}\oint \frac{ds}{2\pi is}
(s^{\pm};q)_{\infty}
\frac{1}
{
(q^{\frac{r}{2}}as;q)_{\infty}^3
(q^{\frac{r}{2}}as^{-1};q)_{\infty}^3
(q^{\frac{r}{2}}a;q)_{\infty}^3
}
\nonumber\\
&\times 
(q^{\frac12}s^{\pm}u^{\pm};q)_{\infty}
(q^{\frac12}s^{\pm}u^{\mp};q)_{\infty}
(q^{\frac12}u^{\pm};q)_{\infty}, \\
\label{so3nf3_Nn-+}
\mathbb{II}_{\mathcal{N},\mathrm{N}+\Psi}^{SO(3)-[3]_{-+}^A}
&=
\frac12 (q)_{\infty}\oint \frac{ds}{2\pi is}
(s^{\pm};q)_{\infty}
\frac{1}
{
(q^{\frac{r}{2}}as;q)_{\infty}^3
(q^{\frac{r}{2}}as^{-1};q)_{\infty}^3
(q^{\frac{r}{2}}a;q)_{\infty}^3
}
\nonumber\\
&\times
(\pm q^{\frac12}s^{\pm};q)_{\infty}
(\mp q^{\frac12}s^{\pm};q)_{\infty}
(\pm q^{\frac12};q)_{\infty}, \\
\label{so3nf3_Nn+-}
\mathbb{II}_{\mathcal{N},\mathrm{N}+\Psi}^{SO(3)-[3]_{+-}^A}
&=
\frac12 (q)_{\infty}\oint \frac{ds}{2\pi is}
(-s^{\pm};q)_{\infty}
\frac{1}
{
(q^{\frac{r}{2}}as;q)_{\infty}^3
(q^{\frac{r}{2}}as^{-1};q)_{\infty}^3
(-q^{\frac{r}{2}}a;q)_{\infty}^3
}
\nonumber\\
&\times
(\pm q^{\frac12}s^{\pm};q)_{\infty}
(\mp q^{\frac12}s^{\pm};q)_{\infty}
(\pm q^{\frac12};q)_{\infty}, \\
\label{so3nf3_Nn--}
\mathbb{II}_{\mathcal{N},\mathrm{N}+\Psi}^{SO(3)-[3]_{--}^A}
&=
\frac12 (q)_{\infty}\oint \frac{ds}{2\pi is}
(-s^{\pm};q)_{\infty}
\frac{1}
{
(q^{\frac{r}{2}}as;q)_{\infty}^3
(q^{\frac{r}{2}}as^{-1};q)_{\infty}^3
(-q^{\frac{r}{2}}a;q)_{\infty}^3
}
\nonumber\\
&\times
(q^{\frac12}s^{\pm}u^{\pm};q)_{\infty}
(q^{\frac12}s^{\pm}u^{\mp};q)_{\infty}
(-q^{\frac12}u^{\pm};q)_{\infty}
\end{align}
with the matching theory B half-indices given by
\begin{align}
\label{so3nf3mag_Ddnd++}
\mathbb{II}_{\mathcal{D},\mathrm{D,N,D}}^{SO(2)-[3]_{++}^B}
&=
\frac{1}{(q)_{\infty}}
\sum_{m\in \mathbb{Z}}
(q^{\frac{1+r}{2}\pm m}au^{\pm};q)_{\infty}^3
(-1)^m q^{\frac{3}{2}m^2}u^{3m}
\frac{1}
{(q^r a^2;q)_{\infty}^6}
(q^{\frac{3r}{2}} a^3;q)_{\infty}, \\
\label{so3nf3mag_Ddnd--}
\mathbb{II}_{\mathcal{D},\mathrm{D,N,D}}^{SO(2)-[3]_{--}^B}
&=
\frac{1}{(-q;q)_{\infty}}
(q^{\frac{1+r}{2}}a;q)_{\infty}^3
(-q^{\frac{1-r}{2}}a;q)_{\infty}^3 
\frac{1}
{
(q^r a^2;q)_{\infty}^6
}
(-q^{\frac{3r}{2}}a^3;q)_{\infty}, \\
\label{so3nf3mag_Ddnd+-}
\mathbb{II}_{\mathcal{D},\mathrm{D,N,D}}^{SO(2)-[3]_{+-}^B}
&=
\frac{1}{(-q;q)_{\infty}}
(q^{\frac{1+r}{2}}a;q)_{\infty}^3
(-q^{\frac{1-r}{2}}a;q)_{\infty}^3 
\frac{1}
{
(q^r a^2;q)_{\infty}^6
}
(q^{\frac{3r}{2}}a^3;q)_{\infty}, \\
\label{so3nf3mag_Ddnd-+}
\mathbb{II}_{\mathcal{D},\mathrm{D,N,D}}^{SO(2)-[3]_{-+}^B}
&=
\frac{1}{(q)_{\infty}}
\sum_{m\in \mathbb{Z}}
(q^{\frac{1+r}{2}\pm m}au^{\pm};q)_{\infty}^3
 q^{\frac{3}{2}m^2}u^{3m}
\frac{1}
{(q^r a^2;q)_{\infty}^6}
(-q^{\frac{3r}{2}} a^3;q)_{\infty} \; .
\end{align}

%%%%%%%%%%%%%%%%%%%%%%%%%%%%%%%%%%%%%%%%%%%
\subsubsection{$SO(3)-[3]$ with $\mathcal{D}, (\mathrm{D,D,D})$}
%%%%%%%%%%%%%%%%%%%%%%%%%%%%%%%%%%%%%%%%%%%

For Dirichlet conditions we have theory A half-indices
\begin{align}
\label{so3nf3_Dd++}
\mathbb{II}_{\mathcal{D},\mathrm{D}}^{SO(3)-[3]_{++}^A}
&=
\frac{1}{(q)_{\infty}}
\sum_{m\in \mathbb{Z}}
\frac{1}{(q^{1\pm m} u^{\pm};q)_{\infty}}
(q^{1-\frac{r}{2}\pm m}a^{-1}u^{\pm};q)_{\infty}^3
(q^{1-\frac{r}{2}}a^{-1};q)_{\infty}^3
q^{m^2}u^{2m}, \\
\label{so3nf3_Dd-+}
\mathbb{II}_{\mathcal{D},\mathrm{D}}^{SO(3)-[3]_{-+}^A}
&=
\frac{1}{(q)_{\infty}}
\sum_{m\in \mathbb{Z}}
\frac{1}{(q^{1\pm m} u^{\pm};q)_{\infty}}
(q^{1-\frac{r}{2}\pm m}a^{-1}u^{\pm};q)_{\infty}^3
(q^{1-\frac{r}{2}}a^{-1};q)_{\infty}^3
(-1)^m q^{m^2}u^{2m}, \\
\label{so3nf3_Dd+-}
\mathbb{II}_{\mathcal{D},\mathrm{D}}^{SO(3)-[3]_{+-}^A}
&=
\frac{1}{(-q;q)_{\infty}}
\sum_{m\in \mathbb{Z}}
\frac{1}{(-q^{1\pm m} u^{\pm};q)_{\infty}}
(q^{1-\frac{r}{2}\pm m}a^{-1}u^{\pm};q)_{\infty}^3
(-q^{1-\frac{r}{2}}a^{-1};q)_{\infty}^3
(-1)^m q^{m^2}u^{2m}, \\
\label{so3nf3_Dd--}
\mathbb{II}_{\mathcal{D},\mathrm{D}}^{SO(3)-[3]_{--}^A}
&=
\frac{1}{(q)_{\infty}}
\sum_{m\in \mathbb{Z}}
\frac{1}{(-q^{1\pm m} u^{\pm};q)_{\infty}}
(q^{1-\frac{r}{2}\pm m}a^{-1}u^{\pm};q)_{\infty}^3
(-q^{1-\frac{r}{2}}a^{-1};q)_{\infty}^3
q^{m^2}u^{2m} \; .
\end{align}
These match the theory B half-indices
\begin{align}
\label{so3nf3mag_Nndn++}
\mathbb{II}_{\mathcal{N},\mathrm{N,D,N}+\Psi}^{SO(2)-[3]_{++}^B}
&=
(q)_{\infty}\oint \frac{ds}{2\pi is}
\frac{1}
{(q^{\frac{1-r}{2}} a^{-1}s^{\pm};q)_{\infty}^3}
(q^{\frac12}s^{\pm}u^{\pm};q)_{\infty}
(q^{\frac12}s^{\pm}u^{\mp};q)_{\infty}
(q^{\frac12}s^{\pm};q)_{\infty}
\nonumber\\
&\times 
(q^{1-r}a^{-2};q)_{\infty}^6 
\frac{1}
{
(q^{1-\frac{3r}{2}} a^{-3};q)_{\infty}
}, \\
\label{so3nf3mag_Nndn--}
\mathbb{II}_{\mathcal{N},\mathrm{N,D,N}+\Psi}^{SO(2)-[3]_{--}^B}
&=
(-q;q)_{\infty}
\frac{1}
{
(q^{\frac{1-r}{2}} a^{-1};q)_{\infty}^3
(-q^{\frac{1-r}{2}} a^{-1};q)_{\infty}^3
}
(q^{\frac12}u^{\pm};q)_{\infty}
(-q^{\frac12}u^{\pm};q)_{\infty}
(\pm q^{\frac12};q)_{\infty}
\nonumber\\
&\times 
(q^{1-r}a^{-2};q)_{\infty}^6 
\frac{1}
{
(-q^{1-\frac{3r}{2}} a^{-3};q)_{\infty}
}, \\
\label{so3nf3mag_Nndn+-}
\mathbb{II}_{\mathcal{N},\mathrm{N,D,N}+\Psi}^{SO(2)-[3]_{+-}^B}
&=
(-q;q)_{\infty}
\frac{1}
{
(q^{\frac{1-r}{2}} a^{-1};q)_{\infty}^3
(-q^{\frac{1-r}{2}} a^{-1};q)_{\infty}^3
}
(q^{\frac12}u^{\pm};q)_{\infty}
(-q^{\frac12}u^{\pm};q)_{\infty}
(\pm q^{\frac12};q)_{\infty}
\nonumber\\
&\times 
(q^{1-r}a^{-2};q)_{\infty}^6 
\frac{1}
{
(q^{1-\frac{3r}{2}} a^{-3};q)_{\infty}
}, \\
\label{so3nf3mag_Nndn-+}
\mathbb{II}_{\mathcal{N},\mathrm{N,D,N}+\Psi}^{SO(2)-[3]_{-+}^B}
&=
(q)_{\infty}\oint \frac{ds}{2\pi is}
\frac{1}
{(q^{\frac{1-r}{2}} a^{-1}s^{\pm};q)_{\infty}^3}
(q^{\frac12}s^{\pm}u^{\pm};q)_{\infty}
(q^{\frac12}s^{\pm}u^{\mp};q)_{\infty}
(-q^{\frac12}s^{\pm};q)_{\infty}
\nonumber\\
&\times 
(q^{1-r}a^{-2};q)_{\infty}^6 
\frac{1}
{
(-q^{1-\frac{3r}{2}} a^{-3};q)_{\infty}
}.
\end{align}

%%%%%%%%%%%%%%%%%%%%%%%%%%%%%%%%%%%%%%%%%%%
\subsection{$N_c=3, N_f=4$ $(SO(3)-[4])$}
%%%%%%%%%%%%%%%%%%%%%%%%%%%%%%%%%%%%%%%%%%%

In the set of examples we have non-Abelian gauge group in both theories.
The theory A full-indices are
\begin{align}
\label{so3nf4_i++}
I^{SO(3)-[4]_{++}^A}
&=
\frac12 \sum_{m\in \mathbb{Z}}
\oint \frac{ds}{2\pi is}
(1-q^{\frac{|m|}{2}} s)
(1-q^{\frac{|m|}{2}} s^{-1})
\nonumber\\
&\times 
\frac
{
(q^{1-\frac{r+|m|}{2}} a^{-1}s^{-1};q)_{\infty}^4
(q^{1-\frac{r+|m|}{2}} a^{-1}s;q)_{\infty}^4
(q^{1-\frac{r}{2}} a^{-1};q)_{\infty}^4
}
{
(q^{\frac{r+|m|}{2}} as;q)_{\infty}^4
(q^{\frac{r+|m|}{2}} as^{-1};q)_{\infty}^4
(q^{\frac{r}{2}} a;q)_{\infty}^4
}
q^{\frac{2(1-r)|m|}{2}-\frac{|m|}{2}}
a^{-4|m|}, \\
\label{so3nf4_i-+}
I^{SO(3)-[4]_{-+}^A}
&=
\frac12 \sum_{m\in \mathbb{Z}}
\oint \frac{ds}{2\pi is}
(1-q^{\frac{|m|}{2}} s)
(1-q^{\frac{|m|}{2}} s^{-1})
\nonumber\\
&\times
\frac
{
(q^{1-\frac{r+|m|}{2}} a^{-1}s^{-1};q)_{\infty}^4
(q^{1-\frac{r+|m|}{2}} a^{-1}s;q)_{\infty}^4
(q^{1-\frac{r}{2}} a^{-1};q)_{\infty}^4
}
{
(q^{\frac{r+|m|}{2}} as;q)_{\infty}^4
(q^{\frac{r+|m|}{2}} as^{-1};q)_{\infty}^4
(q^{\frac{r}{2}} a;q)_{\infty}^4
}
q^{\frac{2(1-r)|m|}{2}-\frac{|m|}{2}}
a^{-4|m|}
(-1)^m, \\
\label{so3nf4_i+-}
I^{SO(3)-[4]_{+-}^A}
&=
\frac12 \sum_{m\in \mathbb{Z}}
\oint \frac{ds}{2\pi is}
(1+q^{\frac{|m|}{2}} s)
(1+q^{\frac{|m|}{2}} s^{-1})
\nonumber\\
&\times
\frac
{
(q^{1-\frac{r+|m|}{2}} a^{-1}s^{-1};q)_{\infty}^4
(q^{1-\frac{r+|m|}{2}} a^{-1}s;q)_{\infty}^4
(-q^{1-\frac{r}{2}} a^{-1};q)_{\infty}^4
}
{
(q^{\frac{r+|m|}{2}} as;q)_{\infty}^4
(q^{\frac{r+|m|}{2}} as^{-1};q)_{\infty}^4
(-q^{\frac{r}{2}} a;q)_{\infty}^4
}
q^{\frac{2(1-r)|m|}{2}-\frac{|m|}{2}}
a^{-4|m|}, \\
\label{so3nf4_i--}
I^{SO(3)-[4]_{--}^A}
&=
\frac12 \sum_{m\in \mathbb{Z}}
\oint \frac{ds}{2\pi is}
(1+q^{\frac{|m|}{2}} s)
(1+q^{\frac{|m|}{2}} s^{-1})
\nonumber\\
&\times
\frac
{
(q^{1-\frac{r+|m|}{2}} a^{-1}s^{-1};q)_{\infty}^4
(q^{1-\frac{r+|m|}{2}} a^{-1}s;q)_{\infty}^4
(-q^{1-\frac{r}{2}} a^{-1};q)_{\infty}^4
}
{
(q^{\frac{r+|m|}{2}} as;q)_{\infty}^4
(q^{\frac{r+|m|}{2}} as^{-1};q)_{\infty}^4
(-q^{\frac{r}{2}} a;q)_{\infty}^4
}
q^{\frac{2(1-r)|m|}{2}-\frac{|m|}{2}}
a^{-4|m|}
(-1)^m \; .
\end{align}
The matching theory B indices are
\begin{align}
\label{so3nf4mag_i++}
I^{SO(3)-[4]_{++}^B}
&=
\frac12 
\sum_{m\in \mathbb{Z}}
\oint \frac{ds}{2\pi is}
(1-q^{\frac{|m|}{2}} s)
(1-q^{\frac{|m|}{2}} s^{-1})
\nonumber\\
&\times 
\frac{
(q^{\frac{1+r+|m|}{2}}as^{-1};q)_{\infty}^4
(q^{\frac{1+r+|m|}{2}}as;q)_{\infty}^4
(q^{\frac{1+r}{2}}a;q)_{\infty}^4
}
{
(q^{\frac{1-r+|m|}{2}}a^{-1}s;q)_{\infty}^4
(q^{\frac{1-r+|m|}{2}}a^{-1}s^{-1};q)_{\infty}^4
(q^{\frac{1-r}{2}}a^{-1};q)_{\infty}^4
}
q^{2r|m|}a^{4|m|}
\nonumber\\
&\times 
\frac{
(q^{1-r}a^{-2};q)_{\infty}^{10}
}
{
(q^{r}a^2;q)_{\infty}^{10}
}
\frac{
(q^{\frac{-1+4r}{2}}a^4;q)_{\infty}
}
{
(q^{\frac{3-4r}{2}} a^{-4};q)_{\infty}
}, \\
\label{so3nf4mag_i--}
I^{SO(3)-[4]_{--}^B}
&=
\frac12 
\sum_{m\in \mathbb{Z}}
\oint \frac{ds}{2\pi is}
(1+q^{\frac{|m|}{2}} s)
(1+q^{\frac{|m|}{2}} s^{-1})
\nonumber\\
&\times 
\frac{
(q^{\frac{1+r+|m|}{2}}as^{-1};q)_{\infty}^4
(q^{\frac{1+r+|m|}{2}}as;q)_{\infty}^4
(-q^{\frac{1+r}{2}}a;q)_{\infty}^4
}
{
(q^{\frac{1-r+|m|}{2}}a^{-1}s;q)_{\infty}^4
(q^{\frac{1-r+|m|}{2}}a^{-1}s^{-1};q)_{\infty}^4
(-q^{\frac{1-r}{2}}a^{-1};q)_{\infty}^4
}
q^{2r|m|}a^{4|m|} (-1)^m 
\nonumber\\
&\times 
\frac{
(q^{1-r}a^{-2};q)_{\infty}^{10}
}
{
(q^{r}a^2;q)_{\infty}^{10}
}
\frac{
(-q^{\frac{-1+4r}{2}}a^4;q)_{\infty}
}
{
(-q^{\frac{3-4r}{2}} a^{-4};q)_{\infty}
}, \\
\label{so3nf4mag_i+-}
I^{SO(3)-[4]_{+-}^B}
&=
\frac12 
\sum_{m\in \mathbb{Z}}
\oint \frac{ds}{2\pi is}
(1+q^{\frac{|m|}{2}} s)
(1+q^{\frac{|m|}{2}} s^{-1})
\nonumber\\
&\times 
\frac{
(q^{\frac{1+r+|m|}{2}}as^{-1};q)_{\infty}^4
(q^{\frac{1+r+|m|}{2}}as;q)_{\infty}^4
(-q^{\frac{1+r}{2}}a;q)_{\infty}^4
}
{
(q^{\frac{1-r+|m|}{2}}a^{-1}s;q)_{\infty}^4
(q^{\frac{1-r+|m|}{2}}a^{-1}s^{-1};q)_{\infty}^4
(-q^{\frac{1-r}{2}}a^{-1};q)_{\infty}^4
}
q^{2r|m|}a^{4|m|} 
\nonumber\\
&\times 
\frac{
(q^{1-r}a^{-2};q)_{\infty}^{10}
}
{
(q^{r}a^2;q)_{\infty}^{10}
}
\frac{
(q^{\frac{-1+4r}{2}}a^4;q)_{\infty}
}
{
(q^{\frac{3-4r}{2}} a^{-4};q)_{\infty}
}, \\
\label{so3nf4mag_i-+}
I^{SO(3)-[4]_{-+}^B}
&=
\frac12 
\sum_{m\in \mathbb{Z}}
\oint \frac{ds}{2\pi is}
(1-q^{\frac{|m|}{2}} s)
(1-q^{\frac{|m|}{2}} s^{-1})
\nonumber\\
&\times 
\frac{
(q^{\frac{1+r+|m|}{2}}as^{-1};q)_{\infty}^4
(q^{\frac{1+r+|m|}{2}}as;q)_{\infty}^4
(q^{\frac{1+r}{2}}a;q)_{\infty}^4
}
{
(q^{\frac{1-r+|m|}{2}}a^{-1}s;q)_{\infty}^4
(q^{\frac{1-r+|m|}{2}}a^{-1}s^{-1};q)_{\infty}^4
(q^{\frac{1-r}{2}}a^{-1};q)_{\infty}^4
}
q^{2r|m|}a^{4|m|} (-1)^m 
\nonumber\\
&\times 
\frac{
(q^{1-r}a^{-2};q)_{\infty}^{10}
}
{
(q^{r}a^2;q)_{\infty}^{10}
}
\frac{
(-q^{\frac{-1+4r}{2}}a^4;q)_{\infty}
}
{
(-q^{\frac{3-4r}{2}} a^{-4};q)_{\infty}
} \; .
\end{align}

%%%%%%%%%%%%%%%%%%%%%%%%%%%%%%%%%%%%%%%%%%%
\subsubsection{$SO(3)-[4]$ with $\mathcal{N}, (\mathrm{N,N,N,N})+\mathrm{Fermis}$}
%%%%%%%%%%%%%%%%%%%%%%%%%%%%%%%%%%%%%%%%%%%

The theory A half-indices with Neumann boundary conditions are
\begin{align}
\label{so3nf4_Nn++}
\mathbb{II}_{\mathcal{N},\mathrm{N}+\Psi }^{SO(3)-[4]_{++}^A}
&=
\frac12 (q)_{\infty}\oint \frac{ds}{2\pi is}
(s^{\pm};q)_{\infty}
\frac{1}
{
(q^{\frac{r}{2}}as;q)_{\infty}^4
(q^{\frac{r}{2}}as^{-1};q)_{\infty}^4
(q^{\frac{r}{2}}a;q)_{\infty}^4
}
\nonumber\\
&\times 
(q^{\frac12}s^{\pm}u^{\pm};q)_{\infty}
(q^{\frac12}s^{\pm}u^{\mp};q)_{\infty}
(q^{\frac12}s^{\pm};q)_{\infty}
(q^{\frac12}u^{\pm};q)_{\infty}
(q^{\frac12};q)_{\infty}, \\
\label{so3nf4_Nn-+}
\mathbb{II}_{\mathcal{N},\mathrm{N}+\Psi }^{SO(3)-[4]_{-+}^A}
&=
\frac12 (q)_{\infty}\oint \frac{ds}{2\pi is}
(s^{\pm};q)_{\infty}
\frac{1}
{
(q^{\frac{r}{2}}as;q)_{\infty}^4
(q^{\frac{r}{2}}as^{-1};q)_{\infty}^4
(q^{\frac{r}{2}}a;q)_{\infty}^4
}
\nonumber\\
&\times
(q^{\frac12}s^{\pm}u^{\pm};q)_{\infty}
(q^{\frac12}s^{\pm}u^{\mp};q)_{\infty}
(-q^{\frac12}s^{\pm};q)_{\infty}
(q^{\frac12}u^{\pm};q)_{\infty}
(-q^{\frac12};q)_{\infty}, \\
\label{so3nf4_Nn+-}
\mathbb{II}_{\mathcal{N},\mathrm{N}+\Psi }^{SO(3)-[4]_{+-}^A}
&=
\frac12 (q)_{\infty}\oint \frac{ds}{2\pi is}
(-s^{\pm};q)_{\infty}
\frac{1}
{
(q^{\frac{r}{2}}as;q)_{\infty}^4
(q^{\frac{r}{2}}as^{-1};q)_{\infty}^4
(-q^{\frac{r}{2}}a;q)_{\infty}^4
}
\nonumber\\
&\times
(q^{\frac12}s^{\pm}u^{\pm};q)_{\infty}
(q^{\frac12}s^{\pm}u^{\mp};q)_{\infty}
(-q^{\frac12}s^{\pm};q)_{\infty}
(-q^{\frac12}u^{\pm};q)_{\infty}
(q^{\frac12};q)_{\infty}, \\
\label{so3nf4_Nn--}
\mathbb{II}_{\mathcal{N},\mathrm{N}+\Psi }^{SO(3)-[4]_{--}^A}
&=
\frac12 (q)_{\infty}\oint \frac{ds}{2\pi is}
(-s^{\pm};q)_{\infty}
\frac{1}
{
(q^{\frac{r}{2}}as;q)_{\infty}^4
(q^{\frac{r}{2}}as^{-1};q)_{\infty}^4
(-q^{\frac{r}{2}}a;q)_{\infty}^4
}
\nonumber\\
&\times
(q^{\frac12}s^{\pm}u^{\pm};q)_{\infty}
(q^{\frac12}s^{\pm}u^{\mp};q)_{\infty}
(q^{\frac12}s^{\pm};q)_{\infty}
(-q^{\frac12}u^{\pm};q)_{\infty}
(-q^{\frac12};q)_{\infty}.
\end{align}
We find that 
these half-indices agree with those for theory B
\begin{align}
\label{so3nf4mag_Ddnd++}
\mathbb{II}_{\mathcal{D},\mathrm{D,N,D}}^{SO(3)-[4]_{++}^B}
&=
\frac{1}{(q)_{\infty}}
\sum_{m\in \mathbb{Z}}
\frac{1}
{(q^{1\pm m}u^{\pm};q)_{\infty}}
(q^{\frac{1+r}{2}\pm m}au^{\pm};q)_{\infty}^4
(q^{\frac{1+r}{2}}a;q)_{\infty}^4
\nonumber\\
&\times 
(-1)^m q^{\frac{3}{2}m^2}u^{3m}
\frac{1}
{(q^r a^2;q)_{\infty}^{10}}
(q^{\frac{-1+4r}{2}} a^4;q)_{\infty}, \\
\label{so3nf4mag_Ddnd--}
\mathbb{II}_{\mathcal{D},\mathrm{D,N,D}}^{SO(3)-[4]_{--}^B}
&=
\frac{1}{(q)_{\infty}}
\sum_{m\in \mathbb{Z}}
\frac{1}
{(-q^{1\pm m}u^{\pm};q)_{\infty}}
(q^{\frac{1+r}{2}\pm m}au^{\pm};q)_{\infty}^4
(-q^{\frac{1+r}{2}}a;q)_{\infty}^4
\nonumber\\
&\times 
(-1)^m q^{\frac{3}{2}m^2}u^{3m}
\frac{1}
{(q^r a^2;q)_{\infty}^{10}}
(-q^{\frac{-1+4r}{2}} a^4;q)_{\infty}, \\
\label{so3nf4mag_Ddnd+-}
\mathbb{II}_{\mathcal{D},\mathrm{D,N,D}}^{SO(3)-[4]_{+-}^B}
&=
\frac{1}{(q)_{\infty}}
\sum_{m\in \mathbb{Z}}
\frac{1}
{(-q^{1\pm m}u^{\pm};q)_{\infty}}
(q^{\frac{1+r}{2}\pm m}au^{\pm};q)_{\infty}^4
(-q^{\frac{1+r}{2}}a;q)_{\infty}^4
\nonumber\\
&\times 
q^{\frac{3}{2}m^2}u^{3m}
\frac{1}
{(q^r a^2;q)_{\infty}^{10}}
(q^{\frac{-1+4r}{2}} a^4;q)_{\infty}, \\
\label{so3nf4mag_Ddnd-+}
\mathbb{II}_{\mathcal{D},\mathrm{D,N,D}}^{SO(3)-[4]_{-+}^B}
&=
\frac{1}{(q)_{\infty}}
\sum_{m\in \mathbb{Z}}
\frac{1}
{(q^{1\pm m}u^{\pm};q)_{\infty}}
(q^{\frac{1+r}{2}\pm m}au^{\pm};q)_{\infty}^4
(q^{\frac{1+r}{2}}a;q)_{\infty}^4
\nonumber\\
&\times 
q^{\frac{3}{2}m^2}u^{3m}
\frac{1}
{(q^r a^2;q)_{\infty}^{10}}
(-q^{\frac{-1+4r}{2}} a^4;q)_{\infty} \; .
\end{align}

%%%%%%%%%%%%%%%%%%%%%%%%%%%%%%%%%%%%%%%%%%%
\subsubsection{$SO(3)-[4]$ with $\mathcal{D}, (\mathrm{D,D,D,D})$}
%%%%%%%%%%%%%%%%%%%%%%%%%%%%%%%%%%%%%%%%%%%

With Dirichlet boundary conditions the theory A half-indices are
\begin{align}
\label{so3nf4_Dd++}
\mathbb{II}_{\mathcal{D},\mathrm{D} }^{SO(3)-[4]_{++}^A}
&=
\frac{1}{(q)_{\infty}}
\sum_{m\in \mathbb{Z}}
\frac{1}{(q^{1\pm m}u^{\pm};q)_{\infty}}
(q^{1-\frac{r}{2}\pm m}a^{-1}u^{\pm};q)_{\infty}^4
(q^{1-\frac{r}{2}}a^{-1};q)_{\infty}^4
(-1)^m q^{\frac{3m^2}{2}} u^{3m}, \\
\label{so3nf4_Dd-+}
\mathbb{II}_{\mathcal{D},\mathrm{D} }^{SO(3)-[4]_{-+}^A}
&=
\frac{1}{(q)_{\infty}}
\sum_{m\in \mathbb{Z}}
\frac{1}{(q^{1\pm m}u^{\pm};q)_{\infty}}
(q^{1-\frac{r}{2}\pm m}a^{-1}u^{\pm};q)_{\infty}^4
(q^{1-\frac{r}{2}}a^{-1};q)_{\infty}^4
q^{\frac{3m^2}{2}} u^{3m}, \\
\label{so3nf4_Dd+-}
\mathbb{II}_{\mathcal{D},\mathrm{D} }^{SO(3)-[4]_{+-}^A}
&=
\frac{1}{(q)_{\infty}}
\sum_{m\in \mathbb{Z}}
\frac{1}{(-q^{1\pm m}u^{\pm};q)_{\infty}}
(q^{1-\frac{r}{2}\pm m}a^{-1}u^{\pm};q)_{\infty}^4
(-q^{1-\frac{r}{2}}a^{-1};q)_{\infty}^4
q^{\frac{3m^2}{2}} u^{3m}, \\
\label{so3nf4_Dd--}
\mathbb{II}_{\mathcal{D},\mathrm{D} }^{SO(3)-[4]_{--}^A}
&=
\frac{1}{(q)_{\infty}}
\sum_{m\in \mathbb{Z}}
\frac{1}{(-q^{1\pm m}u^{\pm};q)_{\infty}}
(q^{1-\frac{r}{2}\pm m}a^{-1}u^{\pm};q)_{\infty}^4
(-q^{1-\frac{r}{2}}a^{-1};q)_{\infty}^4
(-1)^m q^{\frac{3m^2}{2}} u^{3m}.
\end{align}
We find agreement with the theory B
half-indices
\begin{align}
\label{so3nf4mag_Nndn++}
\mathbb{II}_{\mathcal{N},\mathrm{N,D,N}+\Psi}^{SO(3)-[4]_{++}^B}
&=
\frac12 (q)_{\infty} 
\oint \frac{ds}{2\pi is}
(s^{\pm};q)_{\infty}
\frac{1}
{
(q^{\frac{1-r}{2}}a^{-1}s^{\pm};q)_{\infty}^4
(q^{\frac{1-r}{2}}a^{-1};q)_{\infty}^4 
}
\nonumber\\
&\times 
(q^{\frac12}s^{\pm}u^{\pm};q)_{\infty}
(q^{\frac12}s^{\pm}u^{\mp};q)_{\infty}
(q^{\frac12}s^{\pm};q)_{\infty}
(q^{\frac12}u^{\pm};q)_{\infty}
(q^{\frac12};q)_{\infty}
\nonumber\\
&\times 
(q^{1-r}a^{-2};q)_{\infty}^{10}
\frac{1}
{(q^{\frac{3-4r}{2}} a^{-4};q)_{\infty}}, \\
\label{so3nf4mag_Nndn--}
\mathbb{II}_{\mathcal{N},\mathrm{N,D,N}+\Psi}^{SO(3)-[4]_{--}^B}
&=
\frac12 (q)_{\infty} 
\oint \frac{ds}{2\pi is}
(-s^{\pm};q)_{\infty}
\frac{1}
{
(q^{\frac{1-r}{2}}a^{-1}s^{\pm};q)_{\infty}^4
(-q^{\frac{1-r}{2}}a^{-1};q)_{\infty}^4 
}
\nonumber\\
&\times 
(q^{\frac12}s^{\pm}u^{\pm};q)_{\infty}
(q^{\frac12}s^{\pm}u^{\mp};q)_{\infty}
(q^{\frac12}s^{\pm};q)_{\infty}
(-q^{\frac12}u^{\pm};q)_{\infty}
(-q^{\frac12};q)_{\infty}
\nonumber\\
&\times 
(q^{1-r}a^{-2};q)_{\infty}^{10}
\frac{1}
{(-q^{\frac{3-4r}{2}} a^{-4};q)_{\infty}}, \\
\label{so3nf4mag_Nndn+-}
\mathbb{II}_{\mathcal{N},\mathrm{N,D,N}+\Psi}^{SO(3)-[4]_{+-}^B}
&=
\frac12 (q)_{\infty} 
\oint \frac{ds}{2\pi is}
(-s^{\pm};q)_{\infty}
\frac{1}
{
(q^{\frac{1-r}{2}}a^{-1}s^{\pm};q)_{\infty}^4
(-q^{\frac{1-r}{2}}a^{-1};q)_{\infty}^4 
}
\nonumber\\
&\times 
(q^{\frac12}s^{\pm}u^{\pm};q)_{\infty}
(q^{\frac12}s^{\pm}u^{\mp};q)_{\infty}
(-q^{\frac12}s^{\pm};q)_{\infty}
(-q^{\frac12}u^{\pm};q)_{\infty}
(q^{\frac12};q)_{\infty}
\nonumber\\
&\times 
(q^{1-r}a^{-2};q)_{\infty}^{10}
\frac{1}
{(q^{\frac{3-4r}{2}} a^{-4};q)_{\infty}}, \\
\label{so3nf4mag_Nndn-+}
\mathbb{II}_{\mathcal{N},\mathrm{N,D,N}+\Psi}^{SO(3)-[4]_{-+}^B}
&=
\frac12 (q)_{\infty} 
\oint \frac{ds}{2\pi is}
(s^{\pm};q)_{\infty}
\frac{1}
{
(q^{\frac{1-r}{2}}a^{-1}s^{\pm};q)_{\infty}^4
(q^{\frac{1-r}{2}}a^{-1};q)_{\infty}^4 
}
\nonumber\\
&\times 
(q^{\frac12}s^{\pm}u^{\pm};q)_{\infty}
(q^{\frac12}s^{\pm}u^{\mp};q)_{\infty}
(-q^{\frac12}s^{\pm};q)_{\infty}
(q^{\frac12}u^{\pm};q)_{\infty}
(-q^{\frac12};q)_{\infty}
\nonumber\\
&\times 
(q^{1-r}a^{-2};q)_{\infty}^{10}
\frac{1}
{(-q^{\frac{3-4r}{2}} a^{-4};q)_{\infty}} \; .
\end{align}

%%%%%%%%%%%%%%%%%%%%%%%%%%%%%%%%%%%%%%%%%%%
\subsection{Chern-Simons level $k \ne 0$}
\label{SOCSlevel}
%%%%%%%%%%%%%%%%%%%%%%%%%%%%%%%%%%%%%%%%%%%
As shown in \cite{Benini:2011mf} we can induce a non-zero Chern-Simons coupling starting
from the case of vanishing Chern-Simons level with $N_f + |k|$ fundamental
chirals and giving masses to $|k|$ of them. The process is almost identical
to that for symplectic gauge groups described in section~\ref{USpCSlevel}.
Taking the masses to $\pm \infty$
we integrate out these $|k|$ chirals and are left with $N_f$
flavours and Chern-Simons coupling $k = \pm |k|$. Here the $\pm$ signs are the
same for the masses and the sign of $k$. Note that for orthogonal gauge groups
$k \in \Zb$.
In the dual theory B the corresponding chirals get
masses of the opposite sign and consequently theory B gets a Chern-Simons level $-k$. This produces a duality $SO(N_c)_{k} \leftrightarrow SO(|k| + N_f - N_c + 2)_{-k}$ with $N_f$ fundamental chirals in both theories. The matching
with $\Zb_2$ fugacities $\zeta$ and $\chi$ follows in the same way as for $k=0$.

Again for simplicity to avoid  chiral edge modes \cite{Dimofte:2017tpi} we focus on the cases with negative masses for Neumann and positive masses for Dirichlet boundary conditions which don't require additional 2d multiplets other than
those already present for $k=0$. This results in
the matching of half-indices for
\begin{itemize}
\item
$(\Dcal, \mathrm{D})$ b.c.\ in theory A with gauge group $SO(N_c)_{k}$ $\leftrightarrow$
 $(\Ncal, \mathrm{N, D})$ b.c.\ in theory B with gauge group $SO(\tilde{N}_c)_{-k}$
 with $k > 0$.
\item
$(\Ncal, \mathrm{N})$ b.c.\ in theory A with gauge group $SO(N_c)_{k}$ $\leftrightarrow$
 $(\Dcal, \mathrm{D, N})$ b.c.\ in theory B with gauge group $SO(\tilde{N}_c)_{-k}$ 
 with $k < 0$.
\end{itemize}
where we note that again the singlet $V$ is also integrated out for $k \ne 0$.

These half-indices are given by a simple prescription of removing the
$\II_V$ contribution which is present in the half-indices for the $k=0$ cases and noting that
$\tilde{N}_c = |k| + N_f - N_c + 2$.

A derivation again follows from the limit taking
masses to infinity as explained in section~\ref{USpCSlevel} following the
procedure for the full index \cite{Benini:2011mf}.

Again as we have not turned on background fluxes,
our half-indices are not sensitive to other background Chern-Simons levels,
but we note that these again arise from the process of integrating out the chirals. In particular, in addition to the $\pm k \Tr(\bs^2)$ or $\mp k \Tr(\tilde{\bs}^2)$ terms, we have the following contributions from background Chern-Simons levels
\begin{align}
\Acal_{CS \; \Ncal, \mathrm{N}}^{SO(N_c)_{k<0}-[N_f]^A} = & \Acal_{\Ncal, \mathrm{N}}^{SO(N_c)-[N_f + |k|]^A} - \Acal_{\Ncal, \mathrm{N}}^{SO(N_c)_{k<0}-[N_f]^A} \nonumber \\
 & = \frac{1}{2}kN_c\ba^2 + kN_c(r-1)\ba \br + \frac{1}{2}kN(r-1)^2 \br^2, \\
\Acal_{CS \; \Dcal, \mathrm{D,N}}^{SO(\tilde{N}_c)_{k<0}-[N_f]^B} = & \Acal_{\Dcal, \mathrm{D, N, D}}^{SO(\tilde{N}_c)-[N_f + |k|]^B} - \Acal_{\Dcal, \mathrm{D, N}}^{SO(\tilde{N}_c)_{k<0}-[N_f]^B} \nonumber \\
 & = \frac{k}{2} \Tr(\bx^2) + \frac{1}{2} \left(k (N_c + N_f + 3) + N_f^2\right) \ba^2 \nonumber \\ 
 & + (k ((r-1) N_c + r (N_f+3)) - N_f (1 - N_c + N_f (1-r))) \ba \br \nonumber \\
 & + \frac{1}{4} \left(k^2 + k \left(2 N \left(r^2-2 r+2\right)+2 N_f \left(r^2-1\right)+3(2r^2 - 1)\right) \right. \nonumber \\
 & \left. + 2(N+ N_f (r-1)-1)^2\right) \br^2, \\
\Acal_{CS \; \Dcal, \mathrm{D}}^{SO(N_c)_{k>0}-[N_f]^A} = & \Acal_{\Dcal, \mathrm{D}}^{SO(N_c)-[N_f + |k|]^A} - \Acal_{\Dcal, \mathrm{D}}^{SO(N_c)_{k>0}-[N_f]^A} \nonumber \\
 & = \frac{1}{2}kN_c\ba^2 + kN_c(r-1)\ba \br + \frac{1}{2}kN(r-1)^2 \br^2, \\
\Acal_{CS \; \Ncal, \mathrm{N,D}}^{SO(\tilde{N}_c)_{k>0}-[N_f]^B} = & \Acal_{\Ncal, \mathrm{N, \mathrm{D, N}}}^{SO(\tilde{N}_c)-[N_f + |k|]^B} - \Acal_{\Ncal, \mathrm{N, D}}^{SO(\tilde{N}_c)_{k>0}-[N_f]^B} \nonumber \\
 & = \frac{k}{2} \Tr(\bx^2) + \frac{1}{2} \left(k (N_c + N_f + 3) - N_f^2\right) \ba^2 \nonumber \\
 & + (k ((r-1) N_c + r (N_f+3)) + N_f (1 - N_c + N_f (1-r))) \ba \br \nonumber \\
 & + \frac{1}{4} \left(-k^2 + k \left(2 N \left(r^2-2 r+2\right)+2 N_f \left(r^2-1\right) + 3(2r^2 - 1)\right) \right. \nonumber \\
 & \left. - 2(N+ N_f (r-1)-1)^2\right) \br^2. 
\end{align}

%%%%%%%%%%%%%%%%%%%%%%%%%%%%%%%%%%%%%%%%%%%
\subsubsection{$SO(2)_{\pm k}-[2-k]$}
%%%%%%%%%%%%%%%%%%%%%%%%%%%%%%%%%%%%%%%%%%%

Below we have $k \in \{ 1, 2 \}$.

For theory A we have
\begin{align}
\II^{SO(2)_{-k}-[2-k]_{++}^A}_{\Ncal, \mathrm{N}+\Psi} & = (q)_{\infty}
 \oint \frac{ds}{2\pi i s}
 \frac{1}{(q^{\frac{r}{2}} a s^{\pm} ; q)_{\infty}^{2-k}} I_{2d}, \\
I_{2d} & = ( q^{\frac12} s u; q)_{\infty} (q^{\frac12} s u^{-1}; q)_{\infty} (q^{\frac12} s^{-1} u; q)_{\infty} (q^{\frac12}s^{-1} u^{-1}; q)_{\infty}, \\
\II^{SO(2)_{-k}-[2-k]_{+-}^A}_{\Ncal, \mathrm{N}+\Psi} & = (-q; q)_{\infty}
 \frac{1}{(\pm  q^{\frac{r}{2}} a; q)_{\infty}^{2-k}} I_{2d}, \\
I_{2d} & = (q^{\frac12}; q)_{\infty} (-q^{\frac12}; q)_{\infty} (-q^{\frac12}; q)_{\infty} (q^{\frac12}; q)_{\infty}, \\
\II^{SO(2)_{-k}-[2-k]_{-+}^A}_{\Ncal, \mathrm{N}+\Psi} & = (q)_{\infty}
 \oint \frac{ds}{2\pi i s}
 \frac{1}{(q^{\frac{r}{2}} a s^{\pm}; q)_{\infty}^{2-k}} I_{2d}, \\
I_{2d} & = (q^{\frac12}s; q)_{\infty} (-q^{\frac12}s; q)_{\infty} ( q^{\frac12} s^{-1}; q)_{\infty} (-q^{\frac12}s^{-1}; q)_{\infty}, \\
\II^{SO(2)_{-k}-[2-k]_{--}^A}_{\Ncal, \mathrm{N}+\Psi} & = (-q; q)_{\infty}
 \frac{1}{(\pm q^{\frac{r}{2}} a ; q)_{\infty}^{2-k}} I_{2d}, \\
I_{2d} & = (q^{\frac12}u; q)_{\infty} ( q^{\frac12} u^{-1}; q)_{\infty} (-q^{\frac12}u ; q)_{\infty} (-q^{\frac12} u^{-1} ; q)_{\infty},
\end{align}
and
\begin{align}
%\hspace*{-1cm}
\II^{SO(2)_k-[2-k]_{++}^A}_{\Dcal, \mathrm{D}} & = \frac{1}{(q)_{\infty}} \sum_{m \in \Zb}
 q^{m^2} u^{2m}
  (u^{\pm} a q^{1 - r/2 \pm m}; q)_{\infty}^{2-k}, \\
\II^{SO(2)_k-[2-k]_{+-}^A}_{\Dcal, \mathrm{D}} & = \frac{1}{(-q; q)_{\infty}}
  (\pm a q^{1 - r/2}; q)_{\infty}^{2-k}, \\
\II^{SO(2)_k-[2-k]_{-+}^A}_{\Dcal, \mathrm{D}} & = \frac{1}{(q)_{\infty}} \sum_{m \in \Zb}
 (-1)^m q^{m^2} u^{2m}
  (u^{\pm} a q^{1 - r/2 \pm m}; q)_{\infty}^{2-k}, \\
\II^{SO(2)_k-[2-k]_{--}^A}_{\Dcal, \mathrm{D}} & = \frac{1}{(-q; q)_{\infty}}
  (\pm a q^{1 - r/2}; q)_{\infty}^{2-k}, 
\end{align}
while for theory B we have
\begin{align}
%\hspace*{-1cm}
\II^{SO(2)_{-k}-[2-k]_{++}^B}_{\Dcal, \mathrm{D, N, D}} & = \frac{1}{(q)_{\infty}} \sum_{m \in \Zb}
 q^{m^2} u^{2m}
  (u^{\pm} a q^{(1+r)/2 \pm m}; q)_{\infty}^{2-k}
 \frac{1}{(q^r a^2; q)_{\infty}^{(2-k)(3-k)/2}}, \\
\II^{SO(2)_{-k}-[2-k]_{+-}^B}_{\Dcal, \mathrm{D, N, D}} & = \frac{1}{(-q; q)_{\infty}}
  (\pm a q^{(1+r)/2}; q)_{\infty}^{2-k}
 \frac{1}{(q^r a^2; q)_{\infty}^{(2-k)(3-k)/2}}, \\
\II^{SO(2)_{-k}-[2-k]_{-+}^B}_{\Dcal, \mathrm{D, N, D}} & = \frac{1}{(q)_{\infty}} \sum_{m \in \Zb}
 (-1)^m q^{m^2} u^{2m}
  (u^{\pm} a q^{(1+r)/2 \pm m}; q)_{\infty}^{2-k}
 \frac{1}{(q^r a^2; q)_{\infty}^{(2-k)(3-k)/2}}, \\
\II^{SO(2)_{-k}-[2-k]_{--}^B}_{\Dcal, \mathrm{D, N, D}} & = \frac{1}{(-q; q)_{\infty}}
  (\pm a q^{(1+r)/2}; q)_{\infty}^{2-k}
 \frac{1}{(q^r a^2; q)_{\infty}^{(2-k)(3-k)/2}},
\end{align}
and
\begin{align}
\II^{SO(2)_k-[2-k]_{++}^B}_{\Ncal, \mathrm{N, D, N}+\Psi} & = (q)_{\infty}
 \oint \frac{ds}{2\pi i s}
 \frac{1}{(q^{\frac{1-r}{2}} a^{-1} s^{\pm}; q)_{\infty}^{2-k}}
 (q^{1-r} a^{-2} ; q)_{\infty}^{(2-k)(3-k)/2} I_{2d}, \\
I_{2d} & = (q^{\frac12} su; q)_{\infty} (q^{\frac12} s^{-1} u; q)_{\infty} (q^{\frac12} su^{-1}; q)_{\infty} (q^{\frac12} s^{-1} u^{-1} ; q)_{\infty}, \\
\II^{SO(2)_k-[2-k]_{+-}^B}_{\Ncal, \mathrm{N, D, N}+\Psi} & = (-q; q)_{\infty}
 \frac{1}{(\pm q^{\frac{1-r}{2}} a^{-1} ; q)_{\infty}^{2-k}}
 (q^{1-r} a^{-2}; q)_{\infty}^{(2-k)(3-k)/2} I_{2d}, \\
I_{2d} & = (q^{\frac12}; q)_{\infty} (- q^{\frac12}; q)_{\infty} (-q^{\frac12}; q)_{\infty} (q^{\frac12}; q)_{\infty}, \\
\II^{SO(2)_k-[2-k]_{-+}^B}_{\Ncal, \mathrm{N, D, N}+\Psi} & = (q)_{\infty}
 \oint \frac{ds}{2\pi i s}
 \frac{1}{ (q^{\frac{1-r}{2}} a^{-1} s^{\pm} ; q)_{\infty}^{2-k}}
 (q^{1-r} a^{-2} ; q)_{\infty}^{(2-k)(3-k)/2} I_{2d}, \\
I_{2d} & = (q^{\frac12}s; q)_{\infty} ( q^{\frac12} s^{-1}; q)_{\infty} (- q^{\frac12}s; q)_{\infty} (- q^{\frac12}s^{-1} ; q)_{\infty}, \\
\II^{SO(2)_k-[2-k]_{--}^B}_{\Ncal, \mathrm{N, D, N}+\Psi} & = (-q; q)_{\infty}
 \frac{1}{(\pm q^{\frac{1-r}{2}} a^{-1} ; q)_{\infty}^{2-k}}
 (q^{1-r} a^{-2}; q)_{\infty}^{(2-k)(3-k)/2} I_{2d}, \\
I_{2d} & = (q^{\frac12} u; q)_{\infty} (-q^{\frac12} u; q)_{\infty} (q^{\frac12} u^{-1} ; q)_{\infty} (- q^{\frac12} u^{-1}; q)_{\infty}.
\end{align}

%%%%%%%%%%%%%%%%%%%%%%%%%%%%%%%%%%%%%%%%%%%
%%%%%%%%%%%%%%%%%%%%%%%%%%%%%%%%%%%%%%%%%%%
\subsection{Checks on matching of indices}
\label{SO_checks}
%%%%%%%%%%%%%%%%%%%%%%%%%%%%%%%%%%%%%%%%%%%
%%%%%%%%%%%%%%%%%%%%%%%%%%%%%%%%%%%%%%%%%%%
We have performed checks on the claimed matching half-indices using
Mathematica to expand the $q$-series to at least order $q^5$ after first
defining a fugacity $y = q^{r-1/2}$ to replace all terms with dependence on the
parameter $r$ in the R-charge assignment. Details of the expansion to this
order are listed in appendix~\ref{SO_check}.

Some of the identities can also be checked analytically. The cases where the indices do not include any integration or summation over monopole charges are particularly straightforward. This will occur in full-indices and half-indices
with gauge group $SO(1)$ as well as in the case of gauge group $SO(2)$ with
$\chi = -1$. The full-indices and half-indices in both theories will satisfy this property for
precisely the following cases
\begin{align}
SO(1)-[1]_{+-}^A & \leftrightarrow SO(2)-[1]_{+-}^B, \\
SO(1)-[1]_{-+}^A & \leftrightarrow SO(2)-[1]_{--}^B, \\
SO(2)-[2]_{+-}^A & \leftrightarrow SO(2)-[2]_{+-}^B, 
\end{align}
which we have presented and the cases
\begin{align}
SO(2)-[1]_{+-}^A & \leftrightarrow SO(2)-[1]_{+-}^B,  \\
SO(2)-[1]_{--}^A & \leftrightarrow SO(2)-[1]_{-+}^B, 
\end{align}
which we have not presented.
The variations with Chern-Simons levels can also be easily checked directly or
by noting that they arise when taking the appropriate limits of these indices.
For the case of $SO(2)-[2]_{+-}$ this requires first checking the indices
including the flavour fugacities $x_1$ and $x_2$ but this is straightforward.

The analytic checks indicated above rely only on elementary Pochhammer identities
\begin{align}
(\pm x; q)_{\infty} & = (x^2; q^2)_{\infty}, \\
(x; q)_{\infty} = & (x; q^2)_{\infty} (qx; q^2)_{\infty} \; .
\end{align}
When checking the cases with $SO(2)$ and $\chi = -1$ it is useful to note that
setting $x = -q$ in the first of these identities and then using the second
with $x = q$ gives
\begin{align}
(-q; q)_{\infty} & = \frac{(q^2; q^2)_{\infty}}{(q)_{\infty}}
 = \frac{1}{(q; q^2)_{\infty}}. 
\end{align}

We give just one explicit example of $SO(2)-[2]_{+-}$ including the flavour
fugacities.
\begin{align}
\II_{\Ncal, \mathrm{N} + \Psi}^{SO(2)-[2]_{+-}^A} = & (-q;q)_{\infty} \frac{1}{\prod_{\alpha = 1}^2 (\pm q^{r/2}a x_{\alpha}; q)_{\infty}} (\pm q^{1/2}; q)_{\infty}^2 \nonumber \\
 = & (-q;q)_{\infty} \frac{(q; q^2)_{\infty}^2}{\prod_{\alpha = 1}^2 (q^r a^2 x_{\alpha}^2; q^2)_{\infty}}
 = \frac{1}{(-q;q)_{\infty}} \frac{1}{\prod_{\alpha = 1}^2 (q^r a^2 x_{\alpha}^2; q^2)_{\infty}} \; , \\
\II_{\Dcal, \mathrm{D, N, D}}^{SO(2)-[2]_{+-}^B} = & \frac{1}{(-q;q)_{\infty}} \left(\prod_{\alpha = 1}^2 (\pm q^{(1+r)/2}a x_{\alpha}; q)_{\infty} \right) \frac{1}{\prod_{\alpha \le \beta} (\pm q^r a^2 x_{\alpha}x_{\beta}; q)_{\infty}} (\pm q^r a^2; q)_{\infty} \nonumber \\
 = & \frac{1}{(-q;q)_{\infty}} \prod_{\alpha = 1}^2 \frac{(q^{1+r} a^2 x_{\alpha}^2; q^2)_{\infty}}{(q^r a^2 x_{\alpha}^2; q)_{\infty}}
 = \frac{1}{(-q;q)_{\infty}} \frac{1}{\prod_{\alpha = 1}^2 (q^r a^2 x_{\alpha}^2; q^2)_{\infty}}
\end{align}
where for the theory B half-index we used $x_1 x_2 = 1$ for the $SU(2)$ flavour symmetry to see that the contribution for $V$ cancels part of the contribution from $M$.

%%%%%%%%%%%%%%%%%%%%%%%%%%%%%%%%%%%%%%%%%%%
%%%%%%%%%%%%%%%%%%%%%%%%%%%%%%%%%%%%%%%%%%%
\section{Other orthogonal gauge theories}
\label{O_sec}
%%%%%%%%%%%%%%%%%%%%%%%%%%%%%%%%%%%%%%%%%%%
%%%%%%%%%%%%%%%%%%%%%%%%%%%%%%%%%%%%%%%%%%%

For the Lie algebra $\mathfrak{so}(N_c)$, 
there exist other possibilities of the gauge groups, 
$O(N_c)_{\pm}$, $Spin(N_c)$ and $Pin(N_c)_{\pm}$, 
which lead to distinct gauge theories \cite{Aharony:2013kma,Cordova:2017vab}. 
Another possibility is $SO(2N)/\Zb_2$ but we do not consider this case.

%Spin(N)
While the $SO(N_c)$ gauge theory has the monopole operators carrying weights as well as roots of the dual magnetic group, 
the $Spin(N_c)$ gauge theory only contains the monopole operators carrying roots. 
In the $Spin(N_c)$ gauge theory 
the minimal monopole operator which turns on one unit of magnetic flux 
and parametrises the Coulomb branch 
looks semi-classically $V_{Spin}\approx \exp(\frac{2\sigma_1}{\hat{g}_3^2}+2i\gamma_1)$ 
where $\sigma_i$ and $\gamma_i$ are the adjoint scalar fields in the vector multiplet and the dual photons. 
It can be also described as
\begin{align}
V_{Spin}
&=
\begin{cases}
V_1^2 V_2^2 \cdots V_{N-2}^2 V_{N-1} V_{N}&\textrm{for $\epsilon=0$}\cr
V_1^2 V_2^2 \cdots V_{N-2}^2 V_{N-1}^2 V_{N}&\textrm{for $\epsilon=1$}\cr
\end{cases}
\end{align}
in terms of the semi-classical Coulomb branch coordinates
\begin{align}
V_i &\approx 
\exp
\left[
\frac{\sigma_i - \sigma_{i+1}}{\hat{g}_3^2}
+i(\gamma_i -\gamma_{i+1})
\right],\qquad \textrm{for $i=1,\cdots, N-1$}
,\\
V_{N} &\approx 
\begin{cases}
\exp
\left[
\frac{\sigma_{N-1} - \sigma_{N}}{\hat{g}_3^2}
+i(\gamma_{N-1} -\gamma_{N})
\right]&\textrm{for $\epsilon=0$}\cr
\exp
\left[
\frac{2\sigma_{N}}{\hat{g}_3^2}
+i(2\gamma_{N})
\right]&\textrm{for $\epsilon=1$}\cr
\end{cases}
\end{align}
where $\hat{g}_3^2=g_3^2/4\pi$ and $g_3$ is the 3d gauge coupling constant. 
In other words, it is the square of the minimal monopole operator $V\approx \exp(\frac{\sigma_1}{\hat{g}_3^2}+i\gamma_1)$ in the $SO(N_c)$ gauge theory
\begin{align}
V_{Spin}&=V^2. 
\end{align}

%O(N),Pin(N)
One can also consider the gauge theories with disconnected orthogonal gauge groups $O(N_c)$ and $Pin(N_c)$  
which include the reflections along the line bundle or determinant bundle. 
While the baryon $B=Q^{N_c}$ is an independent operator that is charged under $\mathbb{Z}_{2}^{\mathcal{C}}$ for $SO(N_c)$ and $Spin(N_c)$, 
it is not for $O(N_c)$ and $Pin(N_c)$ since $\mathbb{Z}_2^{\mathcal{C}}$ is gauged. 

%O(N)
We can construct $O(N_c)_+$ theories from $SO(N_c)$ theories by gauging the
charge conjugation symmetry $\mathbb{Z}_2^{\mathcal{C}}$ of the $SO(N_c)$.
Alternatively, gauging $\Zb_2^{\mathcal{MC}}$, the diagonal subgroup of
$\Zb_2^{\mathcal{M}} \times \Zb_2^{\mathcal{C}}$, leads to $O(N_c)_-$
theories \cite{Aharony:2013kma, Cordova:2017vab}.
At the level of the Lagrangian description, 
they are distinguished by a discrete theta angle 
which is proportional to $w_1\wedge w_2$ where $w_i\in H^i (X,\mathbb{Z})$, $i=1,2$ are 
$\mathbb{Z}_2$-valued Stiefel-Whitney characteristic classes of $O(N_c)$ bundle on a 3-manifold $X$. 
Both $w_1$ and $w_2$, i.e. theta angle are non-zero only for $O(N_c)$ 
but not for $SO(N_c)$, $Spin(N_c)$ or $Pin(N_c)_{\pm}$ \cite{MR1031992, MR1171915}. 

%O(N)+
The $O(N_c)_+$ gauge theories and their dualities were discussed in 
\cite{Kapustin:2011gh, Benini:2011mf, Hwang:2011ht, Aharony:2011ci, Hwang:2011qt}. 
In particular, the matching of the indices for the $O(N_c)_+$ gauge theories
was tested in \cite{Hwang:2011ht,Hwang:2011qt}.
Similarly to the $SO(N_c)$ gauge theory, 
the $O(N_c)_+$ gauge theory has 
the minimal monopole operator $V=\exp(\frac{\sigma_1}{\hat{g}_3^2}+i\gamma_1)$ 
which is charge-conjugation-even and gauge invariant. 

%O(N)-
On the other hand, in the $O(N_c)_-$ gauge theory 
the monopole operator $V$ is charge-conjugation-odd and not gauge invariant. 
Instead, the product of $V$ with the baryon $B$ as well as the 
monopole operator $V_{Spin}$ and the baryon-monopole operator 
$\beta=Q^{N_c -2}(v_+ - v_-)$ are gauge invariant 
where $v_+\approx \exp(\frac{\sigma}{\hat{g}_3^2}+i\gamma)$ and $v_-\approx \exp(-\frac{\sigma_1}{\hat{g}_3^2}-i\gamma_1)$ 
are the $SO(2)$ monopoles which are even and odd under $\mathbb{Z}_2^{\mathcal{C}}$. 

%Pin(N)
Furthermore, one can obtain the $Pin(N_c)$ gauge theories by gauging the global
symmetry $\Zb_2^{\mathcal{M}} \times \Zb_2^{\mathcal{C}}$, which can also be
viewed as gauging the $\Zb_2^{\mathcal{C}}$ of $Spin(N_c)$ or the
$\Zb_2^{\mathcal{M}}$ of $O(N_c)_{\pm}$
\cite{Aharony:2013kma, Cordova:2017vab}. More precisely this gives $Pin(N_c)_+$
but a modification of the gauging process produces $Pin(N_c)_-$
\cite{Cordova:2017vab}.

%duality
The 4d duality between $Spin(N_c)$ and $O(\tilde{N}_c)_-$ gauge theories gives rise to the 3d IR dualities \cite{Aharony:2013kma}: 
\begin{itemize}

\item Theory A: $Spin(N_c)$ gauge theory with $N_f$ chiral multiplets $Q$ in the vector representation. 

\item Theory B: $O(\tilde{N}_c=N_f-N_c+2)_-$ gauge theory with $N_f$ chiral multiplets $q$ in the vector representation, 
$N_f (N_f+1)/2$ neutral chiral multiplets $M$ in the rank-2 symmetric
representation of $SU(N_f)$ and a chiral multiplet $V$
which has a superpotential. 

\end{itemize}
and similarly with $O(N_c)_-$ for theory A and $Spin(\tilde{N}_c)$ for theory B.

For the $Pin(N_c)_+$ gauge theories, 
the duality can be derived by gauging $\mathbb{Z}_2^{\mathcal{M}}$ of the $O(N_c)_+$ duality \cite{Kapustin:2011gh, Benini:2011mf, Hwang:2011ht, Aharony:2011ci, Hwang:2011qt}. 
It gives rise to the duality between the $Pin(N_c)_+$ gauge theory 
and $Pin(\tilde{N}_c=N_f-N_c+2)_+$ gauge theory. 

The identities (\ref{so_equal}) imply the identities of the $O(N_c)_{\pm}$,
$Spin(N_c)$ and $Pin(N_c)_{+}$ full-indices \cite{Aharony:2013kma} as well as
$Pin(N_c)_{-}$ full-indices:
\begin{align}
\label{o+_o+full}
I^{O(N_c)_{+}-[N_f]^A_{\zeta,\chi'}}&= \left \{ \begin{array}{l} I^{O(\tilde{N}_c)_{+}-[N_f]^B_{\zeta,\chi'}} \; , \; (\zeta, \chi') \ne (-, -) \\ -I^{O(\tilde{N}_c)_{+}-[N_f]^B_{-, -}} \; , \; (\zeta, \chi') = (-, -) \end{array} \right. , \\
\label{o-_Spinfull}
I^{O(N_c)_{-}-[N_f]^{A/B}_{\zeta,\chi'}}&= \left \{ \begin{array}{l} I^{Spin(\tilde{N}_c)-[N_f]^{B/A}_{\chi',\zeta}} \; , \; (\zeta, \chi') \ne (-, -) \\ -I^{Spin(\tilde{N}_c)-[N_f]^{B/A}_{\chi',\zeta}} \; , \; (\zeta, \chi') = (-, -) \end{array} \right. , \\
\label{pin_pinfull}
I^{Pin(N_c)_{\pm}-[N_f]^A_{\zeta',\chi'}}&=I^{Pin(\tilde{N}_c)_{\pm}-[N_f]^B_{\pm\zeta'\chi',\chi'}},
\end{align}
where 
\footnote{See \cite{Aharony:2013kma} for more general expressions with dynamical and background CS couplings. }
the indices can be constructed from the $SO(N_c)_{\zeta\chi}$ indices by summing
over $\zeta$ to gauge $\Zb_2^{\Mcal}$, $\chi$ to gauge $\Zb_2^{\Ccal}$, or
$\chi$ together with a change in sign of $\zeta$ to gauge $\Zb_2^{\Mcal\Ccal}$.
The result is
\begin{align}
I^{O(N_c)_{\pm}-[N_f]_{\zeta, \chi'}}&=
\frac12 
\left(
I^{SO(N_c)-[N_f]_{\zeta,+} }
+\chi'
I^{SO(N_c)-[N_f]_{\pm\zeta,-} }
\right), \\
I^{Spin(N_c)-[N_f]_{\zeta',\chi}}&=
\frac12 
\left(
I^{SO(N_c)-[N_f]_{+,\chi} }
+\zeta'
I^{SO(N_c)-[N_f]_{-,\chi} }
\right), \\
I^{Pin(N_c)_{\pm}-[N_f]_{\zeta',\chi'}}&=
\frac12 
\left(
I^{Spin(N_c)-[N_f]_{\zeta',+}}
+\chi'
I^{Spin(N_c)-[N_f]_{\pm\zeta',-}}
\right),
\end{align}
where $\chi'=\pm1$ corresponds to a projection onto even or odd states under,
$\mathbb{Z}_2^{\mathcal{C}}$ for $O(N_c)_{+}$ and $Pin(N_c)_{+}$ or
$\mathbb{Z}_2^{\mathcal{MC}}$ for $O(N_c)_{-}$,
and $\zeta'=\pm1$ determines a projection onto even or odd states under
$\mathbb{Z}_2^{\mathcal{M}}$ for $Spin(N_c)$ and $Pin(N_c)_{+}$. Such an
interpretation of $\zeta'$ and $\chi'$ is less straightforward for
$Pin(N_c)_{-}$. These parameters
can be viewed as discrete theta angles.

Similarly to the equalities (\ref{o+_o+full})-(\ref{pin_pinfull}) of the full-indices, 
we find the half-index version of the identities: 
\begin{align}
\label{o+_o+half}
\II^{O(N_c)_{+}-[N_f]^{A}_{\zeta,\chi'}}_{\mathcal{B}} &= \left \{ \begin{array}{l} \II^{O(\tilde{N}_c)_{+}-[N_f]^{B}_{\zeta,\chi'}}_{\tilde{\mathcal{B}}} \; , \; (\zeta, \chi') \ne (-, -) \\ -\II^{O(\tilde{N}_c)_{+}-[N_f]^{B}_{-, -}}_{\tilde{\mathcal{B}}} \; , \; (\zeta, \chi') = (-, -) \end{array} \right. , \\
\label{o-_Spinhalf}
\II^{O(N_c)_{-}-[N_f]^{A}_{\zeta,\chi'}}_{\mathcal{B}} &= \left \{ \begin{array}{l} \II^{Spin(\tilde{N}_c)-[N_f]^{B}_{\chi',\zeta}}_{\tilde{\mathcal{B}}} \; , \; (\zeta, \chi') \ne (-, -) \\ -\II^{Spin(\tilde{N}_c)-[N_f]^{B}_{\chi',\zeta}}_{\tilde{\mathcal{B}}} \; , \; (\zeta, \chi') = (-, -) \end{array} \right. , \\
\label{Spin_o-half}
\II^{Spin(N_c)-[N_f]^{A}_{\zeta,\chi'}}_{\mathcal{B}} &= \left \{ \begin{array}{l} \II^{O(\tilde{N}_c)_{-}-[N_f]^{B}_{\chi',\zeta}}_{\tilde{\mathcal{B}}} \; , \; (\zeta, \chi') \ne (-, -) \\ -\II^{O(\tilde{N}_c)_{-}-[N_f]^{B}_{\chi',\zeta}}_{\tilde{\mathcal{B}}} \; , \; (\zeta, \chi') = (-, -) \end{array} \right. , \\
\label{pin_pinhalf}
\II^{Pin(N_c)_{\pm}-[N_f]^{A}_{\zeta',\chi'}}_{\mathcal{B}} &= \II^{Pin(\tilde{N}_c)_{\pm}-[N_f]^{B}_{\pm\zeta'\chi',\chi'}}_{\tilde{\mathcal{B}}},
\end{align}
where we have boundary conditions $(\Bcal; \tilde{\Bcal}) = (\mathcal{N},\mathrm{N}+\Psi; \mathcal{D},\mathrm{D,N,D})$ or $(\Bcal; \tilde{\Bcal}) = (\mathcal{D},\mathrm{D}; \mathcal{N},\mathrm{N,D,N} + \Psi)$.
In both theory A and theory B these half-indices are defined, with boundary conditions $\Bcal$, by gauging the same symmetries as described for the full-indices
\begin{align}
\mathbb{II}^{O(N_c)_{\pm}-[N_f]_{\zeta, \chi'}}_{\mathcal{B}}&=
\frac12 
\left(
\mathbb{II}^{SO(N_c)-[N_f]_{\zeta,+} }_{\mathcal{B}}
+\chi'
\mathbb{II}^{SO(N_c)-[N_f]_{\pm\zeta,-} }_{\mathcal{B}}
\right), \\
\mathbb{II}^{Spin(N_c)-[N_f]_{\zeta',\chi}}_{\mathcal{B}}&=
\frac12 
\left(
\mathbb{II}^{SO(N_c)-[N_f]_{+,\chi} }_{\mathcal{B}}
+\zeta'
\mathbb{II}^{SO(N_c)-[N_f]_{-,\chi} }_{\mathcal{B}}
\right), \\
\mathbb{II}^{Pin(N_c)_{\pm}-[N_f]_{\zeta',\chi'}}_{\mathcal{B}}&=
\frac12 
\left(
\mathbb{II}^{Spin(N_c)-[N_f]_{\zeta',+}}_{\mathcal{B}}
+\chi'
\mathbb{II}^{Spin(N_c)-[N_f]_{\pm\zeta',-}}_{\mathcal{B}}
\right). 
\end{align}
The equalities (\ref{o+_o+half})-(\ref{pin_pinhalf}) indicate the dualities of the basic $\mathcal{N}=(0,2)$ half-BPS boundary conditions 
for 3d $\mathcal{N}=2$ theories with orthogonal gauge groups; $O(N_c)_{\pm}$, $Spin(N_c)$ and $Pin(N_c)_{\pm}$. These can easily be extended to include Chern-Simons levels for the gauge groups by integrating out some of the fundamental chirals as explained for the $SO(N_c)$ half-indices in section~\ref{SOCSlevel}.

%%%%%%%%%%%%%%%%%%%%%%%%%%%%%%%%%%%
\subsection*{Acknowledgements}
The authors would like to thank Dongmin Gang and Chiung Hwang for useful discussions and comments. 
This work is supported by STFC Consolidated Grants ST/P000371/1 and ST/T000708/1.
%%%%%%

\appendix

%%%%%%%%%%%%%%%%%%%%%%%%%%%%%%%%%%
%%%%%%%%%%%%%%%%%%%%%%%%%%%%%%%%%%
\section{General results for 2d anomalies and half-indices}
\label{app_GenAnomInd}
%%%%%%%%%%%%%%%%%%%%%%%%%%%%%%%%%%
%%%%%%%%%%%%%%%%%%%%%%%%%%%%%%%%%%

%%%%%%%%%%%%%%%%%%%%%%%%%%%%%%%%%%
\subsection{Notations}
\label{app_notation}
%%%%%%%%%%%%%%%%%%%%%%%%%%%%%%%%%%
We use the standard notation by defining $q$-shifted factorial
\begin{align}
\label{qpoch_def}
(a;q)_{0}&:=1,\qquad
(a;q)_{n}:=\prod_{k=0}^{n-1}(1-aq^{k}),\qquad 
(q)_{n}:=\prod_{k=1}^{n}(1-q^{k}),\quad 
\quad  n\ge1,
\nonumber \\
(a;q)_{\infty}&:=\prod_{k=0}^{\infty}(1-aq^{k}),\qquad 
(a^{\pm};q)_{\infty}:=(a;q)_{\infty}(a^{-1};q)_{\infty},
\nonumber\\
(a^{\pm}b^{\mp};q)_{\infty}&:=(ab^{-1};q)_{\infty}(a^{-1}b;q)_{\infty}, \qquad 
(q)_{\infty}:=\prod_{k=1}^{\infty} (1-q^k)
\end{align}
where $a$ and $q$ are complex numbers with $|q|<1$.

\subsection{General contributions to 2d anomalies}
\label{Gen2dAnom}
Here we summarise the general results presented
in \cite{Dimofte:2017tpi} to calculate the 2d boundary anomaly polynomial.

The conventions are chosen so that a left-handed 2d complex fermion with
charge $c$ under a
$U(1)$ symmetry with field strength $\bbf$ contributes $+ (c \bbf)^2$ to the anomaly
polynomial. This is therefore the contribution from Fermi multiplet charged under the $U(1)$. A right-handed fermion, and hence a chiral multiplet, contribute
with the opposite sign, i.e.\ $- (c \bbf)^2$.

A key result is that 3d chiral multiplets with Dirichlet (Neumann) boundary
conditions will project out the right-handed (left-handed) fermion, leaving
a boundary left-handed (right-handed) fermion, but that the anomaly
contribution will be half that of a 2d fermion. Therefore a 3d chiral
multiplet with charge $c$ under the $U(1)$ symmetry will contribute $\frac{1}{2} (c \bbf)^2$
for Dirichlet and $-\frac{1}{2} (c \bbf)^2$ for Neumann boundary conditions.
One slight subtlety is that for the $U(1)$ R-symmetry the fermions have a
shifted R-charge so if a chiral multiplet has R-charge $\rho$ there will be
anomaly contributions $\pm \frac{1}{2} ((\rho - 1) \br)^2$.

If we have several $U(1)$ factors we add up the contributions of the form
$c \bbf$ before squaring.

These results generalise to irreducible representations $\bR$ of non-Abelian
simple compact groups $G$ by setting $c = 1$ and replacing
$\bbf^2$ with $\Tr_{\bR}(\bbf^2)$. The trace in representation $\bR$ is given by
$\Tr_{\bR}(\bbf^2) = T_{\bR} \Tr(\bbf^2) = \frac{T_{\bR}}{2h} \Tr_{adjoint}(\bbf^2)$ where $T_{\bR}$ is the quadratic index of $\bR$ (proportional to the quadratic
Casimir multiplied by the dimension of $\bR$) normalised so that
$T_{adjoint} = 2h$ where $h$ is the dual Coxeter number.
We have the following values for $SU(N)$
\begin{align}
T_{Fundamental} = & 1, \\
T_{Adjoint} = & 2h = 2N, \\
T_{Rank \; 2 \; symmetric} = & N+2, \\
T_{Rank \; 2 \; antisymmetric} = & N-2 \; .
\end{align}
For $SO(N)$ we have
\begin{align}
T_{Fundamental} = & 2, \\
T_{Adjoint} = & 2h = 2(N-2)
\end{align}
while for $USp(2N)$
\begin{align}
T_{Fundamental} = & 1, \\
T_{Adjoint} = & 2h = 2(N+1) \; .
\end{align}
More precisely the results for $SO(N)$ apply to $N \ge 4$ but we choose the
same normalisation for $SO(2)$ and $SO(3)$.

In
addition, a 3d Chern-Simons coupling at level $k$ gives a contribution
$k \bbf^2$.

\subsection{General expressions for half-indices}
\label{GenIndices}
Here we summarise the general results presented
in \cite{Gadde:2013wq, Gadde:2013sca, Yoshida:2014ssa, Dimofte:2017tpi} (also see \cite{Benini:2013nda, Benini:2013xpa, Gadde:2013ftv} for 2d indices) to calculate the half-indices.
For a 2d chiral multiplet with R-charge $\rho$ and other $U(1)$ charges
giving fugacity $x$ the contribution to the half-index is
\begin{align}
C(q^{\rho/2}x; q) = & \frac{1}{(q^{\rho/2}x; q)_{\infty} (q^{1 - \rho/2}x^{-1}; q)_{\infty}}
\end{align}
while for a Fermi multiplet we have
\begin{align}
F(q^{(1 + \rho)/2}x; q) = & (q^{(1 + \rho)/2}x; q)_{\infty} (q^{(1 - \rho)/2}x^{-1}; q)_{\infty}. 
\end{align}
A 3d chiral with these charges will give contributions
\begin{align}
\II_{\mathrm{N}} = & \frac{1}{(q^{\rho/2}x; q)_{\infty}}, \\
\II_{\mathrm{D}} = & (q^{1 - \rho/2}x^{-1}; q)_{\infty}
\end{align}
for Neumann or Dirichlet boundary conditions.

If we have a representation $\bR$ of a non-Abelian simple group $G$ we take the product of similar expressions over the weights of $\bR$. So for a 3d chiral
with Neumann boundary conditions we have
\begin{align}
\II_{\mathrm{N}} = & \prod_{\alpha \in weights(\bR)} \frac{1}{(q^{\rho/2}xs^{\alpha}; q)_{\infty}} \; .
\end{align}
For example, the $s^{\alpha}$ could be labelled $s_1, s_2, \ldots , s_N$
with the constraint $s_1 s_2 \cdots s_N = 1$
for the fundamental representation of $SU(N)$ -- for $U(N)$ we simply drop the
constraint $s_1 s_2 \cdots s_N = 1$. These fugacities $s_i$ correspond to
the fundamental weights $\mu_i$, with the constraint $\sum_i \mu_i = 0$ for $SU(N)$.

One point to note is that the weights of the adjoint representation include
zero weights and the vector multiplet fermions have opposite chirality to those from chiral multiplets for the same boundary condition. So, for $U(N)$, the contribution of the vector multiplet with Neumann boundary
conditions is given by the gauginos (with R-charge $1$) in the surviving 2d vector multiplet
\begin{align}
\II_{\Ncal}^{\mathrm{VM}} = & \prod_{\alpha \in weights(\bR)} (qs^{\alpha}; q)_{\infty} = \prod_{i, j} (q s_i s_j^{-1}; q)_{\infty} = (q)_{\infty}^N \prod_{i \ne j} (q s_i s_j^{-1}; q)_{\infty} = (q)_{\infty}^{rank(G)} \prod_{\alpha \in roots(G)} (qs^{\alpha}; q)_{\infty} \; .
\end{align}
In the half-index the projection onto gauge invariant state is imposed by
contour integration over the gauge fugacities $s_i$ and this includes a
Vandermonde determinant which introduces factors $(1 - s^{\alpha})$ in the
product over roots, turning
\begin{align}
\prod_{\alpha \in roots(G)} (qs^{\alpha}; q)_{\infty} \rightarrow &
 \prod_{\alpha \in roots(G)} (1 - s^{\alpha}) (qs^{\alpha}; q)_{\infty}
 = \prod_{\alpha \in roots(G)} (s^{\alpha}; q)_{\infty} \; .
\end{align}

Instead, for Dirichlet boundary conditions the contribution is from the
surviving 2d chiral multiplet (with R-charge $0$) giving contribution
\begin{align}
\hspace*{-1cm}
\II_{\Dcal}^{\mathrm{VM}} = & \prod_{\alpha \in weights(\bR)} \frac{1}{(qs^{\alpha}; q)_{\infty}} = \prod_{i, j} \frac{1}{(q s_i s_j^{-1}; q)_{\infty}} = \frac{1}{(q)_{\infty}^N} \prod_{i \ne j} \frac{1}{(q s_i s_j^{-1}; q)_{\infty}} = \frac{1}{(q)_{\infty}^{rank(G)}} \prod_{\alpha \in roots(G)} \frac{1}{(qs^{\alpha}; q)_{\infty}} \; .
\end{align}
and now the gauge symmetry is broken so $G$ is a global symmetry, and there is
no projection onto gauge invariant states.

However, as explained in
\cite{Dimofte:2017tpi} the complete non-perturbative half-index with Dirichlet
boundary conditions for the vector multiplet
is given after including a sum over monopole charges and including factors
for the effective Chern-Simons levels. Specifically, there is a sum over all
magnetic charges $m \in cochar(G)$. Each term in this sum is given by the same
half-index but with the fugacities $s_i$
shifted by a factor $q^{m_i}$ reflecting the spin induced by magnetic charge
$m_i$ and a factor for the monopole contribution. This monopole factor is
\cite{Dimofte:2017tpi}
\begin{align}
(-q^{1/2})^{k_{eff}[m, m]} s^{k_{eff}[m,-]}
\end{align}
where $k_{eff}$ is a bilinear form defined by the anomaly polynomial. The
interpretation is that for $U(1)$ at Chern-Simons level $k$ we would have
contribution $(-q^{1/2})^{km^2} s^{km}$ reflecting the induced electric
charge $km$ of the monopole due to the Chern-Simons level and consequently
the induced spin $km^2/2$. Note that our convention is that the indices involve
tracing $(-1)^F$, whereas in \cite{Dimofte:2017tpi} the convention was $(-1)^R$.
These conventions are simply related by $q^{1/2} \leftrightarrow (-q)^{1/2}$
and can lead to some sign differences when comparing explicit expressions for
indices and half-indices. In particular, for the Dirichlet half-indices in our convention we
may get sign factors depending on the magnetic charges in the sum, of the
form $(-1)^{km^2}$ as objects with spin $km^2/2$ are fermions for odd $km^2$.
Such terms indeed appear for the orthogonal half-indices,
but not for the symplectic case where in all cases $(-1)^{k_{eff}[m,m]} = 1$.

The results above in terms of $rank(G)$ and product over roots hold also for
$SU(N)$, $USp(2N)$ and $SO(N_c)$.

The details for the symplectic case are very similar to the unitary case. The
fundamental weights for $USp(2N)$ are $\pm \mu_1, \pm \mu_2, \ldots, \pm \mu_N$
giving fugacities $s_i^{\pm 1}$, while the roots are
$\mu_i - \mu_j \; \forall i \ne j$ and $\pm(\mu_i + \mu_j) \; \forall i \le j$.

For $SO(N_c)$ we need to consider separately the cases of $N_c$ even or odd.
For the even case we write $N_c = 2N$ and we have fundamental weights
$\pm \mu_1, \pm \mu_2, \ldots, \pm \mu_{N}$ and the roots are
$\pm \mu_i \pm \mu_j \; \forall i < j$ where all four sign combinations are
taken.

For the case of odd $N_c$ we write $N_c = 2N + 1$ and we have one additional
fundamental weight $\mu_0 = 0$ giving additional roots $\pm \mu_i + \mu_0 = \pm \mu_i \; \forall i$.

The above results, along with details of the monopole sums and effective
Chern-Simons levels define the half-indices, although we found that the
effective Chern-Simons levels need a different normalisation factor for
symplectic gauge groups. However, there is a further complication for the
$SO(N_c)$ indices when the discrete fugacity $\chi$ for the
$\Zb_2^{\Ccal}$ global symmetries is included.

For the case of odd $N_c$ if we have $\chi = +1$ we can consider that we have a gauge fugacity $s_0 = 1$ arising from the fundamental weight $\mu_0 = 0$. If instead we have $\chi = -1$ we set $s_0 = -1$. So, in general we have $s_0 = \chi$
meaning that the fugacities associated to the roots $\pm \mu_i + \mu_0$ are
$\chi s_i^{\pm 1}$.

Note that these four different sectors for $SO(N_c)$, labelled by $\zeta$ and
$\chi$ arise in 3d and so are inherited on the boundary since we have bulk
gauge fields with boundary conditions. In 2d theories on $T^2$ there are
additional sectors as there is a different classification of flat connections
on $T^2$ as clearly explained in \cite{Kim:2014dza}. The point to note here is
that this means the half-index for orthogonal groups is quite different from
the 2d elliptic genus.

For even $N_c = 2N$ if we have $\chi = -1$ we must replace the gauge fugacities
corresponding to fundamental weights $\mu_N$ and $-\mu_N$ with
$s_N \rightarrow 1$ and $s_N^{-1} \rightarrow -1$. In cases of ambiguity, it is
necessary to check whether the factor of $s_N$ originated from fundamental
weight $\mu_N$ or $-\mu_N$. One point to note in this case is that in the derivation of the vector multiplet contribution to the index one of the factors
of $(q)_{\infty}$ (for Neumann or Dirichlet boundary conditions) arose from
the adjoint weight $\mu_N + (-\mu_N) = 0$ giving factor $(qs_N s_N^{-1}; q)_{\infty} = (q)_{\infty}$. In the case of $\chi = -1$ this is replaced, since
$s_N s_N^{-1} \rightarrow 1 \times (-1) = -1$ by the factor
$(-q; q)_{\infty}$. I.e.\ in the half-index we replace $(q)_{\infty}^{rank(G)}$
with $(q)_{\infty}^{rank(G) - 1} (-q; q)_{\infty}$ for $G = SO(2N)$ with $\chi = -1$.

Finally, our notation for fugacities is to label gauge fugacities $s_i$ as above
for half-indices with Neumann boundary conditions for the vector multiplet
and instead $u_i$ if we have Dirichlet boundary conditions. This applies to
both theories A and B in the proposed dualities and is unambiguous as all these
dualities involve Neumann for one theory and Dirichlet for the other. Of course,
gauge invariance is imposed by integrating over the $s_i$ so the results will
depend on the $u_i$ only.

Our notation for other fugacities is $a$ for $U(1)_a$ and $x_{\alpha}$ for
global $SU(N_F)$ flavour symmetries. In numerical checks we often set
$x_{\alpha} = 1$ for simplicity. Note that $\prod_{\alpha} x_{\alpha} = 1$
but we could also use fugacities $\hat{x}_{\alpha}$ which don't obey such a constraint for the global $U(N_f)$
flavour symmetry by combining $SU(N_F)$ and $U(1)_a$. We comment on this
when discussing dualities for non-zero Chern-Simons levels.

%%%%%%%%%%%%%%%%%%%%%%%%%%%%%%%%%%
%%%%%%%%%%%%%%%%%%%%%%%%%%%%%%%%%%
\section{Series expansions of indices}
\label{app_expansion}
%%%%%%%%%%%%%%%%%%%%%%%%%%%%%%%%%%
%%%%%%%%%%%%%%%%%%%%%%%%%%%%%%%%%%

We show several terms in the expansions of indices obtained by using Mathematica.

%%%%%%%%%%%%%%%%%%%%%%%%%%%%%%%%%%%%%%%%%%%
\subsection{$USp(2N_c)-[2N_f]$}
\label{Sp_check}
%%%%%%%%%%%%%%%%%%%%%%%%%%%%%%%%%%%%%%%%%%%

We show the expansions of full-indices and the half-indices in powers of $q$ by defining a fugacity $y:=q^{r-\frac12}$. 
We find the matching at least up to $q^{5}$. 

%%%%%%%%%%%%%%%%%%%%%%%%%%%%%%%%%%%%%%%%%%%
\subsubsection{$USp(2N_c)-[2N_f]$}
\label{Sp_fullcheck}
%%%%%%%%%%%%%%%%%%%%%%%%%%%%%%%%%%%%%%%%%%%

\begin{align}
\begin{array}{c|c|c|c}
N_c
&
N_f
&
\tilde{N}_c
&
%\textrm{series expansions}
I^{USp(2N_c)-[2N_f]^A}
=
I^{USp(2\tilde{N}_c)-[2N_f]^B}
\\ \hline
1
&
3
&
1
&
{\scriptscriptstyle 
\begin{smallmatrix}
1+\sqrt{q} \left(\frac{1}{a^6 y^3}+15 a^2 y\right)+q \left(105 a^4 y^2+\frac{1}{a^{12} y^6}-36\right)
\\
+q^{3/2} \left(490 a^6 y^3+\frac{1}{a^{18} y^9}-384 a^2 y+\frac{21}{a^2 y}\right)
\\
+q^2 \left(1764 a^8 y^4-2100 a^4 y^2+\frac{36}{a^4 y^2}+\frac{1}{a^{24} y^{12}}+558\right)
\\
+q^{5/2} \left(5292 a^{10} y^5-8064 a^6 y^3-\frac{71}{a^6 y^3}+\frac{1}{a^{30} y^{15}}+3690 a^2 y-\frac{384}{a^2 y}\right)
\\
+q^3 \left(13860 a^{12} y^6-24696 a^8 y^4+14526 a^4 y^2+\frac{135}{a^4 y^2}+\frac{36}{a^8 y^4}+\frac{1}{a^{36} y^{18}}-3536\right)
+\cdots
\\
\end{smallmatrix}
}
\\ \hline 
2
&
5
&
2
&
{\scriptscriptstyle 
\begin{smallmatrix}
1+\sqrt{q} \left(\frac{1}{a^{10} y^5}+\left(45 a^2\right) y\right)
\\
+q \left(\left(1035 a^4\right) y^2+\frac{45}{a^8 y^4}+\frac{1}{a^{20} y^{10}}-100\right)
\\+q^{3/2} \left(\left(16005 a^6\right) y^3+\frac{825}{a^6 y^3}-\frac{99}{a^{10} y^5}+\frac{45}{a^{18} y^9}+\frac{1}{a^{30} y^{15}}-\left(4400 a^2\right) y+\frac{55}{a^2 y}\right)
\\
+q^2 \left(\left(186285 a^8\right) y^4-\left(96480 a^4\right) y^2+\frac{9075}{a^4 y^2}-\frac{3200}{a^8 y^4}+\frac{55}{a^{12} y^6}+\frac{825}{a^{16} y^8}-\frac{99}{a^{20} y^{10}}+\frac{45}{a^{28} y^{14}}+\frac{1}{a^{40} y^{20}}+7225\right)+\cdots
\\
\end{smallmatrix}
}
\\ 
\end{array}
\end{align}

%%%%%%%%%%%%%%%%%%%%%%%%%%%%%%%%%%%%%%%%%%%
\subsubsection{$USp(2N_c)-[2N_f]$ with $\mathcal{N}, (\mathrm{N,N,N;N,N,N})+\mathrm{Fermis}$}
\label{Sp_N_check}
%%%%%%%%%%%%%%%%%%%%%%%%%%%%%%%%%%%%%%%%%%%

\begin{align}
\begin{array}{c|c|c|c}
N_c
&
N_f
&
\tilde{N}_c
&
%\textrm{series expansions}
\mathbb{II}^{USp(2N_c)-[2N_f]^A}_{\mathcal{N}, \mathrm{N}+\{\Psi\}}
=
\mathbb{II}^{USp(2\tilde{N}_c)-[2N_f]^B}_{\mathcal{D}, \mathrm{D},\mathrm{N},\mathrm{D}}
\\ \hline
1
&
3
&
1
&
{\scriptscriptstyle 
\begin{smallmatrix}
1+15 a^2 \sqrt{q} y-\frac{6 q^{3/4} \left(a \left(u^2+1\right) \sqrt{y}\right)}{u}
+q \left(105 a^4 y^2+u^2+\frac{1}{u^2}+1\right)-\frac{70 q^{5/4} \left(a^3 \left(u^2+1\right) y^{3/2}\right)}{u}\\
+\frac{a^2 q^{3/2} y \left(u^2 \left(490 a^4 y^2+51\right)+15 u^4+15\right)}{u^2}
-\frac{12 q^{7/4} \left(a \left(u^2+1\right) \sqrt{y} \left(35 a^4 y^2+1\right)\right)}{u}\\
+q^2 \left(u^2 \left(105 a^4 y^2+1\right)+\frac{105 a^4 y^2+1}{u^2}+1764 a^8 y^4+525 a^4 y^2+2\right)
-\frac{2 q^{9/4} \left(a^3 \left(u^2+1\right) y^{3/2} \left(882 a^4 y^2+115\right)\right)}{u}\\
+\frac{a^2 q^{5/2} y \left(u^4 \left(490 a^4 y^2+51\right)+u^2 \left(5292 a^8 y^4+3010 a^4 y^2+117\right)+490 a^4 y^2+51\right)}{u^2}
+\cdots
\\
\end{smallmatrix}
}
\\ \hline 
2
&
5
&
2
&
{\scriptscriptstyle 
\begin{smallmatrix}
1+45 a^2 \sqrt{q} y
-\frac{10 q^{3/4} \left(a \sqrt{y} \left(\text{u1}^2 \text{u2}+\text{u1} \text{u2}^2+\text{u1}+\text{u2}\right)\right)}{\text{u1} \text{u2}}
\\
+q \left(1035 a^4 y^2+\text{u1}^2+\frac{1}{\text{u1}^2}+\text{u1} \left(\text{u2}+\frac{1}{\text{u2}}\right)+\frac{\text{u2}+\frac{1}{\text{u2}}}{\text{u1}}+\text{u2}^2+\frac{1}{\text{u2}^2}+2\right)
\\
-\frac{450 q^{5/4} \left(a^3 y^{3/2} \left(\text{u1}^2 \text{u2}+\text{u1} \text{u2}^2+\text{u1}+\text{u2}\right)\right)}{\text{u1} \text{u2}}
\\
+\frac{5 a^2 q^{3/2} y \left(\text{u1}^2 \left(\text{u2}^2 \left(3201 a^4 y^2+67\right)+18 \text{u2}^4+18\right)+18 \text{u1}^4 \text{u2}^2+29 \text{u1}^3 \left(\text{u2}^3+\text{u2}\right)+29 \text{u1} \left(\text{u2}^3+\text{u2}\right)+18 \text{u2}^2\right)}{\text{u1}^2 \text{u2}^2}
+\cdots
\\
\end{smallmatrix}
}
\\ 
\end{array}
\end{align}

%%%%%%%%%%%%%%%%%%%%%%%%%%%%%%%%%%%%%%%%%%%
\subsubsection{$USp(2N_c)-[2N_f]$ with $\mathcal{D}, (\mathrm{D,D,D; D,D,D})$}
\label{Sp_D_check}
%%%%%%%%%%%%%%%%%%%%%%%%%%%%%%%%%%%%%%%%%%%

\begin{align}
\begin{array}{c|c|c|c}
N_c
&
N_f
&
\tilde{N}_c
&
%\textrm{series expansions}
\mathbb{II}^{USp(2N_c)-[2N_f]^A}_{\mathcal{D}, \mathrm{D}}
=
\mathbb{II}^{USp(2\tilde{N}_c)-[2N_f]^B}_{\mathcal{N}, \mathrm{N},\mathrm{D},\mathrm{N}}
\\ \hline
1
&
3
&
1
&
{\scriptscriptstyle 
\begin{smallmatrix}
1+\left(10 a^2\right) \sqrt{q} y-\frac{q^{3/4} \sqrt{y} \left((5 a) \left(u^2+1\right)\right)}{u}
+q \left(\left(50 a^4\right) y^2+u^2+\frac{1}{u^2}+1\right)
-\frac{q^{5/4} y^{3/2} \left(\left(40 a^3\right) \left(u^2+1\right)\right)}{u}
\\
+\frac{q^{3/2} \left(\left(5 a^2\right) y\right) \left(\left(7 u^2\right) \left(\left(5 a^4\right) y^2+1\right)+2 u^4+2\right)}{u^2}
-\frac{q^{7/4} \left(\left(35 a^4\right) y^2+2\right) \left(\sqrt{y} \left((5 a) \left(u^2+1\right)\right)\right)}{u}
\\
+q^2 \left(u^2 \left(\left(50 a^4\right) y^2+1\right)+\frac{\left(50 a^4\right) y^2+1}{u^2}+\left(490 a^8\right) y^4+\left(250 a^4\right) y^2+2\right)
-\frac{q^{9/4} \left(\left(56 a^4\right) y^2+13\right) \left(y^{3/2} \left(\left(10 a^3\right) \left(u^2+1\right)\right)\right)}{u}
\\+\frac{q^{5/2} \left(a^2 y\right) \left(\left(35 u^4\right) \left(\left(5 a^4\right) y^2+1\right)+\left(2 u^2\right) \left(\left(588 a^8\right) y^4+\left(525 a^4\right) y^2+40\right)+35 \left(\left(5 a^4\right) y^2+1\right)\right)}{u^2}+\cdots
\\
\end{smallmatrix}
}
\\ \hline 
2
&
5
&
2
&
{\scriptscriptstyle 
\begin{smallmatrix}
1+\frac{\sqrt{q}}{a^{10} y^5}
-\frac{10 q^{3/4} \left(\text{u1}^2 \text{u2}+\text{u1} \text{u2}^2+\text{u1}+\text{u2}\right)}{a \text{u1} \text{u2} \sqrt{y}}
\\
+q \left(\frac{1}{a^{20} y^{10}}+\text{u1}^2+\frac{1}{\text{u1}^2}+\text{u1} \left(\text{u2}+\frac{1}{\text{u2}}\right)+\frac{\text{u2}+\frac{1}{\text{u2}}}{\text{u1}}+\text{u2}^2+\frac{1}{\text{u2}^2}+2\right)
\\
-\frac{10 q^{5/4} \left(\text{u1}^2 \text{u2}+\text{u1} \text{u2}^2+\text{u1}+\text{u2}\right)}{a^{11} \text{u1} \text{u2} y^{11/2}}
+\cdots
\\
\end{smallmatrix}
}
\\ 
\end{array}
\end{align}

%%%%%%%%%%%%%%%%%%%%%%%%%%%%%%%%%%%%%%%%%%%
\subsection{$SO(N_c)-[N_f]_{\zeta\chi}$}
\label{SO_check}
%%%%%%%%%%%%%%%%%%%%%%%%%%%%%%%%%%%%%%%%%%%

We show several terms in the expansions of full-indices and half-indices in powers of $q$ by defining a fugacity $y:=q^{r-\frac12}$ or $t:=q^{r-\frac34}$. 
We find the agreement at least up to $q^{5}$. 

%%%%%%%%%%%%%%%%%%%%%%%%%%%%%%%%%%%%%%%%%%%
\subsubsection{$SO(N_c)-[N_f]_{++}$}
\label{SO++_check}
%%%%%%%%%%%%%%%%%%%%%%%%%%%%%%%%%%%%%%%%%%%

%++
\begin{align}
\begin{array}{c|c|c|c}
N_c
&
N_f
&
\tilde{N}_c
&
%\textrm{series expansions}
I^{SO(N_c)-[N_f]^A_{++}}
=
I^{SO(\tilde{N}_c)-[N_f]^B_{++}}
\\ \hline
1
&
1
&
2
&
{\scriptscriptstyle 
\begin{smallmatrix}
1+a \sqrt[4]{q} \sqrt{y}+a^2 \sqrt{q} y+\frac{q^{3/4} \left(a^4 y^2-1\right)}{a \sqrt{y}}
\\
+q \left(a^4 y^2-1\right)+a^5 q^{5/4} y^{5/2}+a^6 q^{3/2} y^3+\frac{q^{7/4} \left(a^8 y^4-1\right)}{a \sqrt{y}}
\\
+q^2 \left(a^8 y^4-2\right)+q^{9/4} \left(a \sqrt{y}\right) \left(a^8 y^4-1\right)+q^{5/2} \left(a^{10} y^5+\frac{1}{a^2 y}\right)
\\
+a^{11} q^{11/4} y^{11/2}+q^3 \left(a^{12} y^6-2\right)+q^{13/4} \left(a \sqrt{y}\right) \left(a^{12} y^6-2\right)
\\
+q^{7/2} \left(a^{14} y^7-a^2 y+\frac{1}{a^2 y}\right)+\frac{q^{15/4} \left(a^{16} y^8+1\right)}{a \sqrt{y}}+q^4 \left(a^{16} y^8-2\right)
\\
+q^{17/4} \left(a \sqrt{y}\right) \left(a^{16} y^8-3\right)+q^{9/2} \left(a^{18} y^9-\left(2 a^2\right) y+\frac{2}{a^2 y}\right)
+\frac{q^{19/4} \left(a^{20} y^{10}-a^4 y^2+3\right)}{a \sqrt{y}}+\cdots
\\
\end{smallmatrix}
}
\\ \hline 
1
&
2
&
3
&
{\scriptscriptstyle 
\begin{smallmatrix}
1+(2 a) q^{3/8} \sqrt{t}-\frac{2 q^{5/8}}{a \sqrt{t}}+\left(3 a^2\right) q^{3/4} t-4 q+\left(4 a^3\right) q^{9/8} t^{3/2}
\\
+\frac{q^{5/4}}{a^2 t}+(-4 a) q^{11/8} \sqrt{t}+\left(5 a^4\right) q^{3/2} t^2+\left(-4 a^2\right) q^{7/4} t
\\
+\left(6 a^5\right) q^{15/8} t^{5/2}-5 q^2+\left(-4 a^3\right) q^{17/8} t^{3/2}+\frac{q^{9/4} \left(\left(7 a^8\right) t^4+4\right)}{a^2 t}
\\
+(-8 a) q^{19/8} \sqrt{t}+\left(-4 a^4\right) q^{5/2} t^2+\frac{8 q^{21/8} \left(a^8 t^4+1\right)}{a \sqrt{t}}
+\left(-8 a^2\right) q^{11/4} t-\frac{2 q^{23/8} \left(\left(2 a^8\right) t^4+1\right)}{a^3 t^{3/2}}+\cdots
\\
\end{smallmatrix}
}
\\ \hline 
1
&
3
&
4
&
{\scriptscriptstyle 
\begin{smallmatrix}
1+(3 a) q^{3/8} \sqrt{t}-\frac{3 q^{5/8}}{a \sqrt{t}}+\left(6 a^2\right) q^{3/4} t-9 q+\left(10 a^3\right) q^{9/8} t^{3/2}
\\
+\frac{3 q^{5/4}}{a^2 t}+(-15 a) q^{11/8} \sqrt{t}+\left(15 a^4\right) q^{3/2} t^2+\frac{6 q^{13/8}}{a \sqrt{t}}+\left(-21 a^2\right) q^{7/4} t
\\
+\frac{q^{15/8} \left(\left(21 a^8\right) t^4-1\right)}{a^3 t^{3/2}}+\left(-27 a^3\right) q^{17/8} t^{3/2}+\frac{q^{9/4} \left(\left(28 a^8\right) t^4+6\right)}{a^2 t}
\\
+(-12 a) q^{19/8} \sqrt{t}+\left(-33 a^4\right) q^{5/2} t^2+\frac{9 q^{21/8} \left(\left(4 a^8\right) t^4
+3\right)}{a \sqrt{t}}+\left(-24 a^2\right) q^{11/4} t-\frac{3 q^{23/8} \left(\left(13 a^8\right) t^4+3\right)}{a^3 t^{3/2}}
+\cdots
\\
\end{smallmatrix}
}
\\ \hline 
2
&
2
&
2
&
{\scriptscriptstyle 
\begin{smallmatrix}
1+\sqrt{q} \left(\left(4 a^2\right) y+\frac{2}{a^2 y}\right)+q \left(\left(9 a^4\right) y^2+\frac{2}{a^4 y^2}-8\right)
\\
+\frac{2 q^{3/2} \left(\left(8 a^{12}\right) y^6-\left(8 a^8\right) y^4+\left(2 a^4\right) y^2+1\right)}{a^6 y^3}
\\
+q^2 \left(\left(25 a^8\right) y^4-\left(24 a^4\right) y^2+\frac{2}{a^8 y^4}+14\right)+q^{5/2} \left(\left(36 a^{10}\right) y^5-\left(32 a^6\right) y^3+\frac{2}{a^{10} y^5}-\frac{16}{a^2 y}-\left(16 a^2\right) y\right)
\\
+q^3 \left(\left(49 a^{12}\right) y^6-\left(40 a^8\right) y^4-\left(32 a^4\right) y^2+\frac{9}{a^4 y^2}+\frac{2}{a^{12} y^6}+64\right)
+\cdots
\\
\end{smallmatrix}
}
\\ \hline 
3
&
3
&
2
&
{\scriptscriptstyle 
\begin{smallmatrix}
1+\frac{\sqrt[4]{q}}{a^3 y^{3/2}}+\sqrt{q} \left(\frac{1}{a^6 y^3}+\left(6 a^2\right) y+\frac{3}{a^2 y}\right)
\\
+\frac{q^{3/4} \left(a^{12} y^6+\left(6 a^8\right) y^4+\left(3 a^4\right) y^2+1\right)}{a^9 y^{9/2}}
\\
+q \left(\left(21 a^4\right) y^2+\frac{3}{a^4 y^2}+\frac{3}{a^8 y^4}+\frac{1}{a^{12} y^6}+1\right)
\\
+\frac{q^{5/4} \left(\left(6 a^{20}\right) y^{10}+\left(6 a^{16}\right) y^8+\left(2 a^{12}\right) y^6+\left(3 a^8\right) y^4+\left(3 a^4\right) y^2+1\right)}{a^{15} y^{15/2}}
\\
+\frac{q^{3/2} \left(a^{12} y^6+\left(56 a^{24}\right) y^{12}-\left(24 a^{20}\right) y^{10}+\left(3 a^8\right) y^4+\left(3 a^4\right) y^2+1\right)}{a^{18} y^9}+\cdots
\\
\end{smallmatrix}
}
\\ 
\end{array}
\end{align}

%%%%%%%%%%%%%%%%%%%%%%%%%%%%%%%%%%%%%%%%%%%
\subsubsection{$SO(N_c)-[N_f]_{-+}$}
\label{SO-+_check}
%%%%%%%%%%%%%%%%%%%%%%%%%%%%%%%%%%%%%%%%%%%

%-+
\begin{align}
\begin{array}{c|c|c|c}
N_c
&
N_f
&
\tilde{N}_c
&
%\textrm{series expansions}
I^{SO(N_c)-[N_f]^A_{-+}}
=
I^{SO(\tilde{N}_c)-[N_f]^B_{--}}
\\ \hline
1
&
1
&
2
&
{\scriptscriptstyle 
\begin{smallmatrix}
1+a \sqrt[4]{q} \sqrt{y}+a^2 \sqrt{q} y+\frac{q^{3/4} \left(a^4 y^2-1\right)}{a \sqrt{y}}+q \left(a^4 y^2-1\right)
\\
+a^5 q^{5/4} y^{5/2}+a^6 q^{3/2} y^3+\frac{q^{7/4} \left(a^8 y^4-1\right)}{a \sqrt{y}}
\\
+q^2 \left(a^8 y^4-2\right)+q^{9/4} \left(a \sqrt{y}\right) \left(a^8 y^4-1\right)
+q^{5/2} \left(a^{10} y^5+\frac{1}{a^2 y}\right)+a^{11} q^{11/4} y^{11/2}+\cdots
\\
\end{smallmatrix}
}
\\ \hline 
1
&
2
&
3
&
{\scriptscriptstyle 
\begin{smallmatrix}
1+(2 a) q^{3/8} \sqrt{t}-\frac{2 q^{5/8}}{a \sqrt{t}}+\left(3 a^2\right) q^{3/4} t-4 q+\left(4 a^3\right) q^{9/8} t^{3/2}
\\
+\frac{q^{5/4}}{a^2 t}+(-4 a) q^{11/8} \sqrt{t}+\left(5 a^4\right) q^{3/2} t^2+\left(-4 a^2\right) q^{7/4} t
\\
+\left(6 a^5\right) q^{15/8} t^{5/2}-5 q^2+\left(-4 a^3\right) q^{17/8} t^{3/2}+\frac{q^{9/4} \left(\left(7 a^8\right) t^4+4\right)}{a^2 t}
\\
+(-8 a) q^{19/8} \sqrt{t}+\left(-4 a^4\right) q^{5/2} t^2+\frac{8 q^{21/8} \left(a^8 t^4+1\right)}{a \sqrt{t}}
+\left(-8 a^2\right) q^{11/4} t-\frac{2 q^{23/8} \left(\left(2 a^8\right) t^4+1\right)}{a^3 t^{3/2}}+\cdots
\\
\end{smallmatrix}
}
\\ \hline 
1
&
3
&
4
&
{\scriptscriptstyle 
\begin{smallmatrix}
1+(3 a) q^{3/8} \sqrt{t}-\frac{3 q^{5/8}}{a \sqrt{t}}+\left(6 a^2\right) q^{3/4} t-9 q+\left(10 a^3\right) q^{9/8} t^{3/2}
\\
+\frac{3 q^{5/4}}{a^2 t}+(-15 a) q^{11/8} \sqrt{t}+\left(15 a^4\right) q^{3/2} t^2+\frac{6 q^{13/8}}{a \sqrt{t}}+\left(-21 a^2\right) q^{7/4} t
\\
+\frac{q^{15/8} \left(\left(21 a^8\right) t^4-1\right)}{a^3 t^{3/2}}+\left(-27 a^3\right) q^{17/8} t^{3/2}+\frac{q^{9/4} \left(\left(28 a^8\right) t^4+6\right)}{a^2 t}
\\
+(-12 a) q^{19/8} \sqrt{t}+\left(-33 a^4\right) q^{5/2} t^2+\frac{9 q^{21/8} \left(\left(4 a^8\right) t^4
+3\right)}{a \sqrt{t}}+\left(-24 a^2\right) q^{11/4} t-\frac{3 q^{23/8} \left(\left(13 a^8\right) t^4+3\right)}{a^3 t^{3/2}}
+\cdots
\\
\end{smallmatrix}
}
\\ \hline 
2
&
2
&
2
&
{\scriptscriptstyle 
\begin{smallmatrix}
1+\sqrt{q} \left(\left(4 a^2\right) y-\frac{2}{a^2 y}\right)+q \left(\left(9 a^4\right) y^2+\frac{2}{a^4 y^2}-8\right)
\\
+\frac{2 q^{3/2} \left(\left(8 a^{12}\right) y^6-\left(8 a^8\right) y^4+\left(2 a^4\right) y^2-1\right)}{a^6 y^3}
\\
+q^2 \left(\left(25 a^8\right) y^4-\left(24 a^4\right) y^2+\frac{2}{a^8 y^4}-2\right)+q^{5/2} \left(\left(36 a^{10}\right) y^5-\left(32 a^6\right) y^3-\frac{2}{a^{10} y^5}-\left(16 a^2\right) y+\frac{16}{a^2 y}\right)
\\
+q^3 \left(\left(49 a^{12}\right) y^6-\left(40 a^8\right) y^4-\left(32 a^4\right) y^2-\frac{7}{a^4 y^2}+\frac{2}{a^{12} y^6}+32\right)
+\cdots
\\
\end{smallmatrix}
}
\\ \hline 
3
&
3
&
2
&
{\scriptscriptstyle 
\begin{smallmatrix}
1-\frac{\sqrt[4]{q}}{a^3 y^{3/2}}+\sqrt{q} \left(\frac{1}{a^6 y^3}-\frac{3}{a^2 y}+\left(6 a^2\right) y\right)
\\
+\frac{q^{3/4} \left(a^{12} y^6-\left(6 a^8\right) y^4+\left(3 a^4\right) y^2-1\right)}{a^9 y^{9/2}}
\\
+q \left(\left(21 a^4\right) y^2+\frac{9}{a^4 y^2}-\frac{3}{a^8 y^4}+\frac{1}{a^{12} y^6}-19\right)
\\
+\frac{q^{5/4} \left(\left(6 a^{20}\right) y^{10}-\left(24 a^{16}\right) y^8+\left(18 a^{12}\right) y^6-\left(9 a^8\right) y^4+\left(3 a^4\right) y^2-1\right)}{a^{15} y^{15/2}}
\\
+\frac{q^{3/2} \left(\left(56 a^{24}\right) y^{12}-\left(66 a^{20}\right) y^{10}+\left(36 a^{16}\right) y^8-\left(19 a^{12}\right) y^6+\left(9 a^8\right) y^4-\left(3 a^4\right) y^2+1\right)}{a^{18} y^9}
+\cdots
\\
\end{smallmatrix}
}
\\ 
\end{array}
\end{align}

%%%%%%%%%%%%%%%%%%%%%%%%%%%%%%%%%%%%%%%%%%%
\subsubsection{$SO(N_c)-[N_f]_{+-}$}
\label{SO+-_check}
%%%%%%%%%%%%%%%%%%%%%%%%%%%%%%%%%%%%%%%%%%%

%-+
\begin{align}
\begin{array}{c|c|c|c}
N_c
&
N_f
&
\tilde{N}_c
&
%\textrm{series expansions}
I^{SO(N_c)-[N_f]^A_{+-}}
=
I^{SO(\tilde{N}_c)-[N_f]^B_{+-}}
\\ \hline
1
&
1
&
2
&
{\scriptscriptstyle 
\begin{smallmatrix}
1-a \sqrt[4]{q} \sqrt{y}+a^2 \sqrt{q} y+\frac{q^{3/4} \left(1-a^4 y^2\right)}{a \sqrt{y}}+q \left(a^4 y^2-1\right)
\\
-a^5 q^{5/4} y^{5/2}+a^6 q^{3/2} y^3+\frac{q^{7/4} \left(1-a^8 y^4\right)}{a \sqrt{y}}+q^2 \left(a^8 y^4-2\right)
\\
+q^{9/4} \left(a \sqrt{y}\right) \left(1-a^8 y^4\right)+q^{5/2} \left(a^{10} y^5+\frac{1}{a^2 y}\right)-a^{11} q^{11/4} y^{11/2}+\cdots
\\
\end{smallmatrix}
}
\\ \hline 
1
&
2
&
3
&
{\scriptscriptstyle 
\begin{smallmatrix}
1+(-2 a) q^{3/8} \sqrt{t}+\frac{2 q^{5/8}}{a \sqrt{t}}+\left(3 a^2\right) q^{3/4} t-4 q+\left(-4 a^3\right) q^{9/8} t^{3/2}
\\
+\frac{q^{5/4}}{a^2 t}+(4 a) q^{11/8} \sqrt{t}+\left(5 a^4\right) q^{3/2} t^2+\left(-4 a^2\right) q^{7/4} t
\\
+\left(-6 a^5\right) q^{15/8} t^{5/2}-5 q^2+\left(4 a^3\right) q^{17/8} t^{3/2}+\frac{q^{9/4} \left(\left(7 a^8\right) t^4+4\right)}{a^2 t}
\\
+(8 a) q^{19/8} \sqrt{t}+\left(-4 a^4\right) q^{5/2} t^2-\frac{8 q^{21/8} \left(a^8 t^4+1\right)}{a \sqrt{t}}+\left(-8 a^2\right) q^{11/4} t
+\frac{q^{23/8} \left(\left(4 a^8\right) t^4+2\right)}{a^3 t^{3/2}}+\cdots
\end{smallmatrix}
}
\\ \hline 
1
&
3
&
4
&
{\scriptscriptstyle 
\begin{smallmatrix}
1+(-3 a) q^{3/8} \sqrt{t}+\frac{3 q^{5/8}}{a \sqrt{t}}+\left(6 a^2\right) q^{3/4} t-9 q+\left(-10 a^3\right) q^{9/8} t^{3/2}
\\
+\frac{3 q^{5/4}}{a^2 t}+(15 a) q^{11/8} \sqrt{t}+\left(15 a^4\right) q^{3/2} t^2-\frac{6 q^{13/8}}{a \sqrt{t}}+\left(-21 a^2\right) q^{7/4} t
\\
+\frac{q^{15/8} \left(1-\left(21 a^8\right) t^4\right)}{a^3 t^{3/2}}+\left(27 a^3\right) q^{17/8} t^{3/2}+\frac{q^{9/4} \left(\left(28 a^8\right) t^4+6\right)}{a^2 t}
\\
+(12 a) q^{19/8} \sqrt{t}+\left(-33 a^4\right) q^{5/2} t^2-\frac{9 q^{21/8} \left(\left(4 a^8\right) t^4+3\right)}{a \sqrt{t}}+\left(-24 a^2\right) q^{11/4} t+\frac{q^{23/8} \left(\left(39 a^8\right) t^4+9\right)}{a^3 t^{3/2}}+\cdots
\\
\end{smallmatrix}
}
\\ \hline 
2
&
2
&
2
&
{\scriptscriptstyle 
\begin{smallmatrix}
1+\left(2 a^2\right) \sqrt{q} y+\left(3 a^4\right) q y^2+\frac{q^{3/2} \left(\left(4 a^8\right) y^4-2\right)}{a^2 y}
\\
+q^2 \left(\left(5 a^8\right) y^4-4\right)+q^{5/2} \left(\left(2 a^2\right) y\right) \left(\left(3 a^8\right) y^4-2\right)
\\
+q^3 \left(\left(7 a^{12}\right) y^6-\left(4 a^4\right) y^2+\frac{1}{a^4 y^2}\right)+q^{7/2} \left(\left(4 a^6\right) y^3\right) \left(\left(2 a^8\right) y^4-1\right)+q^4 \left(\left(9 a^{16}\right) y^8-\left(4 a^8\right) y^4-5\right)
\\
+q^{9/2} \left(\left(2 a^2\right) y\right) \left(\left(5 a^{16}\right) y^8-\left(2 a^8\right) y^4-4\right)+\cdots
\\
\end{smallmatrix}
}
\\ \hline 
3
&
3
&
2
&
{\scriptscriptstyle 
\begin{smallmatrix}
1+\frac{\sqrt[4]{q}}{a^3 y^{3/2}}+\sqrt{q} \left(\frac{1}{a^6 y^3}-\frac{3}{a^2 y}+\left(6 a^2\right) y\right)
\\
+\frac{q^{3/4} \left(a^{12} \left(-y^6\right)+\left(6 a^8\right) y^4-\left(3 a^4\right) y^2+1\right)}{a^9 y^{9/2}}
\\
+q \left(\left(21 a^4\right) y^2+\frac{9}{a^4 y^2}-\frac{3}{a^8 y^4}+\frac{1}{a^{12} y^6}-19\right)
\\
+\frac{q^{5/4} \left(-\left(6 a^{20}\right) y^{10}+\left(24 a^{16}\right) y^8-\left(18 a^{12}\right) y^6+\left(9 a^8\right) y^4-\left(3 a^4\right) y^2+1\right)}{a^{15} y^{15/2}}
\\
+\frac{q^{3/2} \left(\left(56 a^{24}\right) y^{12}-\left(66 a^{20}\right) y^{10}+\left(36 a^{16}\right) y^8-\left(19 a^{12}\right) y^6+\left(9 a^8\right) y^4-\left(3 a^4\right) y^2+1\right)}{a^{18} y^9}
+\cdots
\\
\end{smallmatrix}
}
\\ 
\end{array}
\end{align}

%%%%%%%%%%%%%%%%%%%%%%%%%%%%%%%%%%%%%%%%%%%
\subsubsection{$SO(N_c)-[N_f]_{--}$}
\label{SO--_check}
%%%%%%%%%%%%%%%%%%%%%%%%%%%%%%%%%%%%%%%%%%%

%-+
\begin{align}
\begin{array}{c|c|c|c}
N_c
&
N_f
&
\tilde{N}_c
&
%\textrm{series expansions}
I^{SO(N_c)-[N_f]^A_{--}}
=
I^{SO(\tilde{N}_c)-[N_f]^B_{-+}}
\\ \hline
1
&
1
&
2
&
{\scriptscriptstyle 
\begin{smallmatrix}
1-a \sqrt[4]{q} \sqrt{y}+a^2 \sqrt{q} y+\frac{q^{3/4} \left(1-a^4 y^2\right)}{a \sqrt{y}}+q \left(a^4 y^2-1\right)
\\
-a^5 q^{5/4} y^{5/2}+a^6 q^{3/2} y^3+\frac{q^{7/4} \left(1-a^8 y^4\right)}{a \sqrt{y}}+q^2 \left(a^8 y^4-2\right)
\\
+q^{9/4} \left(a \sqrt{y}\right) \left(1-a^8 y^4\right)+q^{5/2} \left(a^{10} y^5+\frac{1}{a^2 y}\right)-a^{11} q^{11/4} y^{11/2}+\cdots
\\
\end{smallmatrix}
}
\\ \hline 
1
&
2
&
3
&
{\scriptscriptstyle 
\begin{smallmatrix}
1+(-2 a) q^{3/8} \sqrt{t}+\frac{2 q^{5/8}}{a \sqrt{t}}+\left(3 a^2\right) q^{3/4} t-4 q+\left(-4 a^3\right) q^{9/8} t^{3/2}
\\
+\frac{q^{5/4}}{a^2 t}+(4 a) q^{11/8} \sqrt{t}+\left(5 a^4\right) q^{3/2} t^2+\left(-4 a^2\right) q^{7/4} t
\\
+\left(-6 a^5\right) q^{15/8} t^{5/2}-5 q^2+\left(4 a^3\right) q^{17/8} t^{3/2}+\frac{q^{9/4} \left(\left(7 a^8\right) t^4+4\right)}{a^2 t}
\\
+(8 a) q^{19/8} \sqrt{t}+\left(-4 a^4\right) q^{5/2} t^2-\frac{8 q^{21/8} \left(a^8 t^4+1\right)}{a \sqrt{t}}+\left(-8 a^2\right) q^{11/4} t
+\frac{q^{23/8} \left(\left(4 a^8\right) t^4+2\right)}{a^3 t^{3/2}}+\cdots
\\
\end{smallmatrix}
}
\\ \hline 
1
&
3
&
4
&
{\scriptscriptstyle 
\begin{smallmatrix}
1+(-3 a) q^{3/8} \sqrt{t}+\frac{3 q^{5/8}}{a \sqrt{t}}+\left(6 a^2\right) q^{3/4} t-9 q+\left(-10 a^3\right) q^{9/8} t^{3/2}
\\
+\frac{3 q^{5/4}}{a^2 t}+(15 a) q^{11/8} \sqrt{t}+\left(15 a^4\right) q^{3/2} t^2-\frac{6 q^{13/8}}{a \sqrt{t}}+\left(-21 a^2\right) q^{7/4} t
\\
+\frac{q^{15/8} \left(1-\left(21 a^8\right) t^4\right)}{a^3 t^{3/2}}+\left(27 a^3\right) q^{17/8} t^{3/2}+\frac{q^{9/4} \left(\left(28 a^8\right) t^4+6\right)}{a^2 t}
\\
+(12 a) q^{19/8} \sqrt{t}+\left(-33 a^4\right) q^{5/2} t^2-\frac{9 q^{21/8} \left(\left(4 a^8\right) t^4+3\right)}{a \sqrt{t}}+\left(-24 a^2\right) q^{11/4} t+\frac{q^{23/8} \left(\left(39 a^8\right) t^4+9\right)}{a^3 t^{3/2}}+\cdots
\\
\end{smallmatrix}
}
\\ \hline 
2
&
2
&
2
&
{\scriptscriptstyle 
\begin{smallmatrix}
1+\left(2 a^2\right) \sqrt{q} y+\left(3 a^4\right) q y^2+\frac{q^{3/2} \left(\left(4 a^8\right) y^4-2\right)}{a^2 y}
\\
+q^2 \left(\left(5 a^8\right) y^4-4\right)+q^{5/2} \left(\left(2 a^2\right) y\right) \left(\left(3 a^8\right) y^4-2\right)
\\
+q^3 \left(\left(7 a^{12}\right) y^6-\left(4 a^4\right) y^2+\frac{1}{a^4 y^2}\right)+q^{7/2} \left(\left(4 a^6\right) y^3\right) \left(\left(2 a^8\right) y^4-1\right)+q^4 \left(\left(9 a^{16}\right) y^8-\left(4 a^8\right) y^4-5\right)
\\
+q^{9/2} \left(\left(2 a^2\right) y\right) \left(\left(5 a^{16}\right) y^8-\left(2 a^8\right) y^4-4\right)+\cdots
\\
\end{smallmatrix}
}
\\ \hline 
3
&
3
&
2
&
{\scriptscriptstyle 
\begin{smallmatrix}
1-\frac{\sqrt[4]{q}}{a^3 y^{3/2}}+\sqrt{q} \left(\frac{1}{a^6 y^3}+6 a^2 y+\frac{3}{a^2 y}\right)
\\
+\frac{q^{3/4} \left(a^{12} \left(-y^6\right)-6 a^8 y^4-3 a^4 y^2-1\right)}{a^9 y^{9/2}}
\\
+q \left(21 a^4 y^2+\frac{3}{a^4 y^2}+\frac{3}{a^8 y^4}+\frac{1}{a^{12} y^6}+1\right)
\\
+\frac{q^{5/4} \left(-6 a^{20} y^{10}-6 a^{16} y^8-2 a^{12} y^6-3 a^8 y^4-3 a^4 y^2-1\right)}{a^{15} y^{15/2}}
\\
+\frac{q^{3/2} \left(56 a^{24} y^{12}-24 a^{20} y^{10}+a^{12} y^6+3 a^8 y^4+3 a^4 y^2+1\right)}{a^{18} y^9}
+\cdots
\\
\end{smallmatrix}
}
\\ 
\end{array}
\end{align}

%%%%%%%%%%%%%%%%%%%%%%%%%%%%%%%%%%%%%%%%%%%
\subsubsection{$SO(N_c)-[N_f]_{++}$ with $\mathcal{N}, (\mathrm{N,N,N;N,N,N})+\mathrm{Fermis}$}
\label{SO++_N_check}
%%%%%%%%%%%%%%%%%%%%%%%%%%%%%%%%%%%%%%%%%%%

\begin{align}
\begin{array}{c|c|c|c}
N_c
&
N_f
&
\tilde{N}_c
&
%\textrm{series expansions}
\mathbb{II}^{SO(N_c)-[N_f]^A_{++}}_{\mathcal{N}, \mathrm{N}+\{\Psi\}}
=
\mathbb{II}^{SO(\tilde{N}_c)-[N_f]^B_{++}}_{\mathcal{D}, \mathrm{D},\mathrm{N},\mathrm{D}}
\\ \hline
1
&
1
&
2
&
{\scriptscriptstyle 
\begin{smallmatrix}
1+a \sqrt[4]{q} \sqrt{y}+\sqrt{q} \left(a^2 y-u-\frac{1}{u}\right)+\frac{q^{3/4} \left(a \sqrt{y}\right) \left(y \left(a^2 u\right)-u^2-1\right)}{u}
\\
+q \left(-\frac{y \left(a^2 \left(u^2+1\right)\right)}{u}+a^4 y^2+1\right)+q^{5/4} \left(a \sqrt{y}\right) \left(-\frac{y \left(a^2 \left(u^2+1\right)\right)}{u}+a^4 y^2+2\right)
\\
-\frac{q^{3/2} \left(y^2 \left(a^4 \left(u^2+1\right)\right)+y^3 \left(-\left(a^6 u\right)\right)-y \left(\left(2 a^2\right) u\right)+u^2+1\right)}{u}+\cdots
\\
\end{smallmatrix}
}
\\ \hline 
1
&
2
&
3
&
{\scriptscriptstyle 
\begin{smallmatrix}
1+(2 a) q^{3/8} \sqrt{t}-\frac{\sqrt{q} \left(u^2+u+1\right)}{u}+\left(3 a^2\right) q^{3/4} t-\frac{q^{7/8} \left(u^2+u+1\right) \left((2 a) \sqrt{t}\right)}{u}
\\
+q \left(u+\frac{1}{u}+1\right)+\left(4 a^3\right) q^{9/8} t^{3/2}-\frac{q^{5/4} \left(u^2+u+1\right) \left(\left(3 a^2\right) t\right)}{u}
\\
+\frac{q^{11/8} (u+1)^2 \left((2 a) \sqrt{t}\right)}{u}+q^{3/2} \left(\left(5 a^4\right) t^2-\frac{(u+1)^2}{u}\right)+\cdots
\\
\end{smallmatrix}
}
\\ \hline 
1
&
3
&
4
&
{\scriptscriptstyle 
\begin{smallmatrix}
1+(3 a) q^{3/8} \sqrt{t}+\sqrt{q} \left(-\text{u1}-\frac{1}{\text{u1}}-\text{u2}-\frac{1}{\text{u2}}\right)
\\
+\left(6 a^2\right) q^{3/4} t-\frac{q^{7/8} \left((3 a) \sqrt{t}\right) \left(\text{u1}^2 \text{u2}+\text{u1} \text{u2}^2+\text{u1}+\text{u2}\right)}{\text{u1} \text{u2}}
\\
+q \left(\text{u1} \left(\text{u2}+\frac{1}{\text{u2}}\right)+\frac{\text{u2}+\frac{1}{\text{u2}}}{\text{u1}}+2\right)+\left(10 a^3\right) q^{9/8} t^{3/2}
-\frac{q^{5/4} \left(\left(6 a^2\right) t\right) \left(\text{u1}^2 \text{u2}+\text{u1} \text{u2}^2+\text{u1}+\text{u2}\right)}{\text{u1} \text{u2}}+\cdots
\\
\end{smallmatrix}
}
\\ \hline 
2
&
2
&
2
&
{\scriptscriptstyle 
\begin{smallmatrix}
1+\left(4 a^2\right) \sqrt{q} y-\frac{q^{3/4} \sqrt{y} \left((4 a) \left(u^2+1\right)\right)}{u}
\\
+q \left(\left(9 a^4\right) y^2+u^2+\frac{1}{u^2}+1\right)-\frac{q^{5/4} y^{3/2} \left(\left(12 a^3\right) \left(u^2+1\right)\right)}{u}
\\
+\frac{q^{3/2} \left(\left(2 a^2\right) y\right) \left(u^2 \left(\left(8 a^4\right) y^2+9\right)+2 u^4+2\right)}{u^2}
\\
-\frac{q^{7/4} \left(\left(3 a^4\right) y^2+1\right) \left(\sqrt{y} \left((8 a) \left(u^2+1\right)\right)\right)}{u}
+\cdots
\\
\end{smallmatrix}
}
\\ \hline 
3
&
3
&
2
&
{\scriptscriptstyle 
\begin{smallmatrix}
1+\left(6 a^2\right) \sqrt{q} y+\frac{q^{3/4} \left(a \sqrt{y}\right) \left(y \left(a^2 u\right)-3 u^2-3\right)}{u}
\\
+q \left(-\frac{y \left(\left(3 a^2\right) \left(u^2+1\right)\right)}{u}+\left(21 a^4\right) y^2+1\right)
\\
\frac{q^{5/4} \left((3 a) \sqrt{y}\right) \left(y^2 \left(\left(2 a^4\right) u^2\right)-y \left(\left(6 a^2\right) u^3\right)-y \left(\left(6 a^2\right) u\right)+u^4+1\right)}{u^2}
\\
+q^{3/2} \left(\frac{\left(3 a^2\right) y}{u^2}+y \left(\left(3 a^2\right) u^2\right)-\frac{\left(18 a^4\right) y^2}{u}-y^2 \left(\left(18 a^4\right) u\right)+\left(\left(7 a^2\right) y\right) \left(\left(8 a^4\right) y^2+3\right)-u^3-\frac{1}{u^3}\right)+\cdots
\\
\end{smallmatrix}
}
\\ 
\end{array}
\end{align}

%%%%%%%%%%%%%%%%%%%%%%%%%%%%%%%%%%%%%%%%%%%
\subsubsection{$SO(N_c)-[N_f]_{-+}$ with $\mathcal{N}, (\mathrm{N,N,N;N,N,N})+\mathrm{Fermis}$}
\label{SO-+_N_check}
%%%%%%%%%%%%%%%%%%%%%%%%%%%%%%%%%%%%%%%%%%%

\begin{align}
\begin{array}{c|c|c|c}
N_c
&
N_f
&
\tilde{N}_c
&
%\textrm{series expansions}
\mathbb{II}^{SO(N_c)-[N_f]^A_{-+}}_{\mathcal{N}, \mathrm{N}+\{\Psi\}}
=
\mathbb{II}^{SO(\tilde{N}_c)-[N_f]^B_{--}}_{\mathcal{D}, \mathrm{D},\mathrm{N},\mathrm{D}}
\\ \hline
1
&
1
&
2
&
{\scriptscriptstyle 
\begin{smallmatrix}
1+a \sqrt[4]{q} \sqrt{y}+a^2 \sqrt{q} y+a^3 q^{3/4} y^{3/2}+q \left(a^4 y^2-1\right)+a^5 q^{5/4} y^{5/2}
\\
+a^6 q^{3/2} y^3+a^7 q^{7/4} y^{7/2}+a^8 q^2 y^4+a^9 q^{9/4} y^{9/2}+q^{5/2} \left(a^2 y\right) \left(a^8 y^4+1\right)+\cdots
\\
\end{smallmatrix}
}
\\ \hline 
1
&
2
&
3
&
{\scriptscriptstyle 
\begin{smallmatrix}
1+(2 a) q^{3/8} \sqrt{t}+\sqrt{q} \left(-u-\frac{1}{u}+1\right)+\left(3 a^2\right) q^{3/4} t
\\
+q^{7/8} \left(-u-\frac{1}{u}+1\right) \left((2 a) \sqrt{t}\right)+q \left(-u-\frac{1}{u}+1\right)+\left(4 a^3\right) q^{9/8} t^{3/2}
\\
+q^{5/4} \left(-u-\frac{1}{u}+1\right) \left(\left(3 a^2\right) t\right)-\frac{q^{11/8} (u-1)^2 \left((2 a) \sqrt{t}\right)}{u}+q^{3/2} \left(\left(5 a^4\right) t^2-u-\frac{1}{u}+2\right)+\cdots
\\
\end{smallmatrix}
}
\\ \hline 
1
&
3
&
4
&
{\scriptscriptstyle 
\begin{smallmatrix}
1+(3 a) q^{3/8} \sqrt{t}-\frac{\sqrt{q} \left(u^2+1\right)}{u}+\left(6 a^2\right) q^{3/4} t-\frac{q^{7/8} \left(u^2+1\right) \left((3 a) \sqrt{t}\right)}{u}
\\
+\left(10 a^3\right) q^{9/8} t^{3/2}-\frac{q^{5/4} \left(u^2+1\right) \left(\left(6 a^2\right) t\right)}{u}+(3 a) q^{11/8} \sqrt{t}
\\
+\left(15 a^4\right) q^{3/2} t^2-\frac{q^{13/8} \left(u^2+1\right) \left(\left(10 a^3\right) t^{3/2}\right)}{u}+\left(9 a^2\right) q^{7/4} t+\cdots
\\
\end{smallmatrix}
}
\\ \hline 
2
&
2
&
2
&
{\scriptscriptstyle 
\begin{smallmatrix}
1+\left(4 a^2\right) \sqrt{q} y+q \left(\left(9 a^4\right) y^2-1\right)+q^{3/2} \left(\left(2 a^2\right) y\right) \left(\left(8 a^4\right) y^2-1\right)
\\
+q^2 \left(a^4 y^2\right) \left(\left(25 a^4\right) y^2-1\right)+q^{5/2} \left(\left(2 a^2\right) y\right) \left(a^4 y^2+\left(18 a^8\right) y^4+1\right)
+\cdots
\\
\end{smallmatrix}
}
\\ \hline 
3
&
3
&
2
&
{\scriptscriptstyle 
\begin{smallmatrix}
1+6 a^2 \sqrt{q} y+a^3 q^{3/4} y^{3/2}+q \left(21 a^4 y^2-1\right)+6 a^5 q^{5/4} y^{5/2}+a^2 q^{3/2} y \left(56 a^4 y^2-3\right)
\\
+21 a^7 q^{7/4} y^{7/2}+3 a^4 q^2 y^2 \left(42 a^4 y^2-1\right)+a^5 q^{9/4} y^{5/2} \left(56 a^4 y^2+3\right)+a^2 q^{5/2} y \left(252 a^8 y^4+8 a^4 y^2+3\right)+\cdots
\\
\end{smallmatrix}
}
\\ 
\end{array}
\end{align}

%%%%%%%%%%%%%%%%%%%%%%%%%%%%%%%%%%%%%%%%%%%
\subsubsection{$SO(N_c)-[N_f]_{+-}$ with $\mathcal{N}, (\mathrm{N,N,N;N,N,N})+\mathrm{Fermis}$}
\label{SO+-_N_check}
%%%%%%%%%%%%%%%%%%%%%%%%%%%%%%%%%%%%%%%%%%%

\begin{align}
\begin{array}{c|c|c|c}
N_c
&
N_f
&
\tilde{N}_c
&
%\textrm{series expansions}
\mathbb{II}^{SO(N_c)-[N_f]^A_{+-}}_{\mathcal{N}, \mathrm{N}+\{\Psi\}}
=
\mathbb{II}^{SO(\tilde{N}_c)-[N_f]^B_{+-}}_{\mathcal{D}, \mathrm{D},\mathrm{N},\mathrm{D}}
\\ \hline
1
&
1
&
2
&
{\scriptscriptstyle 
\begin{smallmatrix}
1-a \sqrt[4]{q} \sqrt{y}+a^2 \sqrt{q} y-a^3 q^{3/4} y^{3/2}+q \left(a^4 y^2-1\right)-a^5 q^{5/4} y^{5/2}
\\
+a^6 q^{3/2} y^3-a^7 q^{7/4} y^{7/2}+a^8 q^2 y^4-a^9 q^{9/4} y^{9/2}+q^{5/2} \left(a^2 y\right) \left(a^8 y^4+1\right)+\cdots
\\
\end{smallmatrix}
}
\\ \hline 
1
&
2
&
3
&
{\scriptscriptstyle 
\begin{smallmatrix}
1+(-2 a) q^{3/8} \sqrt{t}+\sqrt{q} \left(u+\frac{1}{u}-1\right)+\left(3 a^2\right) q^{3/4} t
\\
+q^{7/8} \left(u+\frac{1}{u}-1\right) \left((-2 a) \sqrt{t}\right)+q \left(-u-\frac{1}{u}+1\right)+\left(-4 a^3\right) q^{9/8} t^{3/2}
\\
+q^{5/4} \left(u+\frac{1}{u}-1\right) \left(\left(3 a^2\right) t\right)+\frac{q^{11/8} (u-1)^2 \left((2 a) \sqrt{t}\right)}{u}+q^{3/2} \left(\left(5 a^4\right) t^2+u+\frac{1}{u}-2\right)+\cdots
\\
\end{smallmatrix}
}
\\ \hline 
1
&
3
&
4
&
{\scriptscriptstyle 
\begin{smallmatrix}
1+(-3 a) q^{3/8} \sqrt{t}+\sqrt{q} \left(u+\frac{1}{u}\right)+\left(6 a^2\right) q^{3/4} t+q^{7/8} \left(u+\frac{1}{u}\right) \left((-3 a) \sqrt{t}\right)
\\
+\left(-10 a^3\right) q^{9/8} t^{3/2}+q^{5/4} \left(u+\frac{1}{u}\right) \left(\left(6 a^2\right) t\right)+(-3 a) q^{11/8} \sqrt{t}
\\
+\left(15 a^4\right) q^{3/2} t^2+q^{13/8} \left(u+\frac{1}{u}\right) \left(\left(-10 a^3\right) t^{3/2}\right)+\left(9 a^2\right) q^{7/4} t
+\cdots
\\
\end{smallmatrix}
}
\\ \hline 
2
&
2
&
2
&
{\scriptscriptstyle 
\begin{smallmatrix}
1+\left(2 a^2\right) \sqrt{q} y+q \left(\left(3 a^4\right) y^2-1\right)+q^{3/2} \left(\left(4 a^6\right) y^3-\left(2 a^2\right) y\right)
\\
+q^2 \left(a^4 y^2\right) \left(\left(5 a^4\right) y^2-3\right)+q^{5/2} \left(\left(2 a^2\right) y\right) \left(\left(3 a^8\right) y^4-\left(2 a^4\right) y^2+1\right)+\cdots
\\
\end{smallmatrix}
}
\\ \hline 
3
&
3
&
2
&
{\scriptscriptstyle 
\begin{smallmatrix}
1+\left(6 a^2\right) \sqrt{q} y-a^3 q^{3/4} y^{3/2}+q \left(\left(21 a^4\right) y^2-1\right)+\left(-6 a^5\right) q^{5/4} y^{5/2}+q^{3/2} \left(a^2 y\right) \left(\left(56 a^4\right) y^2-3\right)
\\
+\left(-21 a^7\right) q^{7/4} y^{7/2}+q^2 \left(\left(3 a^4\right) y^2\right) \left(\left(42 a^4\right) y^2-1\right)
\\
+q^{9/4} \left(-a^5 y^{5/2}\right) \left(\left(56 a^4\right) y^2+3\right)+q^{5/2} \left(a^2 y\right) \left(\left(252 a^8\right) y^4+\left(8 a^4\right) y^2+3\right)+\cdots
\\
\end{smallmatrix}
}
\\ 
\end{array}
\end{align}

%%%%%%%%%%%%%%%%%%%%%%%%%%%%%%%%%%%%%%%%%%%
\subsubsection{$SO(N_c)-[N_f]_{--}$ with $\mathcal{N}, (\mathrm{N,N,N;N,N,N})+\mathrm{Fermis}$}
\label{SO--_N_check}
%%%%%%%%%%%%%%%%%%%%%%%%%%%%%%%%%%%%%%%%%%%

\begin{align}
\begin{array}{c|c|c|c}
N_c
&
N_f
&
\tilde{N}_c
&
%\textrm{series expansions}
\mathbb{II}^{SO(N_c)-[N_f]^A_{--}}_{\mathcal{N}, \mathrm{N}+\{\Psi\}}
=
\mathbb{II}^{SO(\tilde{N}_c)-[N_f]^B_{-+}}_{\mathcal{D}, \mathrm{D},\mathrm{N},\mathrm{D}}
\\ \hline
1
&
1
&
2
&
{\scriptscriptstyle 
\begin{smallmatrix}
1-a \sqrt[4]{q} \sqrt{y}+\sqrt{q} \left(a^2 y+u+\frac{1}{u}\right)-\frac{q^{3/4} \left(a \sqrt{y}\right) \left(y \left(a^2 u\right)+u^2+1\right)}{u}
\\
+q \left(y \left(a^2 \left(u+\frac{1}{u}\right)\right)+a^4 y^2+1\right)+q^{5/4} \left(a \sqrt{y}\right) \left(-y \left(a^2 \left(u+\frac{1}{u}\right)\right)+a^4 \left(-y^2\right)-2\right)
\\
+q^{3/2} \left(y^2 \left(a^4 \left(u+\frac{1}{u}\right)\right)+a^6 y^3+\left(2 a^2\right) y+u+\frac{1}{u}\right)+\cdots
\\
\end{smallmatrix}
}
\\ \hline 
1
&
2
&
3
&
{\scriptscriptstyle 
\begin{smallmatrix}
1+(-2 a) q^{3/8} \sqrt{t}+\sqrt{q} \left(u+\frac{1}{u}+1\right)+\left(3 a^2\right) q^{3/4} t
\\
+q^{7/8} \left(u+\frac{1}{u}+1\right) \left((-2 a) \sqrt{t}\right)+q \left(u+\frac{1}{u}+1\right)+\left(-4 a^3\right) q^{9/8} t^{3/2}
\\
+q^{5/4} \left(u+\frac{1}{u}+1\right) \left(\left(3 a^2\right) t\right)-\frac{q^{11/8} (u+1)^2 \left((2 a) \sqrt{t}\right)}{u}+q^{3/2} \left(\left(5 a^4\right) t^2+u+\frac{1}{u}+2\right)+\cdots
\\
\end{smallmatrix}
}
\\ \hline 
1
&
3
&
4
&
{\scriptscriptstyle 
\begin{smallmatrix}
1+(-3 a) q^{3/8} \sqrt{t}+\sqrt{q} \left(\text{u1}+\frac{1}{\text{u1}}+\text{u2}+\frac{1}{\text{u2}}\right)+\left(6 a^2\right) q^{3/4} t
\\
+q^{7/8} \left((-3 a) \sqrt{t}\right) \left(\text{u1}+\frac{1}{\text{u1}}+\text{u2}+\frac{1}{\text{u2}}\right)+q \left(\text{u1} \left(\text{u2}+\frac{1}{\text{u2}}\right)+\frac{\text{u2}+\frac{1}{\text{u2}}}{\text{u1}}+2\right)
\\
+\left(-10 a^3\right) q^{9/8} t^{3/2}+q^{5/4} \left(\left(6 a^2\right) t\right) \left(\text{u1}+\frac{1}{\text{u1}}+\text{u2}+\frac{1}{\text{u2}}\right)
+\cdots
\\
\end{smallmatrix}
}
\\ \hline 
2
&
2
&
2
&
{\scriptscriptstyle 
\begin{smallmatrix}
1+\left(2 a^2\right) \sqrt{q} y+q \left(\left(3 a^4\right) y^2-u^2-\frac{1}{u^2}+1\right)+\frac{q^{3/2} \left(\left(2 a^2\right) y\right) \left(u^2 \left(\left(2 a^4\right) y^2+1\right)-u^4-1\right)}{u^2}
\\
+q^2 \left(-u^2 \left(\left(3 a^4\right) y^2+1\right)-\frac{\left(3 a^4\right) y^2+1}{u^2}+\left(5 a^8\right) y^4+\left(3 a^4\right) y^2+2\right)+\cdots
\\
\end{smallmatrix}
}
\\ \hline 
3
&
3
&
2
&
{\scriptscriptstyle 
\begin{smallmatrix}
1+\left(6 a^2\right) \sqrt{q} y-\frac{q^{3/4} \left(a \sqrt{y}\right) \left(y \left(a^2 u\right)+3 u^2+3\right)}{u}+q \left(\frac{y \left(\left(3 a^2\right) \left(u^2+1\right)\right)}{u}+\left(21 a^4\right) y^2+1\right)
\\
-\frac{q^{5/4} \left((3 a) \sqrt{y}\right) \left(y^2 \left(\left(2 a^4\right) u^2\right)+y \left(\left(6 a^2\right) u^3\right)+y \left(\left(6 a^2\right) u\right)+u^4+1\right)}{u^2}
\\
+q^{3/2} \left(\frac{\left(3 a^2\right) y}{u^2}+y \left(\left(3 a^2\right) u^2\right)+\frac{\left(18 a^4\right) y^2}{u}+y^2 \left(\left(18 a^4\right) u\right)+\left(\left(7 a^2\right) y\right) \left(\left(8 a^4\right) y^2+3\right)+u^3+\frac{1}{u^3}\right)
+\cdots
\\
\end{smallmatrix}
}
\\ 
\end{array}
\end{align}

%%%%%%%%%%%%%%%%%%%%%%%%%%%%%%%%%%%%%%%%%%%
\subsubsection{$SO(N_c)-[N_f]_{++}$ with $\mathcal{D}, (\mathrm{D,D,D; D,D,D})$}
\label{SO++_D_check}
%%%%%%%%%%%%%%%%%%%%%%%%%%%%%%%%%%%%%%%%%%%

\begin{align}
\begin{array}{c|c|c|c}
N_c
&
N_f
&
\tilde{N}_c
&
%\textrm{series expansions}
\mathbb{II}^{SO(N_c)-[N_f]^A_{++}}_{\mathcal{D}, \mathrm{D}}
=
\mathbb{II}^{SO(\tilde{N}_c)-[N_f]^B_{++}}_{\mathcal{N}, \mathrm{N},\mathrm{D},\mathrm{N}}
\\ \hline
1
&
1
&
2
&
{\scriptscriptstyle 
\begin{smallmatrix}
1-\frac{q^{3/4}}{a \sqrt{y}}-\frac{q^{7/4}}{a \sqrt{y}}+\frac{q^{5/2}}{a^2 y}-\frac{q^{11/4}}{a \sqrt{y}}+\frac{q^{7/2}}{a^2 y}-\frac{q^{15/4}}{a \sqrt{y}}
+\frac{2 q^{9/2}}{a^2 y}-\frac{q^{19/4}}{a \sqrt{y}}-\frac{q^{21/4}}{a^3 y^{3/2}}+\frac{2 q^{11/2}}{a^2 y}-\frac{q^{23/4}}{a \sqrt{y}}-\frac{q^{25/4}}{a^3 y^{3/2}}
+\cdots
\\
\end{smallmatrix}
}
\\ \hline 
1
&
2
&
3
&
{\scriptscriptstyle 
\begin{smallmatrix}
1-\frac{2 q^{5/8}}{a \sqrt{t}}+\frac{q^{5/4}}{a^2 t}-\frac{2 q^{13/8}}{a \sqrt{t}}+\frac{4 q^{9/4}}{a^2 t}-\frac{2 q^{21/8}}{a \sqrt{t}}
-\frac{2 q^{23/8}}{a^3 t^{3/2}}+\frac{5 q^{13/4}}{a^2 t}-\frac{2 q^{29/8}}{a \sqrt{t}}-\frac{4 q^{31/8}}{a^3 t^{3/2}}+\cdots
\\
\end{smallmatrix}
}
\\ \hline 
1
&
3
&
4
&
{\scriptscriptstyle 
\begin{smallmatrix}
1-\frac{3 q^{5/8}}{a \sqrt{t}}+\frac{3 q^{5/4}}{a^2 t}-\frac{3 q^{13/8}}{a \sqrt{t}}-\frac{q^{15/8}}{a^3 t^{3/2}}+\frac{9 q^{9/4}}{a^2 t}-\frac{3 q^{21/8}}{a \sqrt{t}}-\frac{9 q^{23/8}}{a^3 t^{3/2}}
+\cdots
\\
\end{smallmatrix}
}
\\ \hline 
2
&
2
&
2
&
{\scriptscriptstyle 
\begin{smallmatrix}
1+\frac{2 \sqrt{q}}{a^2 y}-\frac{q^{3/4} \left(4 \left(u^2+1\right)\right)}{\sqrt{y} (a u)}+q \left(\frac{2}{a^4 y^2}+u^2+\frac{1}{u^2}+1\right)
\\
-\frac{q^{5/4} \left(4 \left(u^2+1\right)\right)}{y^{3/2} \left(a^3 u\right)}+\frac{2 q^{3/2} \left(y^2 \left(a^4 u^4\right)+u^2 \left(\left(7 a^4\right) y^2+1\right)+a^4 y^2\right)}{y^3 \left(a^6 u^2\right)}
\\
-\frac{q^{7/4} \left(4 \left(u^2+1\right)\right) \left(\left(2 a^4\right) y^2+1\right)}{y^{5/2} \left(a^5 u\right)}+q^2 \left(u^2 \left(\frac{2}{a^4 y^2}+1\right)+\frac{\frac{2}{a^4 y^2}+1}{u^2}+\frac{10}{a^4 y^2}+\frac{2}{a^8 y^4}+2\right)
+\cdots
\\
\end{smallmatrix}
}
\\ \hline 
3
&
3
&
2
&
{\scriptscriptstyle 
\begin{smallmatrix}
1+\frac{\sqrt[4]{q}}{a^3 y^{3/2}}+\frac{\sqrt{q} \left(\left(3 a^4\right) y^2+1\right)}{a^6 y^3}
+\frac{q^{3/4} \left(-y^4 \left(\left(6 a^8\right) u^2\right)-y^4 \left(\left(6 a^8\right) u\right)+y^2 \left(\left(3 a^4\right) u\right)-\left(6 a^8\right) y^4+u\right)}{y^{9/2} \left(a^9 u\right)}
\\
+q \left(u \left(2-\frac{6}{a^4 y^2}\right)+\frac{2-\frac{6}{a^4 y^2}}{u}-\frac{9}{a^4 y^2}+\frac{3}{a^8 y^4}+\frac{1}{a^{12} y^6}+u^2+\frac{1}{u^2}+2\right)
\\
+\frac{q^{5/4} \left(y^6 \left(a^{12} u^4\right)+\left(a^4 y^2-3\right) \left(y^4 \left(\left(2 a^8\right) u^3\right)\right)+u^2 \left(\left(3 a^{12}\right) y^6-\left(9 a^8\right) y^4+\left(3 a^4\right) y^2+1\right)+\left(a^4 y^2-3\right) \left(y^4 \left(\left(2 a^8\right) u\right)\right)+a^{12} y^6\right)}{y^{15/2} \left(a^{15} u^2\right)}
+\cdots
\\
\end{smallmatrix}
}
\\ 
\end{array}
\end{align}

%%%%%%%%%%%%%%%%%%%%%%%%%%%%%%%%%%%%%%%%%%%
\subsubsection{$SO(N_c)-[N_f]_{-+}$ with $\mathcal{D}, (\mathrm{D,D,D; D,D,D})$}
\label{SO-+_D_check}
%%%%%%%%%%%%%%%%%%%%%%%%%%%%%%%%%%%%%%%%%%%

\begin{align}
\begin{array}{c|c|c|c}
N_c
&
N_f
&
\tilde{N}_c
&
%\textrm{series expansions}
\mathbb{II}^{SO(N_c)-[N_f]^A_{-+}}_{\mathcal{D}, \mathrm{D}}
=
\mathbb{II}^{SO(\tilde{N}_c)-[N_f]^B_{--}}_{\mathcal{N}, \mathrm{N},\mathrm{D},\mathrm{N}}
\\ \hline
1
&
1
&
2
&
{\scriptscriptstyle 
\begin{smallmatrix}
1-\frac{q^{3/4}}{a \sqrt{y}}-\frac{q^{7/4}}{a \sqrt{y}}+\frac{q^{5/2}}{a^2 y}-\frac{q^{11/4}}{a \sqrt{y}}+\frac{q^{7/2}}{a^2 y}-\frac{q^{15/4}}{a \sqrt{y}}+\frac{2 q^{9/2}}{a^2 y}-\frac{q^{19/4}}{a \sqrt{y}}-\frac{q^{21/4}}{a^3 y^{3/2}}+\frac{2 q^{11/2}}{a^2 y}-\frac{q^{23/4}}{a \sqrt{y}}-\frac{q^{25/4}}{a^3 y^{3/2}}+\frac{3 q^{13/2}}{a^2 y}+\cdots
\\
\end{smallmatrix}
}
\\ \hline 
1
&
2
&
3
&
{\scriptscriptstyle 
\begin{smallmatrix}
1-\frac{2 q^{5/8}}{a \sqrt{t}}+\frac{q^{5/4}}{a^2 t}-\frac{2 q^{13/8}}{a \sqrt{t}}+\frac{4 q^{9/4}}{a^2 t}-\frac{2 q^{21/8}}{a \sqrt{t}}-\frac{2 q^{23/8}}{a^3 t^{3/2}}+\frac{5 q^{13/4}}{a^2 t}-\frac{2 q^{29/8}}{a \sqrt{t}}-\frac{4 q^{31/8}}{a^3 t^{3/2}}+\cdots
\\
\end{smallmatrix}
}
\\ \hline 
1
&
3
&
4
&
{\scriptscriptstyle 
\begin{smallmatrix}
1-\frac{3 q^{5/8}}{a \sqrt{t}}+\frac{3 q^{5/4}}{a^2 t}-\frac{3 q^{13/8}}{a \sqrt{t}}-\frac{q^{15/8}}{a^3 t^{3/2}}+\frac{9 q^{9/4}}{a^2 t}-\frac{3 q^{21/8}}{a \sqrt{t}}-\frac{9 q^{23/8}}{a^3 t^{3/2}}
+\cdots
\\
\end{smallmatrix}
}
\\ \hline 
2
&
2
&
2
&
{\scriptscriptstyle 
\begin{smallmatrix}
1-\frac{2 \sqrt{q}}{a^2 y}+q \left(\frac{2}{a^4 y^2}-u^2-\frac{1}{u^2}+1\right)+\frac{2 q^{3/2} \left(y^2 \left(a^4 u^4\right)-u^2 \left(\left(3 a^4\right) y^2+1\right)+a^4 y^2\right)}{y^3 \left(a^6 u^2\right)}
\\
+q^2 \left(u^2 \left(-\frac{2}{a^4 y^2}-1\right)+\frac{-\frac{2}{a^4 y^2}-1}{u^2}+\frac{10}{a^4 y^2}+\frac{2}{a^8 y^4}+2\right)
\\
+\frac{2 q^{5/2} \left(u^4 \left(a^4 y^2+\left(3 a^8\right) y^4\right)-u^2 \left(\left(5 a^8\right) y^4+\left(5 a^4\right) y^2+1\right)+a^4 y^2+\left(3 a^8\right) y^4\right)}{y^5 \left(a^{10} u^2\right)}
+\cdots
\\
\end{smallmatrix}
}
\\ \hline 
3
&
3
&
2
&
{\scriptscriptstyle 
\begin{smallmatrix}
1-\frac{\sqrt[4]{q}}{a^3 y^{3/2}}+\frac{\sqrt{q} \left(1-\left(3 a^4\right) y^2\right)}{a^6 y^3}+\frac{q^{3/4} \left(\left(3 a^4\right) y^2-1\right)}{a^9 y^{9/2}}
\\
+q \left(\frac{\left(3 a^8\right) y^4-\left(3 a^4\right) y^2+1}{a^{12} y^6}-u^2-\frac{1}{u^2}\right)
\\
+\frac{q^{5/4} \left(y^6 \left(a^{12} u^4\right)-u^2 \left(a^{12} y^6+\left(3 a^8\right) y^4-\left(3 a^4\right) y^2+1\right)+a^{12} y^6\right)}{y^{15/2} \left(a^{15} u^2\right)}+\cdots
\\
\end{smallmatrix}
}
\\ 
\end{array}
\end{align}

%%%%%%%%%%%%%%%%%%%%%%%%%%%%%%%%%%%%%%%%%%%
\subsubsection{$SO(N_c)-[N_f]_{+-}$ with $\mathcal{D}, (\mathrm{D,D,D; D,D,D})$}
\label{SO+-_D_check}
%%%%%%%%%%%%%%%%%%%%%%%%%%%%%%%%%%%%%%%%%%%

\begin{align}
\begin{array}{c|c|c|c}
N_c
&
N_f
&
\tilde{N}_c
&
%\textrm{series expansions}
\mathbb{II}^{SO(N_c)-[N_f]^A_{+-}}_{\mathcal{D}, \mathrm{D}}
=
\mathbb{II}^{SO(\tilde{N}_c)-[N_f]^B_{+-}}_{\mathcal{N}, \mathrm{N},\mathrm{D},\mathrm{N}}
\\ \hline
1
&
1
&
2
&
{\scriptscriptstyle 
\begin{smallmatrix}
1+\frac{q^{3/4}}{a \sqrt{y}}+\frac{q^{7/4}}{a \sqrt{y}}+\frac{q^{5/2}}{a^2 y}+\frac{q^{11/4}}{a \sqrt{y}}+\frac{q^{7/2}}{a^2 y}+\frac{q^{15/4}}{a \sqrt{y}}+\frac{2 q^{9/2}}{a^2 y}+\frac{q^{19/4}}{a \sqrt{y}}+\frac{q^{21/4}}{a^3 y^{3/2}}+\frac{2 q^{11/2}}{a^2 y}+\frac{q^{23/4}}{a \sqrt{y}}+\frac{q^{25/4}}{a^3 y^{3/2}}+\frac{3 q^{13/2}}{a^2 y}+\cdots
\\
\end{smallmatrix}
}
\\ \hline 
1
&
2
&
3
&
{\scriptscriptstyle 
\begin{smallmatrix}
1+\frac{2 q^{5/8}}{a \sqrt{t}}+\frac{q^{5/4}}{a^2 t}+\frac{2 q^{13/8}}{a \sqrt{t}}+\frac{4 q^{9/4}}{a^2 t}+\frac{2 q^{21/8}}{a \sqrt{t}}+\frac{2 q^{23/8}}{a^3 t^{3/2}}+\frac{5 q^{13/4}}{a^2 t}+\frac{2 q^{29/8}}{a \sqrt{t}}+\frac{4 q^{31/8}}{a^3 t^{3/2}}+\cdots
\\
\end{smallmatrix}
}
\\ \hline 
1
&
3
&
4
&
{\scriptscriptstyle 
\begin{smallmatrix}
1+\frac{3 q^{5/8}}{a \sqrt{t}}+\frac{3 q^{5/4}}{a^2 t}+\frac{3 q^{13/8}}{a \sqrt{t}}+\frac{q^{15/8}}{a^3 t^{3/2}}+\frac{9 q^{9/4}}{a^2 t}+\frac{3 q^{21/8}}{a \sqrt{t}}+\frac{9 q^{23/8}}{a^3 t^{3/2}}
+\cdots
\\
\end{smallmatrix}
}
\\ \hline 
2
&
2
&
2
&
{\scriptscriptstyle 
\begin{smallmatrix}
1-q-\frac{2 q^{3/2}}{a^2 y}+\frac{2 q^{5/2}}{a^2 y}+q^3 \left(\frac{1}{a^4 y^2}-1\right)-\frac{2 q^{7/2}}{a^2 y}+q^4 \left(1-\frac{1}{a^4 y^2}\right)+\frac{4 q^{9/2}}{a^2 y}
+\cdots
\\
\end{smallmatrix}
}
\\ \hline 
3
&
3
&
2
&
{\scriptscriptstyle 
\begin{smallmatrix}
1+\frac{\sqrt[4]{q}}{a^3 y^{3/2}}+\frac{\sqrt{q} \left(1-\left(3 a^4\right) y^2\right)}{a^6 y^3}+\frac{q^{3/4} \left(1-\left(3 a^4\right) y^2\right)}{a^9 y^{9/2}}
\\
+q \left(\frac{3}{a^4 y^2}-\frac{3}{a^8 y^4}+\frac{1}{a^{12} y^6}-u^2-\frac{1}{u^2}-2\right)+\frac{q^{5/4} \left(-y^6 \left(a^{12} u^4\right)-u^2 \left(a^4 y^2-1\right)^3-a^{12} y^6\right)}{y^{15/2} \left(a^{15} u^2\right)}
\\
+\frac{q^{3/2} \left(\frac{y^8 \left(\left(3 a^{16}\right) \left(u^4+1\right)\right)}{u^2}-\frac{y^6 \left(a^{12} \left(u^2+1\right)^2\right)}{u^2}+\left(3 a^8\right) y^4-\left(3 a^4\right) y^2+1\right)}{a^{18} y^9}+\cdots
\\
\end{smallmatrix}
}
\\ 
\end{array}
\end{align}

%%%%%%%%%%%%%%%%%%%%%%%%%%%%%%%%%%%%%%%%%%%
\subsubsection{$SO(N_c)-[N_f]_{--}$ with $\mathcal{D}, (\mathrm{D,D,D; D,D,D})$}
\label{SO--_D_check}
%%%%%%%%%%%%%%%%%%%%%%%%%%%%%%%%%%%%%%%%%%%

\begin{align}
\begin{array}{c|c|c|c}
N_c
&
N_f
&
\tilde{N}_c
&
%\textrm{series expansions}
\mathbb{II}^{SO(N_c)-[N_f]^A_{--}}_{\mathcal{D}, \mathrm{D}}
=
\mathbb{II}^{SO(\tilde{N}_c)-[N_f]^B_{-+}}_{\mathcal{N}, \mathrm{N},\mathrm{D},\mathrm{N}}
\\ \hline
1
&
1
&
2
&
{\scriptscriptstyle 
\begin{smallmatrix}
1+\frac{q^{3/4}}{a \sqrt{y}}+\frac{q^{7/4}}{a \sqrt{y}}+\frac{q^{5/2}}{a^2 y}+\frac{q^{11/4}}{a \sqrt{y}}+\frac{q^{7/2}}{a^2 y}+\frac{q^{15/4}}{a \sqrt{y}}+\frac{2 q^{9/2}}{a^2 y}+\frac{q^{19/4}}{a \sqrt{y}}+\frac{q^{21/4}}{a^3 y^{3/2}}+\frac{2 q^{11/2}}{a^2 y}+\frac{q^{23/4}}{a \sqrt{y}}+\frac{q^{25/4}}{a^3 y^{3/2}}+\frac{3 q^{13/2}}{a^2 y}
+\cdots
\\
\end{smallmatrix}
}
\\ \hline 
1
&
2
&
3
&
{\scriptscriptstyle 
\begin{smallmatrix}
1+\frac{2 q^{5/8}}{a \sqrt{t}}+\frac{q^{5/4}}{a^2 t}+\frac{2 q^{13/8}}{a \sqrt{t}}+\frac{4 q^{9/4}}{a^2 t}+\frac{2 q^{21/8}}{a \sqrt{t}}+\frac{2 q^{23/8}}{a^3 t^{3/2}}+\frac{5 q^{13/4}}{a^2 t}+\frac{2 q^{29/8}}{a \sqrt{t}}+\frac{4 q^{31/8}}{a^3 t^{3/2}}+\cdots
\\
\end{smallmatrix}
}
\\ \hline 
1
&
3
&
4
&
{\scriptscriptstyle 
\begin{smallmatrix}
1+\frac{3 q^{5/8}}{a \sqrt{t}}+\frac{3 q^{5/4}}{a^2 t}+\frac{3 q^{13/8}}{a \sqrt{t}}+\frac{q^{15/8}}{a^3 t^{3/2}}+\frac{9 q^{9/4}}{a^2 t}+\frac{3 q^{21/8}}{a \sqrt{t}}+\frac{9 q^{23/8}}{a^3 t^{3/2}}+\cdots
\\
\end{smallmatrix}
}
\\ \hline 
2
&
2
&
2
&
{\scriptscriptstyle 
\begin{smallmatrix}
1-q-\frac{2 q^{3/2}}{a^2 y}+\frac{2 q^{5/2}}{a^2 y}+q^3 \left(\frac{1}{a^4 y^2}-1\right)-\frac{2 q^{7/2}}{a^2 y}+q^4 \left(1-\frac{1}{a^4 y^2}\right)+\frac{4 q^{9/2}}{a^2 y}+\cdots
\\
\end{smallmatrix}
}
\\ \hline 
3
&
3
&
2
&
{\scriptscriptstyle 
\begin{smallmatrix}
1-\frac{\sqrt[4]{q}}{a^3 y^{3/2}}+\frac{\sqrt{q} \left(\left(3 a^4\right) y^2+1\right)}{a^6 y^3}+\frac{q^{3/4} \left(-y^4 \left(\left(6 a^8\right) u^2\right)+u \left(\left(6 a^8\right) y^4-\left(3 a^4\right) y^2-1\right)+\left(-6 a^8\right) y^4\right)}{y^{9/2} \left(a^9 u\right)}
\\
+q \left(u \left(\frac{6}{a^4 y^2}-2\right)+\frac{\frac{6}{a^4 y^2}-2}{u}-\frac{9}{a^4 y^2}+\frac{3}{a^8 y^4}+\frac{1}{a^{12} y^6}+u^2+\frac{1}{u^2}+2\right)
\\
+\frac{q^{5/4} \left(-y^6 \left(a^{12} u^4\right)+\left(a^4 y^2-3\right) \left(y^4 \left(\left(2 a^8\right) u^3\right)\right)-u^2 \left(\left(3 a^{12}\right) y^6-\left(9 a^8\right) y^4+\left(3 a^4\right) y^2+1\right)+\left(a^4 y^2-3\right) \left(y^4 \left(\left(2 a^8\right) u\right)\right)-a^{12} y^6\right)}{y^{15/2} \left(a^{15} u^2\right)}
+\cdots
\\
\end{smallmatrix}
}
\\ 
\end{array}
\end{align}

\bibliographystyle{utphys}
\bibliography{ref}

\end{document}